\documentclass[a4paper,10pt]{article}

\usepackage[utf8x]{inputenc}
\usepackage[margin=2.8cm]{geometry}
\usepackage{xspace,xcolor,amsmath,amssymb}
\usepackage{graphicx}
\usepackage[FIGTOPCAP,large,sf,bf,nooneline]{subfigure}
\usepackage{float,placeins,xspace}
\usepackage{authblk}
\usepackage{comment}

\title{Stable developmental patterns of gene expression\\ without morphogen gradients}

\author[1,2]{Maciej Majka}
\author[3,4]{Nils B. Becker}
\author[3]{Pieter Rein ten Wolde}
\author[1]{Marcin Zagorski}
\author[3,5*]{Thomas R. Sokolowski}

\affil[1]{Institute of Theoretical Physics and Mark Kac Center for Complex  Systems Research, Jagiellonian  University, ul. prof. Stanis\l{}awa   \L{}ojasiewicza 11, 30-348 Krak\'{o}w, Poland}
\affil[2]{Present address: 
Department of Physics, East Carolina University, Greenville NC, 27858, USA}
\affil[3]{FOM Institute AMOLF, Science Park 104, 1098 XG Amsterdam, The Netherlands}
\affil[4]{Present address: Theoretical Systems Biology, German Cancer Research Center, 69120 Heidelberg, Germany}
\affil[5]{Present address: Frankfurt Institute for Advanced Studies (FIAS), Ruth-Moufang-Stra\ss{}e 1, 60348 Frankfurt am Main, Germany}

\catcode`_=\active
\def_#1{\ensuremath{\sb{\mathrm{#1}}}}

\newcommand{\Drosophila}{{\it Drosophila}\xspace}
\newcommand{\hb}{{\it hb}\xspace}
\newcommand{\kr}{{\it kr}\xspace}
\newcommand{\kni}{{\it kni}\xspace}
\newcommand{\gt}{{\it gt}\xspace}
\newcommand{\Hb}{{\it Hb}\xspace}

\newcommand{\Bcd}{{\it Bcd}\xspace}
\newcommand{\Cad}{{\it Cad}\xspace}
\newcommand{\Tll}{{\it Tll}\xspace}
\newcommand{\GA}{A\xspace}
\newcommand{\GB}{B\xspace}
\newcommand{\GC}{C\xspace}
\newcommand{\GD}{D\xspace}
\newcommand{\kRon}{k^{\rm R}_{on}}
\newcommand{\kRoffS}{k_{\rm s}^{\rm off}}
\newcommand{\kRoffW}{k_{\rm w}^{\rm off}}
\newcommand{\kDon}{k^{\rm D}_{on}}
\newcommand{\kDoff}{k^{\rm D}_{off}}
\newcommand{\vecA}[1]{\vec{#1}}

\newcommand{\avg}[1]{\left\langle #1\right\rangle}
\newcommand{\unit}[1]{{\rm #1}}
\newcommand{\mum}{\unit{\mu m}}
\newcommand{\mumsps}{\unit{\mu m^2/s}}
\newcommand{\SI}{Supporting Information\xspace}
\newcommand{\mysubsubsection}[1]{\vspace{1EM}\noindent{\bf #1}\vspace{1EM}}

\newcommand{\myurl}[1]{\textsf{#1}} 

\newcommand{\changed}[1]{{\color{black}#1}}

\begin{document}

\maketitle
* Corresponding author, e-mail: sokolowski@fias.uni-frankfurt.de

\clearpage
\vspace*{6EM}
\begin{abstract}
Gene expression patterns in developing organisms are established by groups of cross-regulating target genes that are driven by morphogen gradients. As development progresses, morphogen activity is reduced, leaving the emergent pattern without stabilizing positional cues and at risk of rapid deterioration due to the inherently noisy biochemical processes at the cellular level. But remarkably, gene expression patterns remain spatially stable and reproducible over long developmental time spans in many biological systems. Here we combine spatial-stochastic simulations with an enhanced sampling method (Non-Stationary Forward Flux Sampling) and a recently developed stability theory to address how spatiotemporal integrity of a gene expression pattern is maintained in developing tissue lacking morphogen gradients. Using a minimal embryo model consisting of spatially coupled biochemical reactor volumes, we study a prototypical stripe pattern in which weak cross-repression between nearest neighbor expression domains alternates with strong repression between next-nearest neighbor domains, inspired by the gap gene system in the {\it Drosophila} embryo. We find that tuning of the weak repressive interactions to an optimal level can prolong stability of the expression patterns by orders of magnitude, enabling stable patterns over developmentally relevant times in the absence of morphogen gradients. The optimal parameter regime found in simulations of the embryo model closely agrees with the predictions of our coarse-grained stability theory. To elucidate the origin of stability, we analyze a reduced phase space defined by two measures of pattern asymmetry. We find that in the optimal regime, intact patterns are protected via restoring forces that counteract random perturbations and give rise to a metastable basin. Together, our results demonstrate that metastable attractors can emerge as a property of stochastic gene expression patterns even without system-wide positional cues, provided that the gene regulatory interactions shaping the pattern are optimally tuned.
\end{abstract}
\clearpage


\section{Introduction}

Maintaining the integrity of spatial gene expression patterns over time is essential in embryonic development. In early embryo development locally expressed morphogenetic proteins spread through the tissue to form gradients of chemical signals \cite{Gregor2007a,Rogers2011,Lander2011,Balaskas2012,Kicheva2012,Shvartsman2012,Briscoe2015,Bier2015,Stapornwongkul2020,Stapornwongkul2021,Simsek2022}. Inside developing cells, these chemical signals are interpreted by gene regulatory networks to form remarkably precise and reproducible spatial patterns of gene expression that subsequently give rise to different body parts and organs \cite{Gregor2007b,DubuisPNAS2011,Sokolowski2012,Raspopovic2014,Zagorski2017,Petkova2019,Kuzmicz-Kowalska2020,Tkacik2021,Exelby2021,Iyer2022,Sokolowski2023,Minchington2023}. However, as spatial patterns are established by reading out upstream morphogen gradients, their stability is constantly subject to inherently noisy cellular and extracellular processes \cite{Swain2002,Raser2005,Paulsson2005,Raj2008,Chalancon2012,Averbukh2017,Perez-Carrasco2016}. 
Moreover, the activity of morphogenetic gradients that is interpreted by target cells can decrease over developmental time. This decrease in activity can take different forms, including reduction of the relative signaling range as the embryo grows in size \cite{Zagorski2017,Kicheva2014,Almuedo-Castillo2018}, signalling pathway desensitization \cite{Cohen2015}, or complete disappearance of the gradients at later developmental stages \cite{Drocco2011,Durrieu2018}. Together, the inherent cellular stochasticity and reduced role of morphogen gradients at later stages raise the question whether stable patterns can be maintained over sufficiently long developmental times in the absence of morphogen gradients, and, if so, how.

Focusing on the cellular stochasticity, biological cells are facing two types of noise, namely intrinsic and extrinsic noise, with different notions of robustness against the respective noise types \cite{Swain2002,Raser2005,Paulsson2005,Raj2008,Chalancon2012,Averbukh2017,Perez-Carrasco2016}. Intrinsic noise originates from the processes of gene regulation, protein production, and intracellular transport. Thus,  robustness of spatial patterns to intrinsic noise amounts to buffering random fluctuations in the copy numbers of patterning proteins. Extrinsic noise, on the other hand, terms the variations originating from different external conditions including cell size variability \cite{Huh2011,Thomas2019}, cell-to-cell variation in ribosome abundance \cite{Raj2008} or fluctuations in the cellular environment \cite{Pedraza2005,Stockholm2010}. Therefore, the robustness of spatial patterns to extrinsic noise refers to the capability of producing precise patterns in spite of imperfect initial conditions, classically termed ``canalization'' in Waddington's picture of development \cite{Waddington1942,Waddington1959}. Several gene regulatory strategies providing either type of robustness have been studied \cite{Simsek2022,Iyer2022,Averbukh2017}, but our understanding of how nature orchestrates them in the fully interacting wild-type organism is still incomplete.

Among the regulatory mechanisms that drive developmental pattern formation, the regulatory motif in which two genes mutually repress each other is particularly prevalent \cite{Balaskas2012,Alon2007,Vakulenko2009,Cotterell2010,Burda2011,Raspopovic2014,Verd2017,Verd2019,Exelby2021,Perez-Carrasco2016}. Intriguingly, mutual repression can have a dual role in the establishment of spatial patterns. On the one hand, in systems driven by threshold-dependent activation of patterning genes via morphogen gradients, mutual repression is crucial for shaping out expression domains that are bounded from two sides, thus increasing the positional information carried by the expression pattern \cite{Sokolowski2012,Sokolowski2015,Zagorski2017,Exelby2021,Sokolowski2023}. On the other hand, mutual repression can induce bistability leading to stochastic switching between cell fates. Hence, it is a priori unclear to which extent mutual repression supports or counter-acts the formation of stable spatial patterns \cite{Sokolowski2012,Zagorski2017,Exelby2021}. This issue is particularly relevant to systems that lack external cues for symmetry breaking, such as morphogen or maternal gradients, that could force bistable cells into one of their opposing fates.

Here we ask whether a system of mutually interacting genes can maintain an initially arranged expression pattern in the absence of upstream input gradients. To address this question we study a spatially resolved gene regulatory network, conceptually motivated by the gap gene system in the early embryo of the fruit fly {\it Drosophila melanogaster} \cite{Jaeger2004, Jaeger2011, Dubuis2013, Manu2009PlosBiol, Surkova2008, Clyde2003}.
This system implements a particular regulatory architecture, 
in which weak and strong mutual repressive interactions between expression domains of different genes alternate depending on whether the domains are adjacent or not.
This motif, termed ``alternating cushions'', was earlier investigated in terms of stability and robustness against extrinsic noise in initial conditions \cite{Vakulenko2009}. That study employed a reaction-diffusion model with step-like activation functions for representing the underlying gene expression dynamics. Using the so-called ``moving kink approximation'', the study predicted an extensive basin of pattern stability in the parameter space of the model, where the stability could be attributed to repulsive forces between mutually repressing gene expression domains (``cushions''). More recently, an exact solution was obtained in an analogous model for the dynamics of the contact zones between two gene expression domains and for arbitrary combination of activating or repressing interactions between the involved genes \cite{Majka2023}. This work provided exact conditions for stability, leading to a better quantitative understanding of the conditions under which gene expression patterns can survive arbitrary long time. Importantly, it was shown that perfect pattern stability (i.e., a pattern surviving infinitely long) can only be achieved for a very specific combination of system parameters; nevertheless, in the vicinity of these states, there exists a continuity of well-defined but slowly changing gene expression patterns, which can fulfill their biological role for a finite but typically sufficiently long period of time. However, since the reaction-diffusion model considered in \cite{Majka2023} is only a continuous and deterministic limit of the genuinely stochastic microscopic dynamics of gene expression, it remained unclear whether the derived stability conditions provide useful insight into the regime of strong fluctuations. 

In this work, we assess the temporal stability of gene expression patterns interacting via the ``alternating cushions'' motif by numerical simulations of a minimal spatial-stochastic model that features a full microscopic representation of stochastic gene expression and protein diffusion, thus incorporating the relevant intrinsic noise sources.
Using Non-Stationary Forward Flux Sampling (NS-FFS) \cite{BeckerTenWolde2012,BeckerAllenTenWolde2012}, an enhanced biased sampling technique for stochastic systems changing in time, we quantify for how long patterns shaped and maintained only by mutual repression can self-sustain. 
Contrasting with previous approaches \cite{Vakulenko2009,Manu2009PlosCompBiol,Jaeger2004}, NS-FFS allows us to go beyond a local, linear stability analysis of the studied system, and to assess the depth and the width of the emerging basin of stability from large ensembles of stochastic trajectories of the full spatially interacting gene expression pattern.
Moreover, we derive the effective, deterministic model of simulated system that expands the stability theory from \cite{Majka2023} to the case of multiple interfaces and allows us to determine the parameter regime within which the distances between boundaries of adjacent gene expression domains are predicted to remain stable. Eventually, we employ this model to identify the mechanism enhancing the pattern survival time.

Our results show that the stability of patterns arranged in the alternating cushions scheme strongly varies with the strength of mutual repression between adjacent gene expression domains. 
We find that pattern stability time is significantly longer when spatially adjacent genes repress each other with intermediate strength and the next-nearest neighbor genes repress each other strongly. This results in a broad peak of pattern survival time for a range of interaction strength ratios, with a single maximum at the optimal choice. In this enhanced regime, we confirm the existence of robust restoring forces and find signatures of a metastable basin that stabilizes well-ordered patterns (dynamical attractor), in accordance with the previous findings of \cite{Vakulenko2009}. Away from the optimum, forces induced by strong nearest neighbor mutual repression destroy the stripe patterns rapidly, while for weaker nearest neighbor repression the forces are imperceptible when compared to stochastic fluctuations. We manage to explain these observations employing our deterministic, effective model and the recent exact stability theory. We determine the theoretical optimal interaction strength ratio, situated in the vicinity of the numerically predicted optimum. Further analysis reveals a nuanced interplay between fluctuations and a few stabilizing mechanisms present in the deterministic, effective model, leading to enhanced survival time in the vicinity of optimal choice and qualitatively in agreement with numerical observations. In result, we highlight the connection between effective restoration forces seen in simulations, moving-kink approximation model \cite{Vakulenko2009} and exact stability theory \cite{Majka2023}. Going beyond the setting studied in \cite{Vakulenko2009}, we also show that pinning of the pattern at the embryo boundaries, which could be achieved by very short-ranging, peripherally acting maternal inputs, can significantly further enhance the optimal pattern stability.

Taken together, we demonstrate that forces generated in the alternating cushions scheme can maintain the gene expression pattern subject to stochastic production and diffusion of proteins for extremely long times, thanks to the interplay between fluctuations and deterministic dynamics, constituting emergent noise-control mechanism for the close-to-optimal choice of mutual repression parameters. 

\clearpage

\begin{figure}[ht!]
  \centering
  \includegraphics[width=0.9\textwidth]{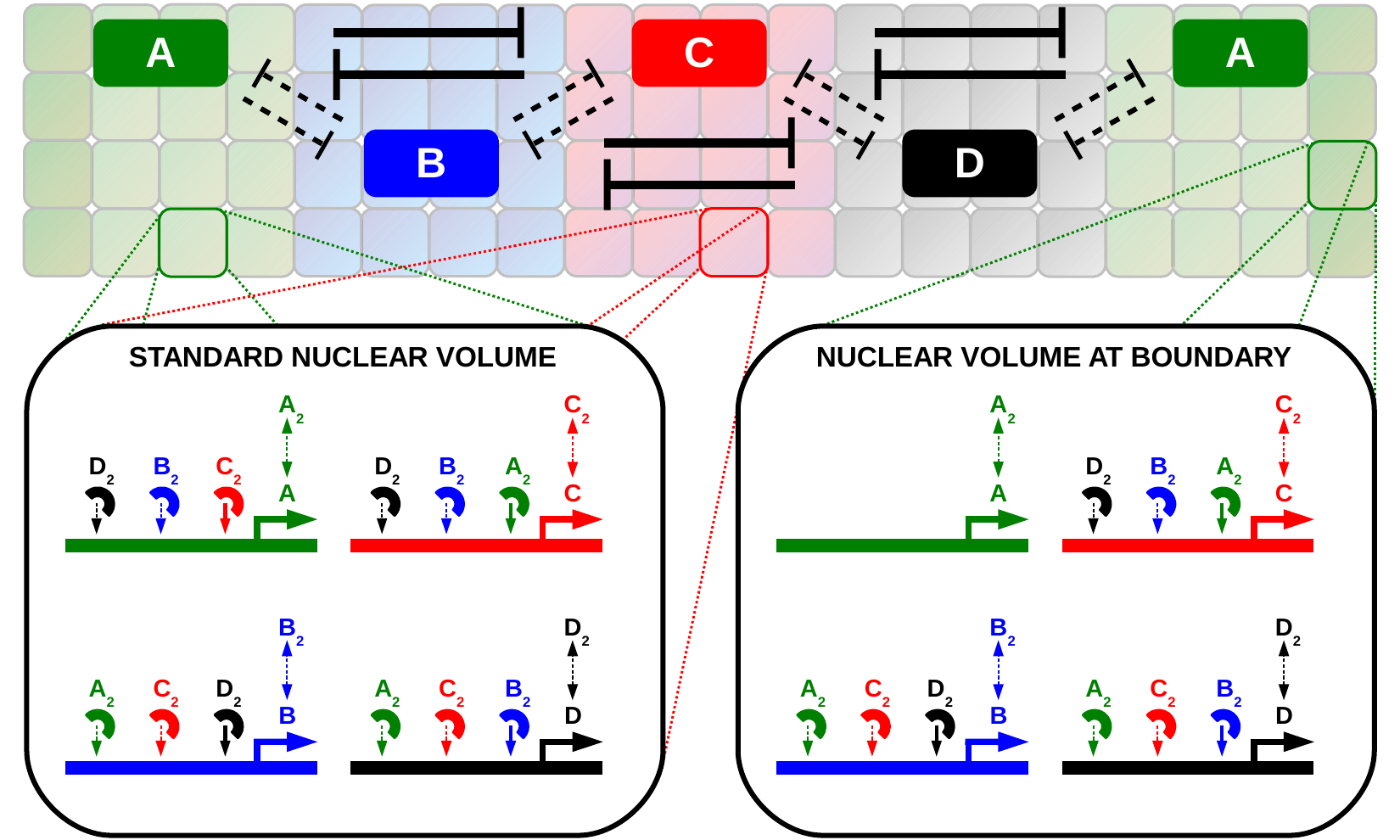}
  \caption{\textbf{Schematic of the spatial gene-regulatory model.}
      We use a cylindrical lattice of reaction volumes
      to mimic the arrangement of cortical nuclei in the posterior \Drosophila embryo at developmental cycle 14.
      In each nuclear volume (shaded squares) we simulate production, degradation, dimerization and mutual repression of the four genes \GA, \GB, \GC and \GD via the \changed{Gillespie} algorithm.
      Each gene is subject to repression by the protein dimers of the other genes,
      as indicated by the schematic promoters. Neighboring nuclei can exchange monomers and dimers via diffusive hopping. The system is initialized in a five-stripe pattern of expression domains in the order \GA--\GB--\GC--\GD--\GA, corresponding to the experimentally observed order in the fly embryo.
      The strength of mutual repression varies among gap gene pairs: genes associated with nearest neighbor (NN) domains repress each other weakly (dashed arrows), while next-nearest neighbors (NNN) domains exhibit strong mutual repression (thick arrows).
      By default, the concentration of \GA is pinned at the system boundary where
      the set of modeled reactions differs from the rest of the system by the fact that
      the \GA promoter can not be repressed.
      Details in the Methods, Sec. \ref{sec-GG-Methods}.
  \label{Fig-ModelSketch}
  }
\end{figure}

\mysubsubsection{Modeling framework}

\noindent
In order to investigate stability of gene expression patterns without external input gradients, we performed stochastic simulations of a spatial pattern of four mutually repressing genes, using NS-FFS.
Here we opted for a minimal spatially resolved stochastic model, shown in schematic Fig. \ref{Fig-ModelSketch}, inspired by the posterior gap gene pattern in \Drosophila development \cite{Jaeger2004}. The model considers four mutually interacting genes \GA, \GB, \GC and \GD, arranged in a five-stripe pattern (with order \GA-\GB-\GC-\GD-\GA) along a cylindrical spatial lattice. The four genes are analogous to the arrangement of the expression domains 
of gap genes \hb, \kr, \kni and \gt in nuclear cycle 14 in the posterior half of the early fly embryo, where \hb is expressed in two \changed{stripes}, in the first (anteriormost) and last (posteriormost) \changed{stripe} \cite{Jaeger2004, Jaeger2011, Dubuis2013, Manu2009PlosBiol, Surkova2008, Clyde2003}
\changed{; in the following, we use the term ``expression domain'' or just ``domain'' of gene \GA for referring to both of the \GA stripes together.}
The spatial lattice consists of $N_z\times N_\phi$ equally spaced and well-stirred reaction volumes with periodic boundary conditions in the circumferential ($\phi$-) direction motivated by the arrangement of cortical nuclei in the developing fly embryo.
Protein diffusion and nuclear exchange are modeled via hopping between neighboring reaction volumes, with a rate proportional to the diffusion coefficient.
In each nucleus, proteins of the genes \GA, \GB, \GC and \GD are produced from their corresponding promoters, dimerize and mutually repress each other by promoter binding.
Each gene can repress the promoter of each other gene.
Repression is non-competitive, i.e., each promoter has binding sites for each of the three other genes' dimers and is inactivated when at least one dimer is bound (``OR''-logics).
The model combines transcription and translation into one production step, neglecting some features of eukaryotic gene expression such as transcriptional bursts and enhancer dynamics, but previous work has shown that this does not alter the results qualitatively \cite{Erdmann2009,Sokolowski2012}.
\changed{
We provide a list of the biochemical equations governing the dynamics of our system in Sec.~S1.1 and and a graphical summary in Fig.~S12 of the Supporting Information.
}

In the anterior-posterior arrangement \GA-\GB-\GC-\GD-\GA, the genes repress each other mutually via the characteristic pattern of strong next-nearest neighbor (NNN) and weaker nearest-neighbor (NN) repression (alternating cushions), as observed in the \Drosophila embryo \cite{Vakulenko2009, Jaeger2004, Clyde2003, Kraut1991a, Kraut1991b, Eldon1991, Hulskamp1990, Jackle1986}.
Specifically, there are two pairs of strongly repressing genes, (\GA,\GC) and (\GB,\GD), and four pairs of genes that repress each other weakly, (\GA,\GB), (\GB,\GC), (\GC,\GD) and (\GD,\GA).
In our model, the difference in repression strength is tuned via the unbinding rates of the repressors from the repressed promoter.
\changed{
The strong-repressor unbinding rate $\kRoffS$ is set to a fixed value such that the NNN gene pairs (\GA,\GC) and (\GB,\GD) are in the bistable regime, while the weak-repressor unbinding rate $\kRoffW$ that tunes the repression between NN gene pairs is varied (ranging from very high unbinding rates, corresponding to very weak repression, towards rates as low as $\kRoffS$, which also brings the NN repressive interactions into the bistable regime). Here being in the bistable regime means that in individual nuclei only one of the two strongly repressing genes can be expressed at a high level, while its counterpart is expressed at very low level, e.g. the state in which \GA is expressed at high and \GC at very low levels, or vice versa. Since the repression between the two genes is assumed to be symmetric, i.e. \GA and \GC unbind from each other's regulatory region with the same rate $\kRoffS$ (and likewise for \GB and \GD), these two mutually exclusive stable expression states are equally probable without any further inputs that could break their symmetry, and therefore form a perfectly symmetric "genetic switch".
Thus, in the absence of external cues capable of forcing the bistable systems into a preferred state, stochastic switching is expected to eventually result in one of the domains to dominate over the respective other domain in the NNN pair, causing its elimination and simultaneous expansion of the dominating gene's domain.} This partial breakdown of the initial pattern can happen independently for both strongly repressing NNN pairs and thus in random temporal order; however, ultimately one of the strong repression partners is eliminated in each of the NNN pairs and the system settles in a new, effectively irreversible state in which only the remaining two genes are co-expressed.

On the one hand, we expect that the presence of the third expression domain in between the NNN pair domains can impede elimination of (one of) the NNN pair domains when additional NN repression is present, because it can spatially move apart the strongly repressing (NNN) expression domains and form a "cushion" domain between them, effectively replacing one interface of strong competition by two interfaces of weak competition that allow for local coexistence of the competitors.
On the other hand, overly strong NN repression is expected to enhance pattern breakdown because then even the overlapping NN expression domains are brought towards the bistable regime.
We therefore study the pattern stability as a function of the \emph{repression strength ratio} $\kappa$, defined as
\begin{align}
 \kappa \equiv \kRoffW / \kRoffS	\;,
 \label{Eq-kappa}
\end{align}
where $\kRoffW$ and $\kRoffS$ are the repressor unbinding rates for weakly repressing NN pairs and strongly repressing NNN pairs, respectively. $\kappa$ is varied through the weak repression unbinding rate $\kRoffW$.
\changed{
For $\kappa = 1$, i.e. $\kRoffW = \kRoffS$, both the NNN and the NN gene pairs are deeply in the bistable regime and repress each other strongly (because both $\kRoffS$ and $\kRoffW$ are low), while in the opposite limit $\kappa \rightarrow \infty$ ($\kRoffW \rightarrow \infty$) only the NNN gene pairs form bistable switches whereas the NN pairs do not affect each other at all.
}

\section{Results}

\mysubsubsection{Pattern stability is quantified by asymmetry factors}

\begin{figure}[ht!]
  \centering
  \includegraphics[width=0.75\textwidth]{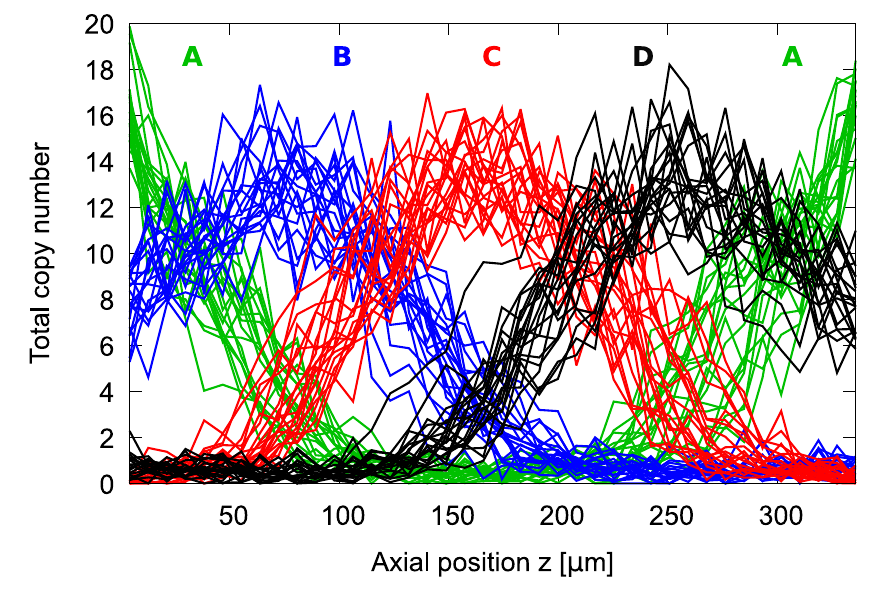}
  \caption{\textbf{Spatial pattern of gene expression.}
      Snapshots of the total copy numbers of all considered patterning proteins
      as a function of the axial coordinate $z$ of the cylinder, averaged over its circumference.
      Colors correspond to Fig. \ref{Fig-ModelSketch} (green = \GA, blue = \GB, red = \GC, black = \GD).
      Snapshots were taken every $60~\unit{min}$ over a total simulated time of $20~\unit{h}$ after
      an initial relaxation phase of $30~\unit{min}$, starting from rectangular domain profiles
      of equal length. No-flux boundary condition at either end.
  \label{Fig-TypicalOutput}
  }
\end{figure}

\noindent
\changed{
In order to quantify pattern stability, here we define how we understand pattern collapse and construct order parameters that track pattern destruction by mapping the pattern dynamics onto a low-dimensional phase space.
A typical ``intact'' spatial pattern of gene expression with (roughly) equally-sized domains is shown in Fig. \ref{Fig-TypicalOutput}.
We consider patterns in which the expression domain of one gene is lost completely as being ``destroyed'' (note that in our terminology the expression domain of gene \GA refers to both stripes at the system boundaries).
In our system, the strong mutual NNN repression and resulting bistability effectively prohibit coexistence of the strongly repressing genes at one location. Hence, an increase in the size of one domain is always accompanied by a reduction in the size of the domain belonging to the strongly interacting partner.
This lead us to introduce the following two order parameters, $\lambda_{AC}$ and $\lambda_{BD}$, here termed {\it asymmetry factors}, that measure the asymmetry of the expression domain sizes for each of the two strongly antagonistic NNN pairs:}
\begin{align}
 \lambda_{AC} \equiv \max([\GA]_{tot},[\GC]_{tot}) / N	\,, \qquad
 \lambda_{BD} \equiv \max([\GB]_{tot}, [\GD]_{tot}) / N	\,. \label{eq:lambdaACBD}
\end{align}
Here $[P]_{tot}$ is the \emph{total} copy number of P proteins (counting dimers twice), and $N=[\GA]_{tot} + [\GB]_{tot} + [\GC]_{tot} + [\GD]_{tot}$ is the total protein number in the system across all species.

In the spatially well-ordered pattern each protein domain occupies roughly the same fraction of the system, such that $\lambda_{AC} \simeq \lambda_{BD} \simeq 0.25$.
As expansion of a domain progresses at the expense of its strong antagonist, $\lambda_{AC}$ (or $\lambda_{BD})$ is enlarged and reaches values around $0.5$ when the shrinking domain is eventually lost.
In order to track progress of complete pattern losing one of its domains, we use the sum $\lambda = \lambda_{AC} + \lambda_{BD}$, with values around $0.5$ for five-stripe patterns and values above $0.75$ indicating pattern breakdown.

Our initial simulations revealed that even for very low protein copy numbers ($\lesssim 20$) the waiting times until one domain is lost are long compared to the duration of the actual breakdown event, and therefore difficult to sample by direct simulation.
\changed{
This lead us to ask whether the pattern breakdown is merely a slow random process akin to an unbiased random walk in configuration space, or whether the patterns have intrinsic restoring capabilities which would counteract the breakdown process; in such scenario, many more (counteracted) random attempts would be required for concluding the pattern breakdown, effectively rendering it a barrier crossing problem in which the transition states towards destroyed patterns form the barrier.
}
In order to resolve which of these alternative mechanisms is responsible for the stabilization of the expression pattern, we combined our stochastic simulations with Non-Stationary Forward Flux Sampling (NS-FFS), which is particularly suited for enhanced sampling of non-equilibrium rare events.
We used $\lambda$ as the progress coordinate for NS-FFS, which aims at generating a branched and weighted trajectory ensemble that, in the most favorable cases, samples the relevant $\lambda$-range uniformly. This allowed us to generate sufficient statistics of rare breakdown events even in the most stable regions of parameter space (see Methods, Sec. \ref{sec-GG-Methods}) .

\changed{
The initial simulations also showed that, in the regime of significant NN repression (small $\kappa$), the expression domains of gene \GA at the boundaries of the system (green expression domains in Fig.~\ref{Fig-TypicalOutput}) are particularly prone to destruction by their opponent domains, as we hypothesized for two reasons:
Firstly, they can expand only in one direction, towards the interior of the system; thus, unlike all the other domains, they cannot compensate shrinkage of the domain at one interface by an expansion at the other interface, and therefore take less effort to get completely destroyed. Secondly, their NN domains (B and D) are not counteracted by any strongly repressing partner at their interfaces with the A domain, which makes it easier for them to invade the A domain. We therefore decided to study a setup in which we "pin" (keep constant) the level of A proteins at the system boundaries by locally disallowing repression in the first and last rings of reaction volumes along the $z$-axis. This setup is not merely an {\it ad hoc} modeling assumption but motivated by experimental findings in the \Drosophila gap gene system, which arguably is the most widely studied example of the alternating cushions arrangement. There, the anterior stripe of the gap gene {\it hunchback} (\hb), which corresponds to gene \GA in our system, not only is under stringent control by {\it Bicoid} (\Bcd) but additionally translated from maternal mRNA localized towards the anterior pole \cite{Driever1989,Porcher2010a}; 
conversely, in the posterior, zygotic \hb expression is driven by a second enhancer under the control of {\it Tailless} (\Tll) \cite{Margolis1995},
which in turn is directly controlled by the maternal terminal system and thus tightly localized \cite{Casanova1990, Weigel1990}.
Our “pinning” prescription mimics this biological situation. To assess how the assumed pinning influences our results, we later compare to simulations in which expression of A can be repressed at the system boundaries, finding our main results hold up also in this less restricted system. In particular, pinning is not necessary for enhanced stability but can further increase the maximal stability time by more than an order of magnitude compared to the system without pinning, as we present further below.}


\mysubsubsection{Long-term pattern stability requires optimal repression strengths}

\begin{figure}[ht!]  
  \centering
  \includegraphics[width=\textwidth]{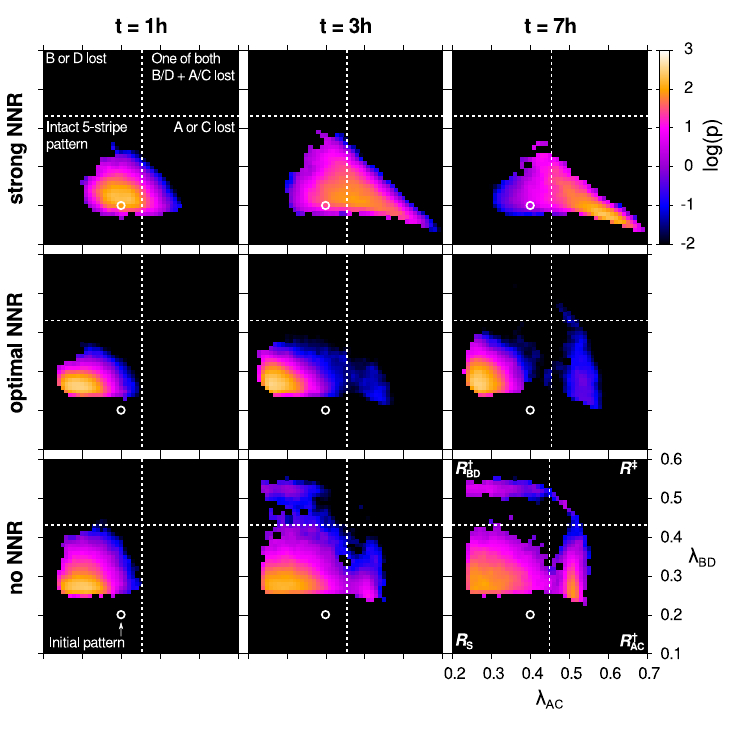}
  \caption{\textbf{Pattern breakdown in the phase space spanned by asymmetry factors.}
      Probability density snapshots of the phase space spanned by asymmetry factors $\lambda_{AC}$ and $\lambda_{BC}$, defined in \eqref{eq:lambdaACBD}, 
      at different times $t$ for varied repression strength ratio $\kappa$. The conditions are following: strong NN repression, $\kappa=3.16$ (top row), optimal NN repression for pattern stability, $\kappa=31.6$ (middle row), and lack of NN repression, $\kappa=\infty$ (bottom row). The simulation was started with the initial rectangular five-stripe \GA-\GB-\GC-\GD-\GA pattern $(\lambda_{AC}, \lambda_{BD})$=$(0.4,0.2)$ (white circle) in the pinned system. All snapshots are normalized histograms of reweighted $(\lambda_{AC}, \lambda_{BD})$-points within $t\pm5~\unit{min}$.
      In the middle and bottom rows we identify three densely populated regions:
      a broad region centered around $(0.30,0.30)$, $R_S$, which contains five-stripe patterns,
      and two smaller regions close to $(0.55, 0.30)$, $R^\dagger_{AC}$, and $(0.30, 0.55)$, $R^\dagger_{BD}$,
      representing patterns with one domain lost (region boundaries (dashed white), details in Methods,  Sec. \ref{sec-GG-Methods}). Ultimately trajectories will converge towards region centered around $(0.55, 0.55)$, $R^\ddagger$, where two domains are lost. 
  \label{Fig-DensityPinned}
  }
\end{figure}

\noindent
In order to see how varied repression strength affects pattern stability, we reweighted histograms of simulated trajectories over the reduced phase space spanned by order parameters $\lambda_{AC}$ and $\lambda_{BD}$ at different times, for different values of $\kappa$ ranging from strong NN repression ($\kappa\simeq 3$) to the limit of non-interacting nearest neighbors ($\kappa=\infty$), see Fig. \ref{Fig-DensityPinned}. We found that there exists a region of stable expression patterns in phase space which is populated rapidly and then remains quasi-stationary, indicating that the system can remain in a metastable state if NN repression is moderate. In particular, the velocity with which the system escapes from the quasi-stationary region strongly depends on $\kappa$, with low and very high $\kappa$ resulting in quick pattern deterioration, and intermediate $\kappa$ values resulting in the most long-lived quasi-stationary states.

\pagebreak

\begin{figure}[ht!]
  \centering
    \includegraphics[width=0.8\textwidth]{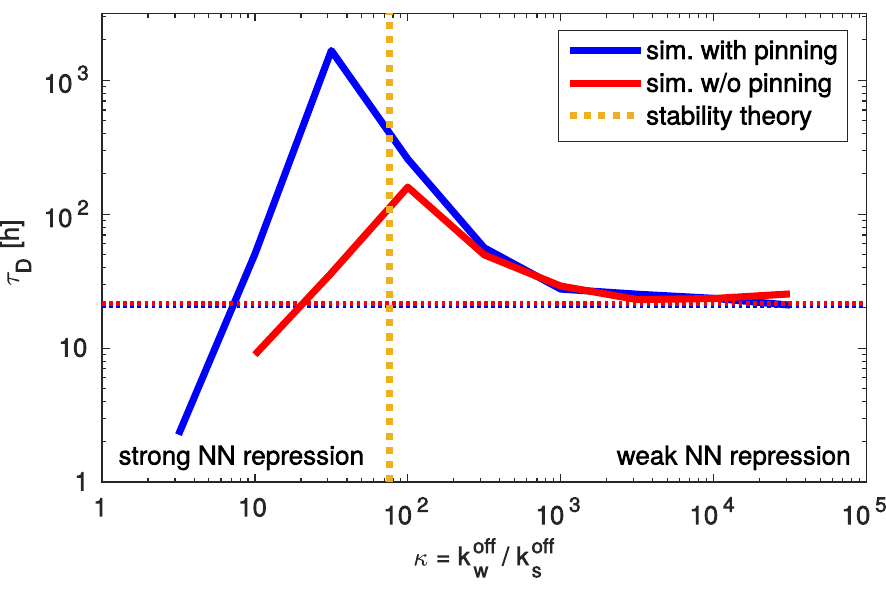}
    \label{Fig-StabilityGrouped}
  \caption{
\textbf{An optimal strength of nearest neighbor repression maximizes pattern stability.}
	  The mean time until pattern destruction $\tau_D$ as a function of $\kappa$, 
	  the ratio between the weak and strong repressor off-rate, for the system in which expression 
	  of gene $\GA$ is fixed at the boundaries.
	  We observe a pronounced maximum of the stability time when the weak repression is about
	  30 times weaker than the strong repression ($\kappa_{\rm opt} = 31.6$) in the system with pinning at the boundaries (blue line).
   When pinning of the pattern at the system boundaries is relaxed (red line), the maximum of stability time moves to $\kappa = 100$.
   The dashed horizontal lines indicate the values for the completely uncoupled systems with $\kappa = \infty$.
   The dashed vertical line (orange) shows the optimal repression strength ratio predicted analytically by our stability theory, $\kappa_{\rm theor} \simeq 76$ (see last part of Results section).
  \label{Fig-Stability}
}
\end{figure}

Stochastic fluctuations can lead to two different events corresponding to partial pattern destruction:
one in which either the \GA or \GC domain is lost first and one in which either the \GB or \GD domain is lost first. Motivated by these observations we defined a region of stable patterns in terms of the asymmetry factors as $R_S \equiv \{(\lambda_{AC}, \lambda_{BD}) \arrowvert \lambda_{AC} \leq 0.45 \text{ and } \lambda_{BD} \leq 0.43 \}$, see Fig.~\ref{Fig-DensityPinned}. States that lie outside of $R_S$ are considered deteriorated patterns, and accordingly we also defined two regions $R^\dagger_{AC}$ and $R^\dagger_{BD}$ and a region $R^\ddagger$ accumulating patterns with one expression domain lost and patterns with two domains lost, respectively. The pattern survival probability $S(t)=\iint_{R_S} p(\lambda_{AC}, \lambda_{BD},t) d\lambda_{AC}d\lambda_{BD}$ is the probability for the system to remain in the region of stable patterns until time $t$. We have never observed re-entry into $R_S$. We found that $S(t)$ is well-described by an exponential decay, $S(t)\propto e^{-k_D t}$, for times $t$ larger than a certain lag-time $t_{lag}$. $k_D$ then defines a deterioration rate, corresponding to average pattern stability time or the mean time until pattern has lost one of its domains,  $\tau_D \equiv 1/k_D$ (see Methods, Sec. \ref{sec-GG-Methods}).

By quantifying pattern stability time, we found that $\tau_D$ depends strongly on the repression strength ratio, with a maximum of $\tau_D$ as a function of $\kappa$ at $\kappa_{opt}\simeq30$, see Fig. \ref{Fig-Stability} (blue curve). For $\kappa$ values close to $\kappa_{opt}$ pattern stability is still markedly enhanced.
While significantly less stable than in the region around the optimum, patterns with stability time on the order of several hours remain possible in the absence of NN repression ($\kappa\rightarrow\infty$). In contrast, when NN and NNN repression have close to equal strength ($\kappa\rightarrow 1$) patterns collapse almost immediately.

\changed{
Examples of individual trajectories leading to (partial) pattern destruction as they proceed in biased simulation time (with increasing $\lambda$) are described in Section~S1.5 of the \SI and shown in SI-Figs.~S7--S9, for the most stable regime, i.e. for $\kappa=\kappa_{\rm opt}$ in the system with pinning. These examples demonstrate that multiple destruction pathways are possible in which the individual domains are destroyed in different order, and that destruction of one domain of a strongly competing gene pair can (but does not need to) facilitate subsequent destruction of a domain in the other strongly competing pair.
}

\mysubsubsection{In the maximally stable regime restoring forces reconstitute perturbed patterns}

\noindent
The observation of a phase space region in which system trajectories persist for long times raises the question whether this region constitutes a true metastable basin of attraction.
We first addressed this question next by analyzing transient behavior of the perturbed patterns.
If enhanced phase space density in certain regions of the $(\lambda_{AC}, \lambda_{BD})$-space were indeed due to the presence of a metastable basin, perturbations that transiently drive the system away from the stable pattern should be counteracted by restoring forces.
To test this hypothesis, we perturbed relaxed five-stripe patterns from the hypothetical basin by artificially enlarging domains in which one gap gene is dominant. Using these perturbed states as initial conditions, we then ran the spatial-stochastic simulator with higher time resolution, and checked whether the perturbed systems relax back into the presumed basin.
We investigated two types of asymmetric perturbations: ``\GC expansion'', in which the central \GC domain is unidirectionally expanded at the expense of the posterior \GA domain, and the converse ``\GA expansion'', in which the anterior \GA domain is enlarged at the expense of the \GC domain. The perturbation experiments are described in detail in the Methods, Sec.~\ref{sec-GG-Methods}, and \changed{Sec.~S1.2 and Fig.~S1 of} the Supporting Information.

We find that at $\kappa=\kappa_{opt}$, for both perturbations the perturbed pattern ensembles relax back to their original positions on a timescale $\sim 10~h$ (see Supporting Fig.~S1). This demonstrates that for optimal repression strength ratio an effective restoring force counteracts deviations from the five-stripe pattern for varied $\lambda_{AC}$.
Moreover, this suggests that the probability-enriched region within $R_S$ is a real metastable state confined by an underlying force field.
In accordance, the timescale of relaxation is orders of magnitude shorter than the timescale of pattern collapse.
Thus, for $\kappa=\kappa_{opt}$ pattern destruction is a Markovian transition between metastable basins with transition waiting times much longer than the timescales of intra-basin dynamics.
In contrast, we could not observe clear restoring behavior in the systems with very weak or no nearest neighbor interaction.
Here perturbations of similar strength tend to result in almost immediate pattern destruction.

In summary, for the repression strengths ratio $\kappa_{opt}\simeq30$ that maximizes stability, pattern breakdown appears to be an activated process characterized by a restoring force towards the initial state.

\begin{figure}[h!]
  \centering
  \makebox[\textwidth][c]{
  \includegraphics[width=1.0\textwidth]{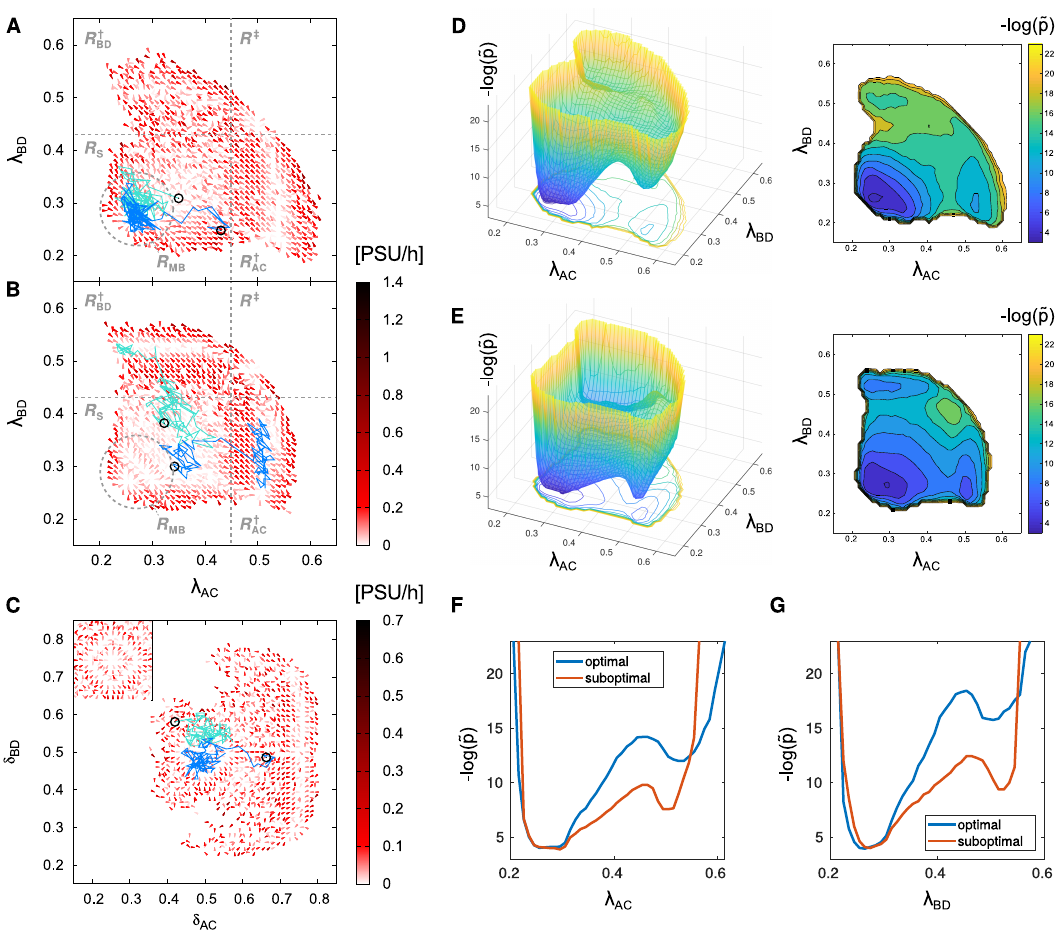}
  }
  \caption{
  \textbf{Phase space velocity fields and passage statistics reveal metastable basins.}
  \textbf{(A, B)} Average phase space velocity fields for the system with optimal ($\kappa=\kappa_{opt}$, A) and suboptimal ($\kappa=1000$, B) repression strength ratio, in the phase space spanned by asymmetry factors $\lambda_{AC}$ and $\lambda_{BD}$ as defined in \eqref{eq:lambdaACBD}. The small subregion with concentrically inwards-pointing velocities towards which perturbed trajectories relax, corresponding to metastable basin of five-stripe patterns, is indicated ($R_{MB}$, dashed circle). Velocity fields were obtained by averaging displacements of all trajectories that exit the local bin (see Methods, Sec. \ref{sec-GG-Methods}). Two examples of trajectories relaxing after perturbations are shown (blue lines = pert. from boundary, turquoise lines = pert. from center) with their starting points (circles). The boundaries of phase space regions (thin dashed lines) are as in Fig.~\ref{Fig-DensityPinned}. Velocity magnitude is indicated with colors.
  \textbf{(C)} The velocity field corresponding to $\kappa=\kappa_{opt}$ for the alternative asymmetry factors (``differences'') $\delta_{AC}$ and $\delta_{BD}$, as in \eqref{eq:deltaACBD}. 
  \changed{
  In C the metastable basin $R_{MB}$ is localized around the ``center point'' $(\delta_{AC},\delta_{BD})=(\frac{1}{2},\frac{1}{2})$, corresponding to an intact pattern with equal proportions of strongly competing genes. 
  For additional clarity, the inset in the upper left corner shows this region without the relaxing trajectories.
  Note the almost concentric pattern of velocity vectors pointing towards the center, highlighting the presence of the metastable basin.
  }
  The magnitude unit ``phase space unit per hour'' (PSU/h) is specific to the chosen asymmetry factors.
  \textbf{(D, E)} The landscapes of the ``pseudopotential'' $-\log \tilde{p}$ computed from the total number of phase space trajectories registered in the respective bin of the phase space.
  The contour plots to the right of the 3D views show a projected view of the same landscapes.
  \textbf{(F, G)} Comparison of sections in $\lambda_{AC}$ and $\lambda_{BD}$ directions, respectively, at $\lambda_{\perp}=0.28$ between the optimal and suboptimal choice of the repression strength ratio $\kappa$. Here the $-\log \tilde{p}$ profile is almost identical in the metastable basin $R_{MB}$, but transitions towards the destroyed pattern states face a higher barrier in the system with optimal $\kappa = \kappa_{opt}$, in both phase space directions.
  \vspace{-1EM}
  \label{Fig-VelocitiesComparison}
  }

\end{figure}


\mysubsubsection{Statistical analysis of phase-space dynamics reveals a metastable basin}

\noindent
We further figured that the existence of a true metastable basin should manifest itself also in the statistics of transient dynamics in phase space.
Here the local velocities in the $(\lambda_{AC}, \lambda_{BD})$ phase space are particularly informative:
forces that drive trajectories back into basins of attraction should translate into local mean phase space velocities with a clear bias towards the bottom of the basin.

To extract the velocity field for our system we modeled the coarse-grained pattern dynamics as overdamped  diffusive motion in the $\vecA \lambda \equiv (\lambda_{AC}, \lambda_{BD})$ plane, assuming that these degrees of freedom capture the slowest time scales of the system and making a Markov approximation for the fast dynamics \cite{Zwanzig1961, Mori1965}.  
This technique has been successfully applied in protein folding \cite{Yang2006, Kopelevich2005, Hummer2003, Plotkin1998}.
The corresponding model equation is
\begin{align}
 \frac{d}{dt} \vecA \lambda  = \avg{\vecA v_\lambda}(\vecA \lambda) + \sqrt{2D_\lambda(\vecA \lambda)} d\vecA W
\label{eq:Langevin}
\end{align}
where $\vecA W$ is uncorrelated (2D) white noise with unit covariance.
We estimated the local drift $\avg{\vecA v_\lambda}(\vecA \lambda)$ and diffusion coefficient $D_\lambda(\vecA \lambda)$ from our reweighed simulated trajectories by averaging local displacements (see Methods, Sec. \ref{sec-GG-Methods}, and Sec. S1.3 in the \SI).
Furthermore, $\avg{\vecA v_\lambda}(\vecA \lambda)$ is proportional to the effective force acting at the reduced phase space point $\vecA \lambda$ in the overdamped Langevin model.
The local mean velocity field $\vecA v_\lambda(\vecA \lambda)$ is determined by 
the conditional transition probabilities $\pi(\vecA \lambda, \vecA \lambda')$ 
between states $\vecA \lambda$ and $\vecA \lambda'$, and thus can be extracted from our transient simulation data. The resulting average velocity field in the reduced phase space of $(\lambda_{AC},\lambda_{BD})$ for the optimal repression strength ratio ($\kappa=31.6$) is in Fig. \ref{Fig-VelocitiesComparison}A, and for suboptimal ($\kappa=1000$) is shown in Fig. \ref{Fig-VelocitiesComparison}B.

Interestingly, in Fig. \ref{Fig-VelocitiesComparison}A one can identify two regions of $(\lambda_{AC},\lambda_{BD})$-space with low average velocities: 
one within the region of stable states $R_S$, the other within the region $R^\dagger_{AC}$ of states in which the \GC expression domain is lost. The region $R^\dagger_{BD}$ in which either \GB or \GD are lost, has no clear boundaries for optimal $\kappa=31.6$, and only for much larger $\kappa=1000$ a low-velocity plateau is clearly seen in this region (Fig.~\ref{Fig-DensityPinned}B).
Notably, in the lower-left corner of the $R_S$ plateau we notice a small region in which average velocities are significantly higher and all pointing inwards. We refer to this region as $R_{MB}$, and identify it as the metastable basin of intact, relaxed five-stripe patterns. In accordance, the two shown exemplary perturbed trajectories relax into $R_{MB}$ after randomly exploring the $R_S$ plateau, and remain confined to the $R_{MB}$ for later times (Fig. \ref{Fig-VelocitiesComparison}A). However, if the system drifts far away from $R_{MB}$, in the direction of $R^\dagger_{AC}$, the trajectories are quickly absorbed into $R^\dagger_{AC}$ once they reach the edge of $R_S$ characterized by high velocity components towards $R^\dagger_{AC}$.

In order to further investigate the low-velocity attraction basins and the high-velocity ridges that separate these basins, we use a different representation of pattern asymmetry, defining the shifted difference coordinates 
\begin{align}
\delta_{AC} \equiv \frac{1}{2}([\GA]_{tot} - [\GC]_{tot})/N + \frac{1}{2} \,, \qquad
\delta_{BD} \equiv \frac{1}{2}([\GB]_{tot} - [\GD]_{tot})/N + \frac{1}{2} \,.
\label{eq:deltaACBD}
\end{align}
These coordinates measure the deviation \changed{from an intact pattern with equal proportions of the strongly competing gene pairs, corresponding to the point $P_0=(\delta_{AC},\delta_{BD})=(\frac{1}{2},\frac{1}{2})$ in the phase space of the shifted difference coordinates,} in a way that retains information about which of the antagonistic genes becomes dominant. Similar latent-space projections have recently proven instrumental in analyzing the temporal dynamics of the emerging gap gene expression pattern \cite{Seyboldt2022}. The corresponding average velocities in $(\delta_{AC},\delta_{BD})$-space are shown in Fig.~\ref{Fig-VelocitiesComparison}C. The low-velocity basin, corresponding to $R_{MB}$, occupies the central part of $(\delta_{AC},\delta_{BD})$-space in Fig.~\ref{Fig-VelocitiesComparison}C.
\changed{Accordingly, perturbed trajectories relax towards the region enclosed by concentric velocity vectors pointing towards $P_0$ (the concentric vector pattern is best seen in the inset), again highlighting the presence of the metastable basin and restoring forces that tend to drive back fluctuations that perturb the intact pattern. Overall, the finding of a metastable basin is in line with the Waddington picture of canalization \cite{Waddington1942, Waddington1959}, in which developmental stages are seen as successive attractors of the underlying dynamics with the intact five-stripe pattern considered here representing such an attractor.}

In Fig.~\ref{Fig-VelocitiesComparison}B  we show the average velocity field for the case with weaker NN repression ($\kappa=1000$). Here the velocity fields are even more plateau-like in the region corresponding to weakly asymmetric patterns, and the characteristic concentric velocity pattern indicative of the basin in the optimal case cannot be clearly discerned any more in this case. In accordance, trajectories starting from perturbed patterns do not relax back and progress towards patterns with at least one domain lost.
\changed{
See also Sec.~S1.4 and Figs.~S3--S6 for the corresponding velocity fields in shifted difference coordinates and additional alternative projections.}

In addition to the average velocity fields of the registered phase space trajectories, the signatures of the metastable basins are also visible in the local phase space density sampled over many trajectories that explored the phase space during the whole sampled time interval, $p(\vecA\lambda)$. A suitable quantity for visualizing the corresponding phase space ``landscape'' is the negative logarithm of $p(\vecA\lambda)$; note that in an equilibrated, stationary system this quantity would be proportional to the energy (landscape) defining the stationary probability distribution of the system. Since our system is genuinely non-stationary, this relationship does not hold. Nevertheless we can consider our most stable systems transiently equilibrated in the metastable basins or origin and akin to stationary systems \emph{until} they irreversibly cross the barrier towards one of the basins corresponding to destroyed patterns.
\changed{
Note that the depth of these destroyed pattern basins grows with the amount of simulated time {\it after} the destruction events, because then the basins continue to be explored by phase space trajectories corresponding to fluctuations of the destroyed patterns; their apparent depth therefore depends on the prescribed maximal duration of the biased simulation trajectories in the NS-FFS scheme, which is a technical simulation parameter.
In contrast, the height difference between the metastable basin of intact patterns and the barrier separating it from the destroyed patterns basin is entirely determined by the biophysical parameters that set the average time scale for the stochastic destruction process. Therefore, the barrier height, as seen from the metastable basin of intact patterns, is a biophysical property that does not depend on technical choices (but obviously the maximal simulation time has to be chosen larger than the fastest barrier crossings in order to make the barrier visible).
}

In Fig.~\ref{Fig-VelocitiesComparison}D and E we plot the ``pseudopotential landscape'' defined as $-\log(\tilde p(\vecA \lambda))$ for optimal $\kappa=31.6$ (D) and suboptimal $\kappa=1000$ (E), where $\tilde p(\vecA \lambda)$ is a locally smoothened version of $p(\vecA\lambda)$ which equalizes out small local spikes in $p(\vecA\lambda)$ but preserves the overall structure of the resulting landscape (see Methods for details). The small plots right of the landscape visualizations show sections through the landscapes in direction of the asymmetry factors $\lambda_{AC}$ and $\lambda_{BD}$ at chosen constant values of the respective orthogonal factor (see Fig.~\ref{Fig-VelocitiesComparison} caption). In both cases we can clearly identify the metastable basin of undestroyed patterns and a barrier separating it from the basins of (half-) destroyed patterns. The basin corresponding to the states in which either the B or D domain is lost is less pronounced for the optimal choice of $\kappa$ due to its lower accessibility, and---more importantly---separated by a higher barrier.
This is best seen in a more detailed explicit comparison of the sections through the landscapes, shown in Fig.~\ref{Fig-VelocitiesComparison}F and G. The comparison clearly reveals that the barrier separating the metastable basin of intact patterns from the basin in which the C domain is lost is both higher and wider for the optimal choice of $\kappa$, overall leading to a markedly lower rate of pattern destruction.

Taken together, the analysis of both the velocity fields and the empirically sampled phase space density demonstrate that the long-time confinement of phase space trajectories close to the five-stripe pattern at optimal NN repression is due to the existence of a metastable basin which impedes progress towards losing one of the domains by restraining the system from leaving the metastable basin. With decreasing strength of NN repression the basin gradually disappears, thus enhancing the probability of pattern deterioration.

\changed{
The finding that pattern stability is enhanced by the emergence of a metastable basin is further supported by the quantification of the diffusion coefficient in $\vecA{\lambda}$ space, $D_\lambda$ (see Eq.~\ref{eq:Langevin} above).
As discussed in detail in Sec.~S1.3 and Fig.~S2 of the \SI, we observe that the average diffusion constant in the metastable basin of intact patterns ($R_S$), $\langle D_\lambda(\vecA \lambda)\rangle_{R_S}$, {\it monotonically} decreases with growing $\kappa$ (i.e., with decreasing NN repression strength), meaning that the estimated average time for leaving the metastable basin by random, ``diffusive'' motion (shown in Fig.~S2B and C) {\it monotonically} increases with increasing $\kappa$. Pattern stabilization around the optimal $\kappa$ therefore cannot be explained by a decrease of a diffusive escape rate, but rather by the emergence of restoring forces that drive deteriorating
patterns back into the metastable basin.
}

\pagebreak
\mysubsubsection{Stability enhancement does not require pinning}

\noindent
To assess whether pinning of the \GA-domains at the system boundaries is necessary for the observed stability enhancement at intermediate NN repression, we repeated our simulations and analysis for a system without pinning.
In contrast to the system with pinning, here the promoters of gene \GA in the nuclei at the system boundaries can be inhibited by the repressors of \GA.
We found that also in the system without pinning, pattern stability is markedly enhanced by the presence of weak interaction partners between two strongly repressing gene domains.
In Figure \ref{Fig-Stability} the red curve shows the mean destruction time $\tau_D$ against the ratio of the repressor off-rates $\kappa$ for the system without pinning.
We again find the highest pattern stability at an optimal repression strength ratio $\kappa^{\rm (np)}_{\rm opt}=100$ (red curve), which is close to the optimum in the system with pinning ($\kappa_{\rm opt}^{\rm (p)}\simeq 30$, blue curve),
albeit with about 10 times lower overall stability times;
yet, these stability times are still about an order of magnitude larger than without fine-tuning of NN interactions.

Overall this demonstrates that enhancement of pattern stability by at least one order of magnitude is possible both with and without pinning of expression at the system boundaries.
\changed{
However, pinning alters the proportion of destruction pathways that the collapsing patterns pursue;
in particular, it prevents the destruction of the peripherical \GA domain, which is the dominant destruction pathway at low and optimal $\kappa \lesssim 100$ in the system without pinning.
We present this effect in more detail in Sec.~S1.6 and Figs.~S10 and S11 of the \SI.}

\mysubsubsection{An analytical model of expression domain competition predicts optimal pattern stability}
\label{Sec-StabTheory}

\noindent
The problem of pattern stability has been recently addressed analytically in \cite{Majka2023}, where general and exact stability conditions for a pattern of two interacting domains were derived. In that work ``stability'' refers not only to the robustness against perturbations, but to the ability of a pattern to survive for infinitely long time. In this section, we show that these stability conditions can be successfully applied to the multi-gene system studied in this work, in order to obtain a coarse-grained prediction of the parameter values leading to pattern stabilization. 

The central result reported in \cite{Majka2023} is the description of the dynamics of a contact zone between two gene-expression domains for various levels of mutual repression between the two expressed genes. A single expression domain can form either by overcoming the ``activation threshold'' in the nearby undifferentiated tissue, resulting in asymptotically constant-velocity expansion, or emerge instantaneously in the entire available tissue, when expression is constitutive (active by default). For two genes in the system (and two respective domains) the scenario depends on the strength of mutual repression. If one gene cannot prevent the expression of the other gene in the bulk of its own domain, the dominating gene overtakes the system exponentially fast, expressing in the entire volume and without forming a meaningful contact zone between domains. For stronger repression, which prevents gene expression deeper in the bulk of its adversary domain, a contact zone emerges, within which both domains of active expression overlap. However, this region of overlap grows indefinitely, albeit with asymptotically constant velocity. When the interaction strength surpasses a critical value, an asymptotically finite-size  contact zone is formed. In this regime one domain can still shrink and the other grow, but in a coordinated manner, preserving the width of the contact zone. Asymptotically, the contact zone drifts with a constant velocity that is determined by the system parameters. This gives rise to a ``travelling'' gene expression pattern. The width and velocity of the contact zone are stable against perturbations in this phase, acting as an attractor of the system dynamics. However, while the travelling pattern is well-organized into two domains, it is not stable in the sense that in finite-size systems it survives only for a limited time, until one domain ``pushes out'' the other. Finally, perfectly stable patterns arise as a special case of travelling patterns with zero-velocity drift. As such, they can survive arbitrarily long. 

The simulations in this work are stochastic, tracking the chemical reactions at single-molecule resolution across the set of reaction-volumes constituting the system. However, in the limit of large particle number and small reaction volumes, this type of spatially discrete and stochastic dynamics approaches the continuous and ultimately deterministic reaction-diffusion dynamics of the type considered in \cite{Majka2023}. The existence of this deterministic limit can be also seen as the manifestation of the emergent noise-control mechanism that overtakes the system. Therefore, we compare the numerically found optimal $\kappa_{\rm opt}$ with the theoretically predicted $\kappa_{\rm theor}$ to assess how well the deterministic theory approximates the dynamics in the highly stochastic regime, and to explain the nature of the emergent noise-control mechanism.


To this end, we mapped the microscopic model used in our stochastic simulations onto the effective reaction-diffusion model analysed in \cite{Majka2023} (see Methods). \changed{We obtained the following continuous model for the expression of each gene $X$ in \{A,B,C,D\}}:
\begin{equation}
\partial_{\it t} X_2(x,t)= D~\partial_{\it xx} X_2(x,t)-\gamma X_2(x,t)+H~\theta\left(1-\sum_{\substack{Y\in\\ \{A,B,C,D\},\\Y\neq X}} \epsilon_{XY} Y_2(x,t) \right) ~,\label{eq:effdyn}
\end{equation}
where $X_2(x,t), Y_2(x,t)$ are the concentration profiles of \changed{the protein dimers (indicated by the subscript 2)}, $D$ is the diffusion constant, $\gamma$ the degradation constant, $H$ a production constant, and $\epsilon_{XY}$ are gene-gene interaction strengths. $\theta(\dots)$ denotes the Heaviside step function, corresponding to steep Hill-type regulatory kinetics.
Note that the derivations in \cite{Majka2023} only apply to systems with size $L\gg\lambda$, where $\lambda \equiv \sqrt{D/\gamma}$ is the characteristic length of gene interaction. For the systems studied here, $\lambda\approx 8.62~{\mum}$, which is much smaller than the system size $L\simeq 340~{\mum}$, warranting application of the theory.

While the original theory in \cite{Majka2023} describes only the contact zone involving exactly two domain boundaries, we can adapt it to the four-gene system studied here. Fig. \ref{Fig-TypicalOutput} shows that in the alternating cushions system there are only two types of contact zones: (i) between two strongly interacting genes (NNN domains) with the third, weakly interacting gene (NN domains), expressed in the background or (ii) between two weakly interacting genes (NN domains), with all other genes having close-to-zero expression level. Thus, we will consider stability of both contact zone types separately.

\noindent
In the type-(i) contact zone, the dynamics of gene expression is described by the effective equations
\begin{equation}
\left\{
\begin{gathered}
\partial_{\it t} X_2(x,t)= D\partial_{\it xx} X_2(x,t)-\gamma X_2(x,t)+H \theta\left(1-K_w^{-1}\frac{H}{\gamma}-K_{s}^{-1} Y_2(x,t) \right) \\
\partial_{\it t} Y_2(x,t)= D\partial_{\it xx} Y_2(x,t)-\gamma Y_2(x,t)+H \theta\left(1-K_w^{-1}\frac{H}{\gamma} -K_{s}^{-1} X_2(x,t) \right)
\end{gathered}\right. \label{eq:eqsi}
\end{equation}
where we approximate that the third ``background gene", has a constant expression level over the contact zone. The equilibrium value of this expression level is $H/\gamma$. $K_w$ and $K_s$ are the weak and strong repression constants, respectively and they satisfy (cf. Eq.~\ref{Eq-kappa} and Methods, Sec. \ref{sec-GG-Methods}):
\begin{equation}
\kappa=\frac{K_w}{K_s}=\frac{k_w^{\rm off}}{k_s^{\rm off}} ~.
\end{equation}
Type-(i) contact zones are established between genes A and C (with B or D in the background) as well as between B and D (with C in the background). In the type-(ii) contact zone, the equations take the form:
\begin{equation}
\left\{
\begin{gathered}
\partial_{\it t} X_2(x,t)= D\partial_{\it xx} X_2(x,t)-\gamma X_2(x,t)+H \theta\left(1-K_{w}^{-1} Y_2(x,t) \right) \\
\partial_{\it t} Y_2(x,t)= D\partial_{\it xx} Y_2(x,t)-\gamma Y_2(x,t)+H \theta\left(1 -K_{w}^{-1} X_2(x,t) \right)
\end{gathered}\right.\label{eq:eqsii}
\end{equation}
This contact zone emerges between gene pairs (A,B), (B,C), (C,D), and (D,A). 

\changed{
Having defined these contact zones, in Method section we adapt the more general derivation of stability conditions from \cite{Majka2023} for the current case. The main idea of this derivation is that the shape of expression profile $X_2(x,t)$, defined by Eq. \eqref{eq:effdyn}, can be found without knowing where the domain boundaries are located. Then, the positions of domain boundaries are sought from a separate set of equations. In \cite{Majka2023}, it is shown that the boundaries asymptotically travel with the common constant velocity $v$, preserving the distance $\Delta r$ between them. This ansatz leads to the algebraic set of equations defining $v$ and $\Delta r$, which can be solved. Eventually, the conditions for pattern stability are equivalent to ensuring that $v=0$ is the correct solution. Let us define two sets of constants, one for each type of contact zone:
\begin{align}
(i)&& \tilde C_{X}=\tilde C_Y=1-K_w^{-1}\frac{H}{\gamma}~, && \epsilon_{XY}=\epsilon_{YX}=K_s^{-1}~, && \epsilon_{XX}=\epsilon_{YY}=0~, \label{eq:type(i)}\\
(ii)&& \tilde C_{X}=\tilde C_Y=1~, && \epsilon_{XY}=\epsilon_{YX}=K_w^{-1}~, && \epsilon_{XX}=\epsilon_{YY}=0~.\label{eq:type(ii)}
\end{align}
and two auxiliary variables:
\begin{align}
R_X=\frac{2\gamma\tilde C_X}{\epsilon_{XY} H}-1~, && R_Y=\frac{2\gamma\tilde C_Y }{\epsilon_{YX} H}-1~,\label{eq:effparams}
\end{align}
Then, the stability conditions, as derived in the Methods section, come down to:
\begin{align}
R_X=R_Y=R~, && -1\le R\le 1~,\label{eq:stabcond}
\end{align}
Additionally, the width of the stable contact zone reads:
\begin{equation}
\Delta r=-\textrm{sgn}(R)\lambda \ln(1-|R|) \label{eq:width}
\end{equation}

In order to determine the range of parameters ensuring the global  stability of pattern, we apply the stability conditions \eqref{eq:stabcond} separately to type-(i) contact zone (Eqs. \eqref{eq:type(i)}) and type-(ii) contact zone (Eqs. \eqref{eq:type(ii)}). One can notice that in each type of contact zone the equality $R_X=R_Y$ is automatically satisfied, due to the common choice of parameters $\gamma$, $H$ and $D$ for both genes, as well as the symmetry in gene interactions. Sharing the same parameters is also the main reason why stability conditions \eqref{eq:stabcond} are much simpler than their general counterpart reported in \cite{Majka2023}. $R_X=R_Y=R$ means that each type of contact zone is characterized by one variable:

\begin{align}
R_{(i)}=\frac{2(1-K_w^{-1}\frac{H}{\gamma})}{K_s^{-1} \frac{H}{\gamma}}-1 && R_{(ii)}=\frac{2\gamma}{K_w^{-1}H}-1 \label{eq:Rs}
\end{align}
The remaining stability condition, $-1<R<1$, applied to $R_{(i)}$ and $R_{(ii)}$, results in the following inequalities: 
}
\begin{equation}
\begin{aligned}
(i) && K_w\ge \frac{H}{\gamma}~,~ && K_s\le \frac{1}{(\frac{H}{\gamma})^{-1}-K_w^{-1}}~, \\
(ii) && K_w\le \frac{H}{\gamma}~.
\end{aligned}\label{eq:condfin}
\end{equation}
These conditions show that the addition of weak interactions is instrumental for increasing system stability. On the one hand, the type-(i) contact zone is stable (i.e., immobile) provided that the weak interaction strength $K_w^{-1}$ does not exceed $\left(H/\gamma\right)^{-1}$; otherwise it would prevent the expression of strongly interacting genes in this region. On the other hand, for the type-(ii) contact zone it is necessary that $K_w^{-1}>\left(H/\gamma\right)^{-1}$, as this minimal strength of repression is required to prevent co-expression of both weakly interacting genes in the same region. 
In order to simultaneously stabilize both types of contact zones, one needs to negotiate between these two largely opposite goals. This trade-off can be achieved only for the most marginal value in both parameter ranges, $K_w=H/\gamma$, which highlights why in the alternating cushions architecture the weak interactions have to be fine-tuned for pattern stability. In contrast, but in line with the numerical findings, the strong interactions characterized by $K_s$ can be arbitrarily large, $K_s\le+\infty$. 

The simulations in this work were performed for $K_s\simeq0.003~{\rm \mu m^{-3}}$ with $K_w$ varied to obtain different values of $\kappa$, see Methods, Sec. \ref{sec-GG-Methods}. Calculated from these microscopic parameters, $H/\gamma\simeq0.23~{\rm \mu m^{-3}}$. The resulting theoretical value of $\kappa$ that ensures stability is then $\kappa_{\rm theor} \simeq 76$. This number is of the same order of magnitude as the optimal $\kappa$ in the simulated stochastic systems, showing slightly better agreement with the no-pinning case ($\kappa_{opt}^{\rm (np)}\simeq 100$) than with the case with pinning at the boundaries ($\kappa_{opt}^{\rm (p)}\simeq 30$), see Fig. \ref{Fig-Stability}.

\mysubsubsection{The emergent noise-control mechanism can be understood via the analytical model}

\begin{figure}[ht!]
\includegraphics[width=\textwidth]{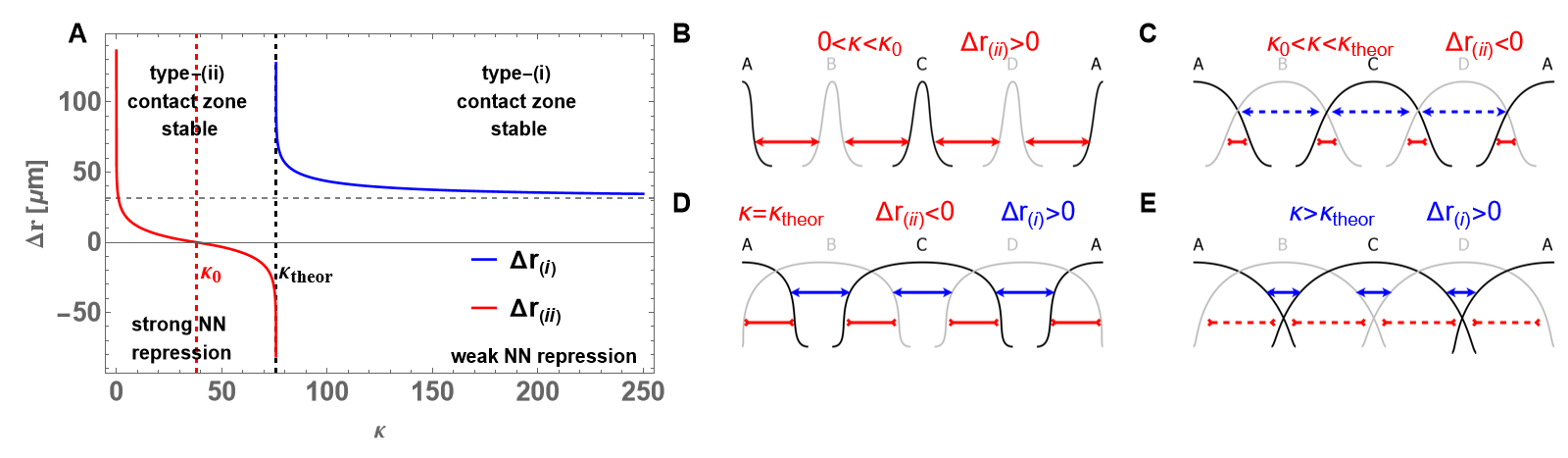}
\caption{\textbf{Analysis of theoretical contact zone widths uncovers a deterministic stabilization mechanism.} \textbf{(A)} Plot of theoretical contact zone widths in the approximation of separate interfaces, $\Delta r_{(i)}$ and $\Delta r_{(ii)}$, for type-(i) contact zones (between two NNN domains with third interacting gene expressed in the background) and type-(ii) contact zones (between NN domains), calculated from Eq.~\eqref{eq:width} and Eqs.~\eqref{eq:Rs} with $\kappa=K_w/K_s$ and $K_s$ kept constant. In their respective regimes of stability, the widths are restored by deterministic dynamics if perturbed. Vertical lines: $\kappa_0\simeq38$ (red, dashed) at which $\Delta r_{(ii)}$ changes sign; critical $\kappa_{\rm{theor}}\simeq 76$ (black, dashed) ensuring simultaneous stability of type-(i) and type-(ii) contact zones. Horizontal line (gray, dashed): limit of $\Delta r_{(i)}\simeq 31.3$ [$\mum$] without any weak interactions ($\kappa\to+\infty$). 
\textbf{(B-E)} Schematic representations of system states in various regimes of $\kappa$, predicted by the deterministic model. Solid arrows: stable contact zones (restorable width); dashed arrows: unstable contact zones (non-restorable width); inward arrowheads indicate $\Delta r_{(ii)}<0$; contact zones of type-(ii) (red), and type-(i) (blue). 
\textbf{(B)} $0<\kappa<\kappa_0$: type-(ii) contact zone stable, $\Delta r_{(ii)}>0$, no type-(i) contact zones, domain widths lack stabilization against fluctuations. 
\textbf{(C)} $\kappa_0<\kappa<\kappa_{\rm{theor}}$: type-(ii) contact zone stable, partial overlap of domains, $\Delta r_{(ii)}<0$, provides minimal domain width stabilization against fluctuations, but fluctuations can shift entire contact zones. 
\textbf{(D)} $\kappa=\kappa_{\rm{theor}}$: type-(i) and type-(ii) contact zones stable, maximizing their widths ($\Delta r_{(i)}$ and $\Delta r_{(ii)}$ tend to $\pm\infty$ in the approximation of separate contact zones). Pattern is restored after any perturbation. 
\textbf{(E)} $\kappa>\kappa_{\rm{theor}}$ type-(i) contact zones stable, but $\Delta r_{(i)}\ll L$, fluctuations can shift entire contact zones.
\label{fig:deltaR}}
\end{figure}

\noindent
The analytical deterministic model can be employed to obtain further insights into the mechanism of increased pattern robustness against noise in the vicinity of optimal $\kappa$. For this, we must first consider the width of type-(i) and type-(ii) contact zones in their stability regions predicted by the theory from \cite{Majka2023}. Inserting $R_{(i)}$ and $R_{(ii)}$ into Eq. \eqref{eq:width} with $K_w=\kappa K_s$, we obtain $\Delta r_{(i)}$ and $\Delta r_{(ii)}$ as functions of $\kappa$, shown in Fig. \ref{fig:deltaR}A. Here, $\Delta r>0$ indicates a no-expression region between the domains (a gap), while $\Delta r<0$ means that active expression regions overlap. One can instantly notice that $\Delta r_{(i)}\to+\infty$ and $\Delta r_{(ii)}\to-\infty$ at $\kappa=\kappa_{\rm{theor}}$. Tending to infinite values is an artefact of our analysis, in which we treat each contact zone as a separate region, disconnected from the others. However, this behaviour conveys an important message. At $\kappa=\kappa_{\rm{theor}}$ the system attempts to maximize the size of each contact zone, forming five contact zones tightly filling the entire system. In this state, any pattern perturbation distorts at least two contact zones. Since each contact zone is stable, their maximized widths are attractors for the deterministic dynamics \cite{Majka2023}, and consequently the system tends to remove the perturbation. This is the origin of increased survival time of patterns at optimal $\kappa$. This restoration behavior is qualitatively similar to the model of repulsive forces between domain boundaries (kinks), discussed in \cite{Vakulenko2009}. Although the analytical stability theory \cite{Majka2023} does not rely on the concept of explicit restoration forces, these forces arise effectively, leading to the occurrence of the pseudopotential in our phase space analysis, in Fig. \ref{Fig-VelocitiesComparison}. Thus, the effective restoration forces form a link between the exact stability theory \cite{Majka2023} and the approximation of interacting kinks in \cite{Vakulenko2009}.

The existence of a rigorously sharp stability condition, $K_w=H/\gamma$, raises the question about the deterministic dynamics for suboptimal choice of $\kappa$ and its influence on the stochastic system. Let us first consider the case $\kappa<\kappa_{\rm{theor}}$, in which type-(ii) contact zones are stable. In this regime, the system forms a pattern of domains A-B-C-D-A, but the NNN domains are so distant from each other that strong interactions are not yet important. There are four type-(ii) contact zones in this system. For $\kappa\simeq 0$, the weak interactions are extremely repressive and $\Delta r_{(ii)}\to+\infty$. Thus, the pattern collapses. For somewhat larger $\kappa$, a finite-size gap ($\Delta r_{(ii)}>0$) between NN domains emerges (see Fig. \ref{fig:deltaR}B) and is reduced to zero width ($\Delta r_{(ii)}=0$) at $\kappa_0\simeq 38$. In this regime, the pattern can survive arbitrarily long in the absence of fluctuations, but the domain widths are not stabilized in any way. Thus, in the presence of noise, the survival time of the domain depends on its size (which grows with $\kappa$), as larger domains take longer to be destroyed.
For $\kappa>\kappa_0$, NN domains begin to overlap, as $\Delta r_{(ii)}$ becomes negative (see Fig. \ref{fig:deltaR}C). This marks the first emergence of the additional stabilizing mechanism, as the deterministic dynamics will tend to restore $\Delta r_{(ii)}$ in each contact zone if perturbed. This means that, in the presence of noise, $\Delta r_{(ii)}$ would keep returning to its deterministic value, but fluctuations can still shift a stable contact zone as one entity. If, as a result, two contact zones meet or one is pushed to the system boundary, this causes the collapse of a domain and partial desintegration of the pattern. As $\kappa$ further approaches $\kappa_{theor}$, the overlap becomes large enough such that NNN domains begin to interact and type-(i) contact zones are formed. These contact zones have a certain minimal width, but they are not stable, in the sense that this width would not be restored if increased. The occurrence of type-(i) contact zones imposes a barrier for the further growth of $\Delta r_{(ii)}$ with $\kappa$ (see Fig. \ref{fig:deltaR}D). At this stage, a major enhancement of pattern stability occurs, as stable type-(ii) contact zones and type-(i) contact zones tightly fill the system. At $\kappa=\kappa_{theor}$ also the type-(i) contact zones gain stability, due to the further increase of overlap between NN domains. This results in the maximal global robustness of the pattern against noise. For $\kappa>\kappa_{theor}$, the width of the now-stable type-(i) contact zones quickly decreases and eventually saturates at $\Delta r_{(i)}=31.3$ [$\mum$] in the limit of completely absent weak interactions ($\kappa\to+\infty$). In this regime, with type-(i) contact zones width $\Delta r_{(i)}<L/3$ and no mechanism restoring the width of type-(ii) contact zones (see Fig. \ref{fig:deltaR}E), the pattern gradually loses stability against noise. Fluctuations can shift each type-(i) contact zone as one entity (analogous behaviour to the type-(ii) interfaces in $\kappa_0<\kappa<\kappa_{\rm{theor}}$ regime), eventually leading to pattern destruction.


In summary, the emergent deterministic dynamics, described in \cite{Majka2023}, is crucial for stabilizing the highly stochastic system simulated in this work. The increase in survival time $\tau_D$, towards $\kappa_{opt}$, illustrated in Fig. \ref{Fig-Stability}, is directly associated with the gradual activation of deterministic stabilization mechanisms, described in the paragraphs above. General principles of pattern stabilization, outlined in \cite{Majka2023} for two genes, apply also to the four-gene system studied here, but many-gene competition and stochasticity results in a more nuanced picture of stabilization. A more detailed investigation would require considering the full spatial variability of all expression profiles together, but the approximated effective model proves useful in predicting optimal parameters.  

\section{Discussion}

In many developing organisms, morphogen gradients provide a long-range positioning system by activating downstream patterning genes in a concentration-dependent manner.
Prominent examples are the gap gene system in \Drosophila, whose main maternal regulators are the
morphogen gradients of \Bcd and \Cad spreading along the embryo axis \cite{Surkova2008, Jaeger2004, Driever1988a, Driever1988b, Frohnhofer1986, Struhl1989, Macdonald1986, Schulz1995, Mlodzik1987},
and the vertebrate neural tube with \emph{Shh} and \emph{BMP}/\emph{Wnt} secreted from the opposite sides of the neural tube \cite{Balaskas2012,Briscoe2015,Bier2015,Kuzmicz-Kowalska2020,Exelby2021,Minchington2023}.
For the Drosophila embryo, multiple studies have shown that mutual interactions between gap genes play a crucial role in abdominal segmentation \cite{Sokolowski2012,Jaeger2004,Dubuis2013,Manu2009PlosBiol,Surkova2008,Sokolowski2015,Kozlov2012,Crombach2012, Manu2009PlosCompBiol}, 
leading to the formation of stable domains with slow effective dynamics \cite{Vakulenko2009}. 
However, it remains unclear how such a system could be robust given the stochastic nature of gene expression and regulation if the emergent interactions are not fine-tuned to mitigate the resulting noise.
Moreover, it is observed that maternal regulators such as the Bcd gradient disappear while the expression patterns invoked downstream persist \cite{Surdej1998,Spirov2009}.
In support of the view that self-coordination properties emerge in the gap gene system after maternal activation, a more recent study which found that the gap gene expression pattern scales with the size of the embryo with high precision, while---surprisingly---the Bcd gradient does not display any scaling properties \cite{Nikolic2023}.
Similar emerging self-organizing properties have been observed in other developmental systems \cite{Raspopovic2014,Almuedo-Castillo2018,Morales2021}.

Here we asked whether a system of mutually repressing developmental patterning genes arranged in successive expression domains can indeed be stable over developmentally relevant time intervals without upstream morphogen gradients while facing unavoidable fluctuations in the expressed gene products. Such copy number fluctuations can induce bistable switching at the domain boundaries, resulting in stochastic movement of the boundary which ultimately can lead to destruction of one of the gene expression domains.
We quantified the mean stability time of a five-stripe expression pattern formed by four interacting genes
in a stochastic model conceptually inspired by the posterior \Drosophila embryo in cycle 14
as a function of the repression strength between neighboring stripes.
To be able to simulate the breakdown of very stable patterns we employed Non-Stationary Forward Flux Sampling (NS-FFS), an enhanced sampling scheme for simulating rare events in non-stationary systems with transient dynamics \cite{BeckerAllenTenWolde2012}.
We find that for an optimal value of the repression strength between adjacent expression domains 
the stability of the pattern is increased by about an order of magnitude. 
This stability optimum can be traced back to the fact that bistable switching at the boundary between domains of strongly mutually repressing genes is inhibited 
by an intervening cushion domain of a gene that weakly represses both strong partners. 
This stabilizing mechanism works best if the spacer gene represses its nearest neighbors (NN) with moderate strength: 
very weak NN repression has no effect while strong NN repression globally destabilizes overlapping domains.
At the optimal repression strength ($\kappa=\kappa_{\rm opt}$) the cushion thus slows down the random motion of the domain boundary and subsequent pattern destruction.

Stability is enhanced even more, by one more order of magnitude, if expression of the outermost gene is pinned at the system boundaries,
which effectively anchors the whole expression pattern.
Such a situation may emerge when the outermost gene remains under control of maternal cues, such as maternally deposited mRNA,
while the other gene stripes form only by zygotic interactions.
Furthermore, it resembles the late stages of neural tube development in which the Shh and BMP morphogen gradients are acting close to system boundaries \cite{Kicheva2014, Zagorski2017}. 
In the system considered here we find that five-stripe patterns form a metastable attractor of the dynamics with a restoring force that counteracts perturbations, such as non-perfect initial conditions. 
In the optimal stability regime, 
our observations are consistent with the Waddington picture \cite{Waddington1942,Waddington1959} of development as canalization into successive metastable states, with the ordered initial gap gene pattern representing one of the metastable states in this succession.
Earlier work already demonstrated that developmental attractors may emerge as an intrinsic property of the gene expression pattern established through mutual interactions \cite{ Manu2009PlosBiol,Manu2009PlosCompBiol}.
Here, we demonstrate that even without morphogen gradients metastable basins can arise and protect expression patterns against stochastic fluctuations.


Further insight comes from the application of the stability theory derived in \cite{Majka2023} to the model of four genes interacting in the alternating cushions scheme. In agreement with the simulations, these analytical calculations reveal that the presence of weak interactions is necessary for stabilizing the system and establishing long-surviving patterns. More specifically, theoretical analysis shows that requirements for stability of type-(i) contact zones (i.e. two strongly interacting genes with the third weakly interacting in the background) and type-(ii) contact zones (i.e. two weakly interacting genes with other genes at very low expression level) are to certain degree incompatible, and agreement between them can be achieved only for the most marginal value of $\kappa=\kappa_{\rm theor}$ in the respective stability range for each type. As a consequence, simultaneously ensuring perfect stability of both contact zone types requires fine-tuning of the weak repression strength, quantified by the corresponding dissociation constant $K_w$. 
This analytical prediction of one optimal value of $\kappa$ is in qualitative agreement with the numerical simulations, which show a very sharp rise in the survival time of expression pattern near one particular value of $\kappa = \kappa_{\rm opt}$, see Fig. \ref{Fig-Stability}. 

Quantitatively, the numerical $\kappa_{opt}$ and theoretical $\kappa_{theor}$ agree particularly well in the no-pinning case (for pinning: $|\kappa_{\rm opt}^{\rm (p)}-\kappa_{\rm theor}|/\kappa_{\rm opt}^{\rm (p)}=139\%$, for no-pinning: $|\kappa_{\rm opt}^{\rm (np)}-\kappa_{\rm theor}|/\kappa_{\rm opt}^{\rm (np)}=24\%$). This is in line with the assumptions of \cite{Majka2023}, where an open system was considered and system boundary effects, such as pinning, were neglected. Differences between $\kappa_{\rm theor}$ and $\kappa_{num}$ are expected due to the nature of approximations employed in the mapping of microscopic model on its effective representation \eqref{eq:effdyn}. It is plausible that this discrepancy could be resolved by constructing an even higher-level stability theory that takes into account the spatial variability of all four genes in each contact zone.

Further, using the division into type-(i) and type-(ii) contact zones, we investigated the behaviour of effective deterministic model Eq. \eqref{eq:effdyn}, in the entire range of $\kappa$. We found that the preference of the system to form possibly large contact zones, combined with the stability of at least one type of interfaces between domains, results in the increased robustness of the pattern against fluctuations, in the vicinity of optimal $\kappa$. These observations are in agreement with our highly stochastic and microscopically detailed simulations, for which the deterministic model is only the continuous-limit approximation. Yet, the approximate agreement between $\kappa_{\rm opt}$ and $\kappa_{\rm theor}$ as well as the broad peak of increased survival time (Fig. \ref{Fig-Stability}), suggest that the deterministic dynamics of model \eqref{eq:effdyn} is still remarkably important for this system. The interplay between deterministic and stochastic component of dynamics in simulations results in the emergent noise-control mechanism, significantly increasing survival time of patterns. We also found that the shifting of the stable contact zones by fluctuations is the major reason of pattern destruction for $\kappa$ away from the optimal value.

The observed stability times appear sufficient for early fly embryogenesis  ($\simeq 2 h$ until cycle 14) for all NN repression strengths weaker than the optimal value, 
with or without pinning, even for the reduced system size considered here for computational feasibility.
\changed{
We expect that the stability times will systematically increase when a larger system size is chosen in a more realistic description. Note that the system size can increase in two ways, either by increasing the considered lattice of reaction volumes (nuclei) or by allowing for a larger maximal copy number per reaction volume. In the first case, stability is enhanced because the expression states of more reaction volumes need to be switched in order to destroy the now larger expression domains, while the local molecular noise level (which is a key determinant for the speed of this process) remains the same. In the second case, the molecular noise is reduced, such that detrimental cell switching events are impeded, leading to longer average destruction times for the individual expression domains and consequently overall longer stability.
}
Nevertheless, based on the theoretical and numerical evidence we believe that the stability enhancing mechanisms uncovered in this study will also apply to biologically relevant system sizes. Other factors potentially affecting stability are autoactivation interactions and interactions with other genes not included in the simplified regulatory network studied here, which will likely affect the dynamics of the gene expression pattern.
\changed{
Moreover, in our system the parameters do not differ between the four interacting genes (with the exception of the repressor unbinding rate from the \GA promoter at the boundaries in the system with pinning); as expected, we therefore did not observe any directional preference for domain shifts and destruction events, unlike in \Drosophila, where the gap genes exhibit a systematic anterior shift in early developmental cycle 14 \cite{Jaeger2008,Jaeger2011,Verd2017} (which appear to require the action of shadow enhancers not considered here \cite{ElSherif2016}). The possible effect of asymmetric regulatory interactions on pattern stability is an interesting open question that could be assessed in future iterations of the model and theory presented here.
Note, however, that the adapted stability theory clearly identifies the weak nearest-neighbor repression strength as the key parameter for enhancing stability, while the strong repressive interactions are found not to affect stability as long as they are chosen strong enough.
Furthermore, since in the theory the predicted value of the optimal repression strength ratio is entirely determined from the properties of the contact zones between the expression domains, this prediction does not depend on the spatial system size, provided that it is large enough as to accommodate all the contact zones. We therefore expect that increasing the size and realism of our spatial-stochastic model in the described ways would alter the recorded stability times, but at the same time retain the observed key property of strong stability enhancement at an optimal repression strength ratio.}


Our work puts an interesting perspective on the role of maternal gradients in establishing and maintaining developmental patterns. We show that sufficiently stable patterns can exist without morphogen gradients, but at the same time that their stability is significantly enhanced by pinning the patterns at the embryo boundaries. Taken together, this suggests that morphogens do not act deep inside the embryo interior, which could explain why the patterns remain stable even when the morphogen inputs disappear \cite{Surdej1998,Spirov2009}. Instead, they may predominantly act at the embryo boundaries as to break symmetry, by selecting the desired pattern from the larger set of patterns that, by permutation, would also be stable. By acting only at the periphery, the morphogens, which themselves do not exhibit scaling, still would allow scaling of the downstream pattern with embryo length, in line with recent findings \cite{Nikolic2023}.

\pagebreak
The stabilizing mechanism arising from fine-tuning nearest-neighbor interactions in the alternating cushions scheme can be also considered in the broader class of regulatory mechanisms providing pattern stability against intrinsic and extrinsic noise \cite{Simsek2022,Iyer2022,Averbukh2017}. In future studies, it may be instrumental to further numerically and analytically explore the proposed model by including other biologically relevant features. Possible extensions include growth of the tissue by cell divisions, self-correcting mechanisms through cell-to-cell communication other than diffusive exchange of proteins, or inclusion of more specific noise types. \changed{Another interesting scenario would be studying the alternating cushions 
system considered here but with periodic boundary conditions, possibly 
relevant to sea urchins and sea stars that develop pentaradial symmetry 
in later stages of development \cite{McClay2011}; this likely would allow for further 
increase of pattern stability in the optimal repression strength regime 
without any pinning, as it would provide an alternative way of 
stabilizing the A domain.} These extensions could further test the validity of our stability theory under more realistic biological conditions. However, due to the remarkable agreement between our adapted stability theory and the numerical simulations of the minimal model studied in this work, we believe that more realistic variants of it will result in quantitative but not qualitative changes in our predictions.

\section{Methods}
\label{sec-GG-Methods}
\subsection*{Details of the model}
Our model is inspired by arguably the most paradigmatic developmental system
in which development of distinct cell fates is determined by local protein expression
patterns driven by external morphogen gradients, the early embryo of the fruit fly {\it Drosophila melanogaster}.
We model the egg-shaped embryo with its cortical layer of nuclei as
a cylindrical array of reaction volumes coupled by diffusion of proteins.
Every volume (nucleus) contains four individual promoters for each of the genes
\GA, \GB, \GC and \GD.
Each promoter can be repressed by the products of the three others with
different affinities; this system of four mutually inhibiting genes represents the gap gene system in the early fly embryo, formed by the four genes \hb, \kr, \kni and \gt, and comprises its essential regulatory interactions.
For combined repressive interactions, we employ OR-logic, i.e. whenever one of the three repressor sites is
occupied expression of the gene is completely blocked.
There is no competition for repressor sites on the promoters.
In the unrepressed state the promoters exhibit constitutive
protein production, i.e. no external activator signal is required.
This deliberately mimics a situation in which activation of the genes
is not provided by external morphogen gradients but by either an omnipresent
master activator or auto-activation with a low activation threshold.
Consequently, our model explicitly does not include morphogen gradients.
As a simplifying assumption, we treat the whole production process, 
i.e. transcription, elongation and translation, as one step governed by a single rate $\beta$.
Proteins however can form (homo)dimers and dedimerize again \cite{McCarty2003,Sauer1993},
and only in their dimeric form they act as repressors.
This is to ensure that antagonistic genes form bistable pairs for sufficiently strong
mutual repression.
Initially, all simulations are set up in a stripe pattern similar to the
experimentally observed order in the embryo posterior, i.e.
\GA-\GB-\GC-\GD-\GA \cite{Surkova2008, Jaeger2004, Clyde2003}.
This implies a fixed definition of ``gene neighborhood'' to which we refer throughout
this paper: 
by nearest neighbors (NN) we mean the pairs (\GA, \GB), (\GB, \GC), etc.,
while the pairs (\GA, \GC) and (\GB, \GD) are considered next-nearest neighbors (NNN).
A key ingredient of our model is that nearest-neighbor repression is
weaker than repression between next-nearest neighbor domains 
(see ``Parameter choice'' in Methods).
By default we pin the expression of \GA at the system boundaries,
i.e. in nuclei on the two outermost rings of the cylinder the \GA promoter
is irrepressible, and therefore constitutively produces \GA proteins.
This is motivated by the fact that in the real \Drosophila embryo the gene \Hb is under strict control by
the maternal morphogen \Bcd throughout the anterior half \cite{Driever1989}, 
while in the posterior a second enhancer exposes \Hb 
to positive regulation by the maternal terminal system \cite{Margolis1995, Casanova1990, Weigel1990}.
We compare this system to a system in which there is no pinning and all nuclei are identical.

\subsection*{Simulations}
To perform rare-event sampling of the spatially resolved system we integrate our
``Gap Gene Gillespie'' (GGG) simulator used in previous work \cite{Sokolowski2012, Erdmann2009} 
with the NS-FFS scheme \cite{BeckerAllenTenWolde2012}.
NS-FFS is used to monitor and process a progress coordinate written out by GGG at regular simulation interrupts,
at which GGG trajectories are cloned and restarted in a way that sampling is enhanced in the direction of increased 
progress coordinate, i.e. towards pattern destruction. \index{GGG} \index{Gap Gene Gillespie}

\subsubsection*{Spatially resolved stochastic simulations (GGG)}
In GGG, the model is implemented via the Stochastic Simulation Algorithm by Gillespie \cite{Gillespie1976, Gillespie1977}
on a cylindrical 2D lattice of reaction volumes at constant distance $l=8.5~\unit{\mum}$, with periodic boundary
conditions in the circumferential direction of the array. \index{Gillespie algorithm} \index{Stochastic Simulation Algorithm} \index{SSA}
An abstract graph of the reaction network that displays the set of reactions for any of the simulated promoters
is shown in Figure S12 in the \SI.
Diffusive chemical species (patterning gene proteins and their dimers) hop between neighboring volumes via the next-subvolume method 
\cite{Hattne2005} which integrates diffusion into the Gillespie algorithm by annihilation of a species copy in the volume of origin 
and instantaneous insertion of that copy in a randomly chosen neighboring volume with a rate $k_{diff}=4D_P/l^2$,
where $D_P$ corresponds to the protein diffusion coefficient.
The source code of GGG can be downloaded from \myurl{https://github.com/TheSokoLab/Pabra-GGG}.

\subsubsection*{Forward flux sampling}
We employ the recently developed non-stationary forward flux sampling (NS-FFS) method \cite{BeckerAllenTenWolde2012, BeckerTenWolde2012, Allen2005}
to enhance stochastic sampling of system realizations that increase a (reaction) progress coordinate $\lambda$ while retaining correct statistical weight.
NS-FFS achieves this by branching off multiple child trajectories upon crossing predefined interfaces in 
undersampled regions of $(\lambda,t)$-space and pruning trajectories that cross interfaces in oversampled regions.
The NS-FFS scheme aims at equilizing the flux of simulated trajectories in the reaction coordinate direction among the time bins.
The rate of branching and pruning is calculated from the temporal trajectory crossing statistics collected during runtime.
To that purpose the time domain is subdivided into equidistant time intervals.
For a detailed account of the reweighting procedure we refer to \cite{BeckerAllenTenWolde2012}.
\index{forward flux fampling} \index{Non-Stationary Forward Flux Sampling} \index{NS-FFS}

\subsubsection*{Progress coordinates}
The choice of a suitable progress coordinate is a critical step of the FFS technique.
Here, we seek to enhance progress of the simulated patterns towards their destroyed state.
The destruction events are in particular characterized by the disappearance of one of the partners 
within each of the strongly repressing gene pairs.
Progress towards destruction thus is accompanied by increasing pair asymmetry,
which can be quantified for each pair separately by the following two asymmetry factors:
\begin{align}
 \lambda_{AC} &\equiv \max([\GA]_{tot},[\GC]_{tot}) / N \\
 \lambda_{BD} &\equiv \max([\GB]_{tot}, [\GD]_{tot}) / N
\end{align}
where $N = [\GA]_{tot} + [\GB]_{tot} + [\GC]_{tot} + [\GD]_{tot}$ is the number of all proteins in the system.
Based on this we define our progress coordinate, which increases whenever asymmetry among any of the pairs is augmented, via
\begin{align}
 \lambda &\equiv \lambda_{AC} + \lambda_{BD} = \left[ \max([\GA]_{tot},[\GB]_{tot}) + \max([\GB]_{tot},[\GD]_{tot}) \right] / N	\;.
\end{align}

Since NS-FFS features multi-dimensional reaction coordinates we compared our standard choice to a setup in which the two 
components $\lambda_{AC}$, $\lambda_{BD}$ of the reaction coordinate $\lambda$ are treated as two separate reaction coordinates 
with an own set of interfaces each.
While an orthogonal pair of reaction coordinates captures the principal reaction paths in our system more accurately,
the acquisition of crossing statistics is prolongated because of the increased number of bins in these simulations,
and we did not find any substantial advantage of this choice in terms of branching behavior.
We therefore preferred the standard definition.
\index{progress coordinate} \index{reaction coordinate}

\subsubsection*{Combination of simulation methods}
In order to wrap NS-FFS around the GGG simulator we run GGG for a predefined
simulation time $t_{GGG}=60~\unit{s}$. 
At the end of the simulation the reaction coordinates are calculated
and passed on to the NS-FFS module, and the end state of the simulation is recorded.
The NS-FFS module then determines whether an interface crossing has occurred and, if so,
decides on whether the trajectory shall be branched or pruned.
In case of branching NS-FFS will prompt $n_B \geq 1$ restarts of the GGG simulator
with the recorded end state as initial condition, different random seeds and with new statistical weights.
At each crossing and at measuring times spaced by a regular interval $\Delta t$ the time, branch weight
and reaction coordinate values are stored in a tree-like data structure that facilitates later analysis.

Trajectory trees are started from a standardized, regular-stripe initial condition passed to the first call of GGG.
Propagation of the tree stops when all child branches have either reached the end of the time histogram or have been pruned.
Subsequently a new tree is started with a different random seed.
NS-FFS monitors the cumulative simulated time $T_{cum}$ and terminates simulation
when $T_{cum}$ exceeds a predefined maximal simulation time $T_{max}$ and the last 
trajectory tree has been propagated towards the end.
Typically, $T_{max}=3-7~\unit{h}$ and $T_{cum}=2-5\cdot10^{7}~s$,
which usually results in several thousand independent starts from the initial condition.
By default we start from an artificial pattern consisting of five non-overlapping stripes with rectangular profiles
occupying an equal part of the total system length $L/5$ each and equal number of monomers (no dimers) in each nucleus
close to the expected total copy numbers.
We find that these initial patterns quickly relax towards typical metastable patterns, i.e. into the metastable main basin of attraction,
which justifies our approach a posteriori.

The source code of the NS-FFS path-branching algorithm (Pabra) combined with GGG can be downloaded from \myurl{https://github.com/TheSokoLab/Pabra-GGG}.

\subsection*{Parameter choice}

\subsubsection*{Repression}
We are mainly concerned about the importance of distinct repression strength of nearest-neighbor (NN)
as compared to next-nearest neighbor (NNN) interaction.
We assume repressor binding-rates to be diffusion-limited via $\kRon=4\pi\sigma_R D_N$,
where $D_N$ is the \emph{intranuclear} diffusion constant and $\sigma_R$ an effective target radius.
Repression strength therefore is varied by changing the unbinding rates of the repressing dimers.
The main parameter in our simulations is $\kappa=\kRoffW/\kRoffS$, the ratio between NN and NNN repressor off-rate.
In this work only $\kRoffW$ is varied, while $\kRoffS$ is chosen sufficiently low to guarantee bistability between next-nearest neighbor genes,
which is a precondition for the formation of individual stripe domains in the first place, see Table S1 in the \SI.
For $\kappa=1$ NN and NNN repressive interactions are equally strong,
while for large $\kappa$ values NN repression is much weaker than NNN repression.
In the ``uncoupled limit'' $\kappa\rightarrow\infty$ the two bistable pairs coexist without sensing each other.
We do not consider cases with $\kappa<1$.

\subsubsection*{Dimerization}
We set the dimerization forward rate $\kDon$ to be equal to two times the diffusion-limited repressor binding rate,
which is accounting for the fact that both reaction partners are diffusing.
The dimerization backward rate is set via $\kDoff=\kDon/V_N$ ($V_N$ = nuclear volume) as in \cite{Warren2004,Warren2005,Morelli2008,Sokolowski2012}
to ensure that at any moment most of the proteins are dimerized.

\subsubsection*{Production and degradation}
In our model both monomers and dimers are degraded.
This leads to a nontrivial dependence of the total copy number on production, degradation and (de)dimerization rates,
as we discuss with more detail in \cite{Sokolowski2012}.
Since we did not find any experimental reports of gap protein lifetimes, we chose equal monomeric ($\mu_M$) and equal dimeric degradation rate ($\mu_D$) 
for all genes and set these quantities to values that lead to a reasonable effective lifetime of the corresponding proteins of $t_{eff} \simeq 100~\unit{s}$.
The steady-state copy number is tuned via the production rate $\beta$.
By default, we consider copy numbers as low as possible ($\simeq 15$) to minimize computational effort.
The effect of increasing the average copy number is discussed in the ``Discussion'' section.

\subsubsection*{Geometry and internuclear transport}
The choice of our geometric parameters, in particular of the lattice constant, is inspired by experimental measurements
in the \Drosophila embryo by Gregor et al. \cite{Gregor2007a}.
Information on the diffusion constants of proteins involved in early \Drosophila patterning is scarce.
The diffusion constant of the morphogen \Bcd has been measured by several groups, 
yet its true value is still under debate \cite{Gregor2007b, Abu-Arish2010}.
In our model we therefore set for all patterning proteins an effective internuclear diffusion constant $D_P=1~\unit{\mumsps}$,
which comprises both protein import/export and actual diffusion.
This value is a reasonable cytoplasmic diffusion coefficient and well within the bounds reported for \Bcd.

The simulated lattice is 40 nuclei long so that the total system length $L$ roughly corresponds
to the posterior 2/3 of the \Drosophila embryo in cycle 14.
To reduce computation effort we simulate a system with smaller circumference (8 nuclei) as compared to the living embryo.
This is justified by the fact that for our standard diffusion constant $D_P$ and effective protein lifetime $\mu_{eff}$ 
the diffusive correlation length $l_{corr}=\sqrt{D_P/\mu_{eff}}$ is $\leq 2$ nuclei.
A larger circumference therefore is not expected to introduce new features into the system,
but might alter the timescales of expression boundary movement and domain desintegration.
We discuss the effect of reduced system size on measured stability times in the ``Discussion'' section.

A complete overview of the specific numerical values of our model parameters is found in Table~S1 of the \SI.

\subsection*{Data analysis}
\subsubsection*{Quantification of pattern stability}

In order to analyse pattern stability we represent each simulated pattern as a point in ($\lambda_{AC}$, $\lambda_{BD}$) phase space. For every pattern simulation from time $t=0$ until time $t=t_{\rm end}$ the temporal sequence of these points corresponds to a trajectory in the ($\lambda_{AC}$, $\lambda_{BD}$) space. For each parameter choice and pinning scenario, we restarted the simulations with 6000 trajectories started from the relaxed initial patterns at $t=0$; the trajectories ensemble is then further enriched by the branching process at the NS-FFS interfaces. Next, the trajectories are binned with the statistical weight assigned by NS-FFS, and then the histograms are normalized. As a result, we can identify a few distinct regions that accumulate probability.

In order to formally define these regions we define rectangular boundaries that enclose accumulated probability regions corresponding to different types of patterns:
\begin{itemize}
 \item the metastable main basin with five-stripe pattern:\\
       $R_S\equiv\{(\lambda_{AC}, \lambda_{BD}) | {\lambda_{AC}\leq 0.45 \wedge \lambda_{BD}\leq 0.43} \}$
 \item the basin in which either the \GA or \GC protein domain was lost:\\
       $R^\dagger_{AC}\equiv\{(\lambda_{AC}, \lambda_{BD}) | {\lambda_{AC} > 0.45 \wedge \lambda_{BD}\leq 0.43} \}$
 \item the basin in which either the \GB or \GD protein domain was lost:\\
       $R^\dagger_{BD}\equiv\{(\lambda_{AC}, \lambda_{BD}) | {\lambda_{AC} \leq 0.45 \wedge \lambda_{BD} > 0.43} \}$
 \item the basin in which either \GA or \GC and one of \GB or \GD were lost:\\
       $R^\ddagger\equiv\{(\lambda_{AC}, \lambda_{BD}) | {\lambda_{AC} > 0.45 \wedge \lambda_{BD} > 0.43} \}$
\end{itemize}

Note that the location of the regions slightly changes for different values of $\kappa$.
We found that the above boundary definitions constitute a good compromise.
For each basin we compute the fraction of total probability as a function of time by integrating the weights of
trajectories that are within the basin at time $t$.
We define the \emph{pattern survival probability} to be the integrated probability in $R_S$ at time $t$ after initialization:
$S(t)=\iint_{R_S} p(t) d\lambda_{AC} d\lambda_{BD}$.
As expected, $S(t)$ displays roughly exponential decay behavior after a certain lag phase that can be attributed to initial relaxation.
To obtain the \emph{pattern destruction rate} $k_D$ we fit a function $f(x)\equiv \exp(-k_D(t-t_{lag}))$ to $S(t)$.
This only yields satisfactory results if the fitting range is adapted accordingly, i.e. only $S(t)$ values for $t>t_{lag}$ are taken into account.
Since $t_{lag}$ itself is a fitting parameter we adopted the following protocol:
Starting from a value of $t_{start}$ that is clearly in the relaxation regime we perform the fit on the interval $[t_{start}, t_{end}]$
where $t_{end}$ is the largest time recorded.
We then choose the fitted values $k_D$ and $t_{lag}$ for which $|t_{lag}-t_{start}|$ is minimal.
From this we compute the pattern stability time (average time until pattern has lost one of the domains) via $\tau_D \equiv 1/k_D$.
In most considered cases the patterns are very stable, i.e. $k_D$ very small, and we can expand
$S(t)\simeq 1 - k_D(t-t_{lag})$.
As a control, we therefore also fitted $g(t)\equiv k_D(t-t_{lag})$ to $1-S(t)$ for a fixed $t_{lag}$ clearly in the
exponential regime and obtained almost identical results.

\subsubsection*{Computation of average probability fluxes}
To quantify which destruction pathways are dominant we computed the average fluxes $J_{avg}$ into the regions of (partly) destroyed patterns.
Here the average flux is defined as the average rate of increase in time of the fractional probability in the region and
obtained by fitting a linear function $h(t)\equiv J_{avg} t + P_0$ to $P_R(t) \equiv \iint_R p(t) d\lambda_{AC} d\lambda_{BD}$
for $R \in \{R^\dagger_{AC}, R^\dagger_{BD}, R^\ddagger \}$ over the interval $[t_{start},t_{end}]$ with $t_{start}$ chosen such that $\partial_{\it t} P_R(t)\neq0$ for $t>t_{start}$.
$P_0$ depends on the particular choice of $t_{start}$ and is discarded.

\subsubsection*{Computation of average flux velocities}
The average local drift velocity and diffusion constant of the trajectories in the $(\lambda_{AC},\lambda_{BD})$ phase space
are computed by averaging displacements $\Delta \lambda_{AC(BD)}\equiv\lambda_{AC(BD)}(t+\Delta t)-\lambda_{AC(BD)}(t)$
and squared displacements $\Delta \lambda^2 \equiv \Delta \lambda_{AC}^2 + \Delta \lambda_{BD}^2$ on a two-dimensional lattice of bins
covering the whole phase space.
Displacements $\Delta \lambda_{AC(BD)}$ are assigned to the bin at $\vecA{\lambda} \equiv (\lambda_{AC}, \lambda_{BD})$,
i.e. we are averaging outgoing displacements and the averaged vector $\langle\vecA{\Delta\lambda}\rangle (\vecA{\lambda})$ 
therefore will represent the average velocity with which trajectories \emph{leave} this bin.
The local phase space diffusion constant is calculated as
$D_\lambda (\vecA{\lambda}) \equiv \frac{1}{4\Delta t} \left[ \langle\Delta \lambda^2\rangle (\vecA{\lambda})
  - \left( \langle\Delta\lambda_{AC}\rangle^2 (\vecA{\lambda}) + \langle\Delta\lambda_{AC}\rangle^2 (\vecA{\lambda}) \right) \right]$.
This is done in the same way for other combinations of phase space coordinates.
The diffusion-drift decomposition is explained in more detail in the \SI.

\subsubsection*{Computation of ``pseudopotential'' landscapes}
The trajectory binning procedure used for computing the average flux velocities as described above was at the same time used for computing the ``pseudopotential'' $-\log(\tilde p(\vecA \lambda)$. Herein $\tilde p(\vecA \lambda)$ is the local density calculated from the reweighed number of trajectories \emph{leaving} the bin at $\vecA \lambda = (\lambda_{AC},\lambda_{BD})$, and smoothened afterwards by 2D median filtering over $n_{filt}$ neighboring bins. For the 2D median filtering we used the \texttt{medfilt2} function from the MATLAB Image Processing Toolbox. We empirically chose $n_{filt}=4$ as we found that this choice efficiently removes local spikes in $p(\vecA \lambda)$ without changing the overall shape of the landscape.

\subsection*{Perturbation experiments}
Simulations starting from perturbed initial conditions were performed directly via the GGG simulator.
First the systems were relaxed to representative states within the metastable basin for a simulated time
of $t_{relax}=30~\unit{min}$.
The final states of these runs then were post-modified according to the following two protocols:
\begin{enumerate}
 \item ``\GC expansion'': starting from mid-embryo the central \GC protein domain was expanded as follows:
	the configurations in the nuclei just posterior to mid-embryo were
	copied and used to overwrite configurations in the subsequent $\Delta$ rows in the axial 
	($z$-) direction of the cylinder. The original configurations were stored and for each nucleus at
	row $z_i > N_z/2+\Delta$ (counting from the anterior) the configuration was overwritten by 
	the original configuration at $z_i-\Delta$. The posterior-most nucleus was exempted from overwriting to preserve pinning.

 \item ``\GA expansion'': here the anterior \GA protein domain was enlarged at the expense of the \GC protein domain.
	To this purpose we applied the same copy-paste procedure as above starting form $z_i=5$,
	however only nuclei up to mid-embryo ($z_i \leq N_z/2$) were overwritten by the original configurations at $z_i-\Delta$.
\end{enumerate}
$\Delta$ quantifies the severity of perturbation. We found that $\Delta<4$ results in
changes to the pattern that were hard to distinguish from noise,
while for $\Delta > 12$ perturbations were large enough to induce immediate pattern destruction with high probability.
We therefore limited systematic tests to perturbations with $\Delta\in\lbrace 4, 8, 12 \rbrace$.
Starting from the perturbed initial conditions simulations were continued for $t_{sim}=20~h$ and
snapshots of the current configurations in all nuclei were written out with an acquisition interval 
of $10~\unit{min}$ (simulated time).
10 samples starting from 10 different perturbed initial conditions were produced for each set of parameters.

In order to overcome the difficulties of boundary detection we quantified the motion of protein domains
by tracking their center of mass (CoM) along the $z$-axis of the cylinder.
For each considered gene $G$ we define the CoM $z_G$ as
\begin{align}
 z_G &\equiv \frac{\int_z \int_r z G_{tot}(r,z) dr dz }{\int_z \int_r G_{tot}(r,z) dr dz}
\end{align}
where $G_{tot}=[G]+2[G_2]$ is the total copy number.
Since our system features two \GA domains we calculate $z_\GA$ separately for the anterior ($\GA_{ant}$)
and the posterior ($\GA_{post}$) part of the embryo by restricting $z$-integration adequately.
While the CoM remains unchanged upon symmetric changes of the domain boundaries or global copy number increase,
it is well-suited to indicate relaxations from the asymmetric perturbations that we apply.
To find general trends in the time-evolution of the domains CoM trajectories were averaged over the 10 samples.

\subsection*{Effective reaction-diffusion dynamics of dimer expression}

In this section we map the fully microscopic model defined in Fig.~\ref{Fig-ModelSketch} onto the effective model from \cite{Majka2023}. First, we postulate that the stochastic dynamics of gene expression studied in this paper corresponds to the following effective dynamical equations:
\begin{equation}
\begin{gathered}
\partial_{\it t} X_1(x,t)=-\mu_M X_1(x,t)-k_{on}^DX_1^2(x,t)+k_{off}^DX_2(x,t)+\beta f(\{Y_2(x,t)\}_{Y\neq X})~,\\
\partial_{\it t} X_2(x,t)=D_0\partial_{\it xx}X_2(x,t)-(k_{off}^D+\mu_D)X_2(x,t)+k_{on}^DX_1^2(x,t)~,
\end{gathered}\label{eq:start}
\end{equation}
where $X,Y\in\{A,B,C,D\}$ denotes the expressed protein species, $X_1$, $Y_1$ are the concentrations of its monomer, and $X_2$, $Y_2$ are concentrations of its dimers. The synthesis and decay of dimers is described by rates $k_{on}$ and $k_{off}$. Both, monomers and dimers degrade with rates $\mu_D$ and $\mu_M$. In this system, only dimers are allowed to diffuse (with diffusivity $D_0$) and only monomers are primarily synthesized, with maximal production rate $\beta$ and production kinetics described by function $f(\{Y_2(x,t)\}_{Y\neq X})$, which we specify later. However, we assume that in the absence of other dimers $Y_2\neq X_2$ the production is active by default, so $f(\{0\})=1$. Since the system has cylindrical symmetry, we will treat the axis $x$ as distinguished and treat the system as effectively one-dimensional.

The fact that the model defined by eqs. \eqref{eq:start} involves monomers and dimers complicates its mapping onto the model in \cite{Majka2023}. We therefore translate it into a simplified model, tracking the effective dynamics of dimers only. To this end, we will first determine the ratio between stationary concentrations $\tilde X_1$ and $\tilde X_2$ in the absence of other dimers ($Y_2(x,t)=0$) and assuming system homogeneity. In this case, equations \eqref{eq:start} turn into:
\begin{equation}
\begin{gathered}
0=-\mu_M \tilde X_1-k_{on}^D \tilde X_1^2+k_{off}^D \tilde X_2+\beta ~,\\
0=-(k_{off}^D+\mu_D)\tilde X_2+k_{on}^D \tilde X_1^2 ~.
\end{gathered}
\end{equation}
Solving for $\tilde X_1$ and $\tilde X_2$, we obtain:
\begin{equation}
\begin{gathered}
\tilde X_1=\frac{1}{2k_{on}^D}\left(-\frac{k_{off}^D+\mu_D}{\mu_D}+\sqrt{\frac{(k_{off}^D+\mu_D)^2}{\mu_D^2}\mu_M^2+4k_{on}^D \beta \frac{k_{off}^D+\mu_D}{\mu_D}} \right) ~,\\
\tilde X_2=\frac{\beta}{\mu_D}-\frac{\mu_M}{\mu_D}\tilde X_1 ~.
\end{gathered}
\end{equation}
We will now sum both equations in \eqref{eq:start} to obtain
\begin{equation}
\partial_{\it t}(X_1(x,t)+X_2(x,t))=D\partial_{\it xx}X_2(x,t)-\mu_D X_2(x,t)-\mu_M X_1(x,t)+\beta f(\{Y_2\}_{Y\neq X}) ~,
\end{equation}
and approximate
\begin{equation}
X_1(x,t)\approx \frac{\tilde X_1}{\tilde X_2} X_2(x,t) ~.
\end{equation}
In other words, we assume that $X_1(x,t)$ follows strictly $X_2(x,t)$. The advantage of this approximation is that it becomes exact in the stationary state. This procedure results in the following effective equation for $X_2(x,t)$:
\begin{equation}
\partial_{\it t} X_2(x,t)= D_X\partial_{\it xx} X_2(x,t)-\gamma_X X_2(x,t)+H_X f(\{Y_2\}_{Y\neq X}) ~, \label{eq:effectiveform}
\end{equation}
where the rescaled constants are:
\begin{align}
D_X=\frac{D_0}{1+\frac{\tilde X_1}{\tilde X_2}} ~, &&\gamma_X=\frac{\mu_D+\mu_M\frac{\tilde X_1}{\tilde X_2}}{1+\frac{\tilde X_1}{\tilde X_2}} ~, &&H_X=\frac{\beta}{1+\frac{\tilde X_1}{\tilde X_2}} ~.
\end{align}

We can now specify the kinetics function. The microscopic dynamics is such that each gene $X$ is produced, unless it is blocked by the biding of any other dimer to its repressor site on the promoter. In the averaged-out description, we expect that a sufficiently high concentration of free repressor particles effectively shuts down the production of $X$. Similarly to \cite{Majka2023}, we will assume that this transition is steep, so we can choose the functional form of the regulatory Hill function in \eqref{eq:effectiveform} to have the overall shape of Heaviside step function:
\begin{equation}
f(\{Y_2\}_{Y\neq X})=\theta\left(\sum_{Y\neq X}\epsilon_{XY} Y_2(x,t)-C_X\right) ~. \label{eq:theta}
\end{equation}

Finally, we relate the effective gene interaction constants $\epsilon_{XY}$ to microscopic parameters by the following reasoning: In the microscopic simulations, the attachment of $Y_2$ to the repressor site is described by the constant $k_{on}^R$ and detachment by $k^{\rm off}_{\rm s}$ or $k^{\rm off}_{\rm w}$. Assuming that the repressor production speed can be approximated by Michealis-Menten kinetics, with the repressor site acting like a ``catalyst'', we know that
\begin{equation}
K_{Y}=k^{\rm off}_{\rm w,s}/k_{\rm on}^R
\end{equation}
where $k^{\rm off}_{\rm w,s}$ (standing for either $k^{\rm off}_{\rm w}$ or $k^{\rm off}_{\rm s}$) is the concentration of repressor dimers $Y_2$ at which the velocity of production of $Y_1$ is at the half of its maximal value. We postulate that at this point $Y_2$ effectively switches off the production of $X$, and we equate this point with reaching the threshold for production in \eqref{eq:theta}. Hence, the following is satisfied:
\begin{equation}
\epsilon_{XY}K_{Y}-C_X=0 ~.
\end{equation}
Solving for $\epsilon_{XY}$ we obtain:
\begin{align}
\epsilon_{XY}=\frac{C_X}{K_{Y}} ~.
\end{align}
We choose $C_X<0$ to ensure that the production of $X$ is active by default, in the absence of repressive dimers ($Y_2(x,t)=0$). Since $C_X$ is now present in every term in \eqref{eq:theta}, we can factor it out and neglect. Taken together, and assuming that diffusion, degradation and production constants are the same for all genes, that is: $D_X=D$, $\gamma_X=\gamma$ and $H_X=H$ for all $X\in\{A,B,C,D\}$; the microscopic dynamics of gene expression mapped onto the effective model results in eq. \eqref{eq:effdyn}.

\changed{
\subsection*{Derivation of stability conditions for a contact zone between two domains}

This section outlines the origin of stability conditions for the effective continuous model, derived in the previous section, which is employed for the analysis of the four-gene pattern in the approximation of separate contact zones. We discuss the major steps leading to stability conditions in the current case, taking advantage of specific setting of the system studied in this work. For a detailed and more general derivation we refer the reader to our earlier work \cite{Majka2023}.

Let us consider a pair of genes (X,Y), whose expression dynamics is described by the effective equations:
\begin{equation}
\begin{gathered}
\partial_{\it t} X_2(x,t)= D~\partial_{\it xx} X_2(x,t)-\gamma X_2(x,t)+H~\theta\left(\tilde C_X- \epsilon_{XY} Y_2(x,t) \right)\\
\partial_{\it t} Y_2(x,t)= D~\partial_{\it xx} Y_2(x,t)-\gamma Y_2(x,t)+H~\theta\left(\tilde C_Y- \epsilon_{YX} X_2(x,t) \right)
\end{gathered}
\end{equation}
These equations describe both type-(i) and type-(ii) contact zones, albeit for different values of constants (see Eqs. \eqref{eq:type(i)} and \eqref{eq:type(ii)}). We assume that the system is open ($L\to+\infty$) and two respective expression domains occupy the opposite `ends' of the system. That is, initially, the expression profiles of respective dimers read:
\begin{align}
X_2(x,0)=A_X\theta(q_X(0)-x)&&Y_2(x,0)=A_Y\theta(x-q_Y(0))
\end{align}
Here $A_X$ and $A_Y$ are initial amplitudes, sufficiently high to initiate auto-activation, while $q_X(t)$ and $q_Y(t)$ denote the positions of expression domain boundaries.

The effective equations in the form \eqref{eq:effdyn} can be solved analytically in this system, providing the spatio-temporal profile of expression for both dimers:
\begin{equation}
\begin{gathered}
X_2(x,t)=\int_{-\infty}^{+\infty}dx'G(x-x',0)X_2(x,0)+H\int_0^t dt' \int_{q_X(t')}^{+\infty}dx' G(x-x',t-t') \\
Y_2(x,t)=\int_{-\infty}^{+\infty}dx'G(x-x',0)Y_2(x,0)+H\int_0^t dt' \int_{-\infty}^{q_Y(t')}dx' G(x-x',t-t')
\end{gathered}\label{eq:expprofsol}
\end{equation}
where the Green's function of eq. \eqref{eq:effdyn} in the open system reads:
\begin{equation}
G(x-x',t-t')=\frac{e^{-\gamma t-\frac{(x-x')^2}{4D(t-t')}}}{\sqrt{4\pi D (t-t')}}
\end{equation}
However, the solution \eqref{eq:expprofsol} is known only up to the position of domain boundaries, $q_X(t)$ and $q_Y(t)$. In order to determine these positions, the "Free Boundary Problem" must be solved, defined by the activation conditions at each boundary:
\begin{equation}
\begin{gathered}
\tilde C_X=\epsilon_{XY} Y_2(q_X(t),t)\\
\tilde C_Y=\epsilon_{YX} X_2(q_Y(t),t)\\
\end{gathered}
\end{equation}
Inserting the solution \eqref{eq:expprofsol} into these equations leads to a system of coupled nonlinear integral equations. For $t$ large enough that the system `forgets' its initial conditions, these equations simplify into:
\begin{equation}
\begin{gathered}
\frac{\tilde C_X}{\epsilon_{XY}H}=\int_0^t dt' \int^{q_Y(t')}_{-\infty}dx' G(q_X(t)-x',t-t') \\
\frac{\tilde C_Y}{\epsilon_{YX}H}=\int_0^t dt' \int^{+\infty}_{q_X(t')}dx' G(q_Y(t)-x',t-t')
\end{gathered}\label{eq:inteqs}
\end{equation}
In \cite{Majka2023} we show that the asymptotic solution of these equations is provided by the constant velocity ansatz, $q_X(t)=v_Xt+q_X^{\infty}$ and $q_Y(t)=v_Yt+q_Y^{\infty}$, where $q_X^{\infty}$ and $q_Y^{\infty}$ are constants. For this choice, the right-hand side integrals in Eqs. \eqref{eq:inteqs} saturate at constant values, though corresponding to domain boundaries travelling with constant velocities. Moreover, for sufficiently strong interactions between the genes, these velocities must be equal, $v_X=v_Y=v$, meaning that the domains change their size in a coordinated manner. For the constant velocity ansatz, the integrals can be analytically computed. This turns the system of integral equations system into an algebraic system, defining the common velocity $v$ and the distance between the boundaries $\Delta r=q_X^{\infty}-q_Y^{\infty}$. In the $t\to+\infty$ limit, this system reads:
\begin{equation}
\begin{gathered}
\frac{2\tilde C_X\gamma}{\epsilon_{XY}H}=\textrm{sgn}(\Delta r)-e^{\frac{v\Delta r}{D}-\frac{|\Delta r|\sqrt{4D\gamma+v^2}}{2D}}\left(\frac{v}{\sqrt{4D\gamma+v^2}}+\textrm{sgn}(\Delta r) \right)\\
\frac{2\tilde C_Y\gamma}{\epsilon_{YX}H}=\textrm{sgn}(\Delta r)+e^{\frac{v\Delta r}{D}-\frac{|\Delta r|\sqrt{4D\gamma+v^2}}{2D}}\left(\frac{v}{\sqrt{4D\gamma+v^2}}-\textrm{sgn}(\Delta r) \right)
\end{gathered}\label{eq:integrated}
\end{equation}
Finally, the patterns for which $v=0$ are stable, that is, they do not change in the long-time limit. Substituting $v=0$ turns Eqs. \eqref{eq:integrated} into:
\begin{equation}
\begin{gathered}
R_X=\textrm{sgn}(\Delta r)\left(1-e^{-\frac{|\Delta r|}{\lambda}}\right)\\
R_Y=\textrm{sgn}(\Delta r)\left(1-e^{-\frac{|\Delta r|}{\lambda}}\right)
\end{gathered}\label{eq:integrated2}
\end{equation}
where $R_X$ and $R_Y$ are defined as in Eqs. \eqref{eq:effparams}. In order to make $v=0$ the solution of Eqs. \eqref{eq:integrated}, Eqs. \eqref{eq:integrated2} must be solvable and satisfied by the same $\Delta r$, as the system is over-defined. This happens, provided that:
\begin{align}
R_X=R_Y&&-1<R_X<1&&-1<R_X<1
\end{align}
which constitutes the stability conditions utilized in this work. Further, we can also derive:
\begin{equation}
|\Delta r|=-\lambda \ln\left(1-\frac{R_X}{\textrm{sgn}(\Delta r)}\right)
\end{equation}
However, one can notice in Eqs. \eqref{eq:integrated2} that as $1-e^{-|\Delta r|/\lambda}>0$ for any $\Delta r$, then: 
\begin{equation}
\textrm{sgn}(\Delta r)=\textrm{sgn}(R_X)=\textrm{sgn}(R_Y)
\end{equation}
This allows us to obtain $\Delta r$, as provided by the formula \eqref{eq:width}.
}

\section*{Acknowledgments}
M.M. and M.Z. were supported by the Polish National Agency for Academic Exchange and by a grant from the Priority Research Area DigiWorld under the Strategic Programme Excellence Initiative at Jagiellonian University. M.Z. was supported by National Science Center, Poland, no. 2021/42/E/NZ2/00188. T.R.S., N.B.B. and P.R.t.W. were supported by the Foundation for Fundamental Research on Matter (FOM), which is part of the Netherlands Organisation for Scientific Research (NWO). T.R.S. was supported by the Center for Multiscale Modelling in Life Sciences (CMMS), sponsored by the Hessian Ministry of Science and Art. The funding agencies did not play any role in the study design, data collection and analysis, decision to publish, or preparation of the manuscript.


\bibliographystyle{plos2015}
\bibliography{GapGenes}

\end{document}


\begin{flushleft}
{\Large
\textbf{SUPPLEMENTARY INFORMATION}\\
\vspace{1em}
\textbf{Stable developmental patterns of gene expression without morphogen gradients}
}
\\
\vspace{1em}
Maciej Majka$^{1}$, 
Nils B. Becker$^{2,3}$,
Pieter Rein ten Wolde$^{2}$, 
Marcin Zagorski$^{1}$,
and Thomas R. Sokolowski$^{2,4,\ast}$
\\
\bf{1} Institute of Theoretical Physics and Mark Kac Center for Complex  Systems Research, Jagiellonian  University, ul. prof. Stanis\l{}awa   Lojasiewicza 11, 30-348 Krak\'{o}w, Poland
\\
\bf{2} FOM Institute AMOLF, Science Park 104, 1098 XG Amsterdam, The Netherlands
\\
\bf{3} Present address: Theoretical Systems Biology, German Cancer Research Center, \\69120 Heidelberg, Germany
\\
\bf{4} Present address: Frankfurt Institute for Advanced Studies (FIAS), Ruth-Moufang-Straße 1, 60348 Frankfurt am Main, Germany
\\
\vspace{1ex}
$\ast$ Corresponding author, e-mail: sokolowski@fias.uni-frankfurt.de
\end{flushleft}

\vspace*{1cm}

\changed{
\section{Chemical equations describing the gene regulatory network}
The biochemical system that we model is governed by the following set of chemical equations:

\subsection*{Monomer production, dimerization, monomer and dimer degradation}
\ce{P^A ->[$\beta$] A}
\hfill
\ce{A <=>[$\kDon$][$\kDoff$] A2}
\hfill
\ce{A ->[$\mu_M$] $\varnothing$}
\hfill
\ce{A2 ->[$\mu_D$] $\varnothing$}

\subsection*{Promoter repression and derepression (1st repressor dimer binding)}
\ce{P^A <=>[{$\kRon$[C_2]}][$\kRoffS$] P^A_{C_2}}
\hfill
\ce{P^A <=>[{$\kRon$[B_2]}][$\kRoffW$] P^A_{B_2}}
\hfill\ce{P^A <=>[{$\kRon$[D_2]}][$\kRoffW$] P^A_{D_2}}

\subsection*{Promoter repression and derepression (2nd repressor dimer binding)}
\ce{P^A_{B_2} <=>[{$\kRon$[C_2]}][$\kRoffS$] P^A_{B_2C_2}}
\hspace{3.8cm}
\ce{P^A_{C_2} <=>[{$\kRon$[B_2]}][$\kRoffW$] P^A_{B_2C_2}}\\
\\
\\
\ce{P^A_{B_2} <=>[{$\kRon$[D_2]}][$\kRoffW$] P^A_{B_2D_2}}
\hspace{3.8cm}
\ce{P^A_{D_2} <=>[{$\kRon$[B_2]}][$\kRoffW$] P^A_{B_2D_2}}\\
\\
\\
\ce{P^A_{C_2} <=>[{$\kRon$[D_2]}][$\kRoffW$] P^A_{C_2D_2}}
\hspace{3.8cm}
\ce{P^A_{D_2} <=>[{$\kRon$[C_2]}][$\kRoffS$] P^A_{C_2D_2}}\\

\subsection*{Promoter repression and derepression (3rd repressor dimer binding)}
\ce{P^A_{B_2D_2} <=>[{$\kRon$[C_2]}][$\kRoffS$] P^A_{B_2C_2D_2}}
\hfill
\ce{P^A_{B_2C_2} <=>[{$\kRon$[D_2]}][$\kRoffW$] P^A_{B_2C_2D_2}}
\hfill
\ce{P^A_{C_2D_2} <=>[{$\kRon$[B_2]}][$\kRoffW$] P^A_{B_2C_2D_2}}\\

\pagebreak
Note that here, for brevity, we only show the complete set of reactions for the promoter of gene \GA, and its products (A monomers and dimers A_2).
An analogous set of equations holds for the promoters of genes \GB, \GC and \GD, with $\rm P^A$ replaced by $\rm P^B$, $\rm P^C$ or $\rm P^D$, respectively, all occurrences of the corresponding dimers (B_2, C_2 or D_2) replaced by A_2 and corresponding monomers A replaced by B, C, or D, respectively, and the unbinding rates adjusted to the respective strength of mutual interactions.

In addition to the shown set of equations, we provide a graph-like summary of the reaction network in Fig.~\ref{Fig-GG-SI-ReactionNetwork}.
}
\section{Analysis of perturbation experiments for assessing pattern restoring forces}

\begin{figure}[ht!]
  \subfigure{
    \label{Fig-SI-RestoringForcePinnedFromCenter}
    \includegraphics[width=0.48\textwidth]{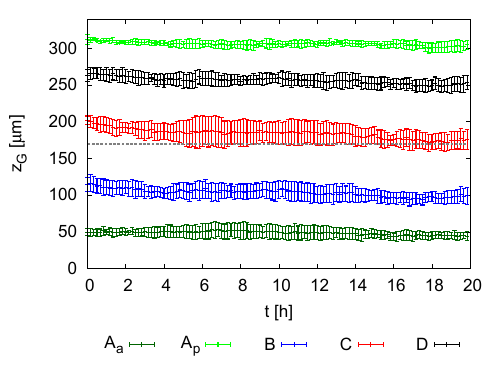}
  }
  \subfigure{
    \label{Fig-SI-RestoringForcePinnedFromBoundary}
    \includegraphics[width=0.48\textwidth]{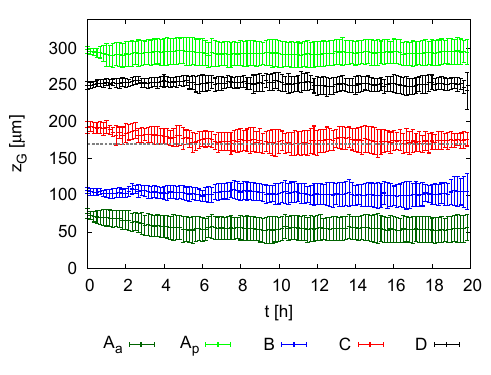}
  }
\caption{\textbf{Perturbed trajectories are restored to their origin at optimal NN repression strength.}
    Shown are averaged time traces of the copy number center-of-masses $z_G$ for the five domains
    of the stable pattern (${\rm \GA}_a$ = anterior \GA domain, ${\rm \GA}_p$ = posterior \GA domain) at $\kappa=\kappa_{opt}$
    for two different perturbations: 
    \mysubref{Fig-SI-RestoringForcePinnedFromCenter} ``\GC expansion'', i.e. prolongation of the 
    central \GC domain by $\Delta=8$ nuclei into the posterior
    and \mysubref{Fig-SI-RestoringForcePinnedFromBoundary} ``\GA expansion'', 
    i.e. prolongation of the anterior \GA domain by $\Delta=8$ nuclei towards the center of the embryo.
    The gray-dashed line marks the center of the system.
    In both cases we observe a restoration of the metastable state on a timescale $\lesssim 10~\unit{h}$.\\
  \label{Fig-SI-RestoringForcePinned}
}
\end{figure}

Perturbation experiments on patterns initially relaxed into their long-persisting intact state were carried out by expanding expression domains at the expense of the respective antagonistic gene's domains and simulating the subsequent dynamics of the system, as described in the main text and Methods, Sec. 4. The stochastic time trajectories of such perturbation experiments that were repeated many times were then analysed as follows:

In order to quantify the spatial properties of a given domain $G$ we used its center of mass, $z_G$, where $G$ stands for one of the five domains $\{ \GA_a, \GB, \GC, \GD, \GA_p \}$, with $\GA_a$ and $\GA_p$ marking the anterior and posterior parts of $\GA$. Advantageously, $z_G$ is a robust measure, as it avoids ambiguity associated with determining domain boundaries in the presence of gene expression noise.
Figure \ref{Fig-SI-RestoringForcePinned} shows, for $\kappa=\kappa_{opt}$,
time traces of $z_G$ for the two types of perturbations, averaged over 10 independent samples in each case. For both perturbations the average centers of copy number relax back to their original positions on a timescale $\sim 10~h$. This demonstrates that for optimal repression strength ratio an effective restoring force counteracts deviations from the five-stripe pattern for varied $\lambda_{AC}$.

\section{Estimation of phase space diffusion coefficient from overdamped Langevin dynamics}
\label{Sec-SI-OverdampedLangevin} \index{overdamped Langevin dynamics}

Let $f(X,Y,t)$ be a twice differentiable real function depending on a two-dimensional diffusion-drift processes $\vekA X=(X,Y)$ and time $t$ (explicitly).
In the overdamped Langevin limit, i.e. assuming that the displacements of the random walker are governed only
by forces that stem from an underlying force field and by Gaussian noise, and that its accelerations and inertia are negligible,
we can describe this random processes via
\begin{align}
d\vekA X & = \vekA v dt + \sigma d\vekA W
\end{align}
where $\vekA W$ is a (two-dimensional) Wiener processes and $\vekA v = (v_X, v_Y)$ a (local) drift velocity resulting from the potential forces.

We then can calculate the differential of $f$ with \textsc{It\={o}}'s Lemma (as a generalization of Taylor expansion) as follows:
\newcommand{\pder}[2]{\frac{\partial #1}{\partial #2}}
\newcommand{\pderII}[2]{\frac{\partial^2 #1}{\partial #2^2}}
\begin{align}
 df(X,Y,t) & = \sigma \pder{f}{X} dW_X + \sigma \pder{f}{Y} dW_Y  \nn\\
	   & \quad + \left[ \pder{f}{t} + v_X \pder{f}{X} + v_Y \pder{f}{Y}
			    + \frac{\sigma^2}{2}\pderII{f}{X} + \frac{\sigma^2}{2}\pderII{f}{Y} 
			    + \zeta \sigma^2 \frac{\partial^2 f}{\partial X \partial Y} \right] dt	\nn\\
\end{align}
Here $\zeta$ measures the correlation between $X$ and $Y$.

In order to apply this general formula to the specific diffusion-drift problem for the phase space coordinates $(\lambda_{AC}, \lambda_{BD})$ defined in the main text Eq. (2), we assign for brevity $\lambda_x = \lambda_{AC}$ and $\lambda_y = \lambda_{BD}$, and then
we set $X=\lambda_x, Y=\lambda_y$ and $f(X,Y,t)=f(\lambda_x,\lambda_y)=(\lambda_x-\lambda_{x_0})^2 + (\lambda_y-\lambda_{y_0})^2 \equiv \Delta\lambda^2$
(the squared displacement function).

It\={o}'s Lemma now reads (note that the time and mixed derivatives vanish):
\begin{align}
 d (\Delta\lambda^2) &= d \left[ (\Delta\lambda_x)^2 + (\Delta\lambda_y)^2 \right] = d \left[ (\lambda_x-\lambda_{x_0})^2 + (\lambda_y-\lambda_{y_0})^2 \right] \nn\\
		     &\simeq 2(\lambda_x-\lambda_{x_0}) (v_{\lambda_x} dt + \sigma dW_x) + 2(\lambda_y-\lambda_{y_0}) (v_{\lambda_y} dt + \sigma dW_y) \nn\\
		     & \quad + \frac{\sigma^2}{2} 2 dt + \frac{\sigma^2}{2} 2 dt
\end{align}

To relate the above formula to the displacements sampled in our simulations with a fixed acquisition time interval $\Delta t$
we shall integrate the infinitesimal contributions over this interval.
At the same time we take the ensemble average to account for the averaging of independent samples, which causes the
Gaussian terms $\sigma dW_x$ and $\sigma dW_y$ to vanish.
We further assume that, to a good approximation, the drift velocities and diffusion coefficients are constant 
over the time interval $\Delta t$ and diffusion isotropic in $\lambda_x$ and $\lambda_y$ direction,
i.e. $D_{\lambda_x}=D_{\lambda_y}=D_\lambda(\vekA{\lambda})$.
Finally, using $\sigma = \sqrt{2 D_\lambda}$, we obtain:
\begin{align}
 \avg{\Delta\lambda^2} &= \avg{\int_{\Delta t} d(\Delta\lambda^2) } 	\nn\\
		       &= \avg{\int_{0}^{\Delta t} 2 \underbrace{\left[ \lambda_x(t) - \lambda_{x}(0) \right]}_{\simeq\avg{v_{\lambda_x}}t} \underbrace{v_{\lambda_x}}_{\simeq \avg{v_{\lambda_x}}} dt}
			+ \avg{\int_{0}^{\Delta t} 2 \underbrace{\left[ \lambda_y(t) - \lambda_{y}(0) \right]}_{\simeq\avg{v_{\lambda_y}}t} \underbrace{v_{\lambda_y}}_{\simeq \avg{v_{\lambda_y}}} dt} \nn\\
		       &\quad + \avg{\int_0^{\Delta t} 4D_\lambda dt} 
			      + \int_{\Delta t} \underbrace{\avg{2\Delta\lambda_x\sigma dW_x}}_{0} + \int_{\Delta t} \underbrace{\avg{2\Delta\lambda_y\sigma dW_y } }_{0} \nn\\
		       &\simeq \avg{ \avg{v_{\lambda_x}}^2 \int_{0}^{\Delta t} 2t dt} + \avg{ \avg{v_{\lambda_y}}^2 \int_{0}^{\Delta t} 2t dt}
			  + 4 \avg{D_\lambda} \Delta t										\nn\\
		       &\simeq \avg{v_{\lambda_x}\Delta t}^2 + \avg{v_{\lambda_y}\Delta t}^2 + 4 \avg{D_\lambda} \Delta t	\nn\\
		       &= \avg{\Delta{\lambda_x}}^2 + \avg{\Delta{\lambda_y}}^2 + 4 \avg{D_\lambda} \Delta t
\end{align}

The final result shows that, knowing the average displacements $\avg{\Delta{\lambda_x}}$ and $\avg{\Delta{\lambda_x}}$ and average
squared displacements $\avg{\Delta\lambda^2}$ at $\vekA{\lambda}$, we can compute
the average diffusion coefficient $\avg{D_\lambda}(\vekA{\lambda})$ via:
\begin{align}
 \avg{D_\lambda}(\vekA{\lambda}) &= \frac{1}{4\Delta t} \left[ \avg{\Delta\lambda^2}(\vekA{\lambda}) - \left( \avg{\Delta{\lambda_x}}^2(\vekA{\lambda}) + \avg{\Delta{\lambda_y}}^2(\vekA{\lambda}) \right)\right]
				 = \frac{1}{4\Delta t} \mathcal{V}_{\avg{\lambda}}(\vekA{\lambda})
\end{align}
The bracket term containing the first moments corrects the mean squared displacement for the contributions coming from the deterministic drift
and tends to zero as the process becomes purely diffusive.

\begin{figure}[H]
  \centering
  \subfigure[][]{
    \includegraphics[width=0.45\textwidth]{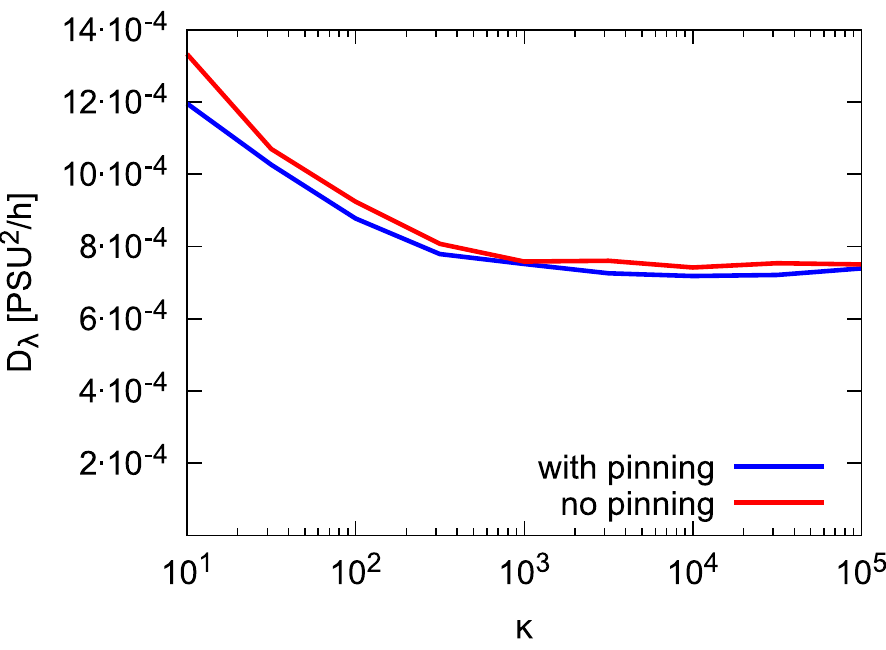}
    \label{Fig-GG-SI-DiffusionCoefficients}
  }
  \subfigure[][]{
    \includegraphics[width=0.45\textwidth]{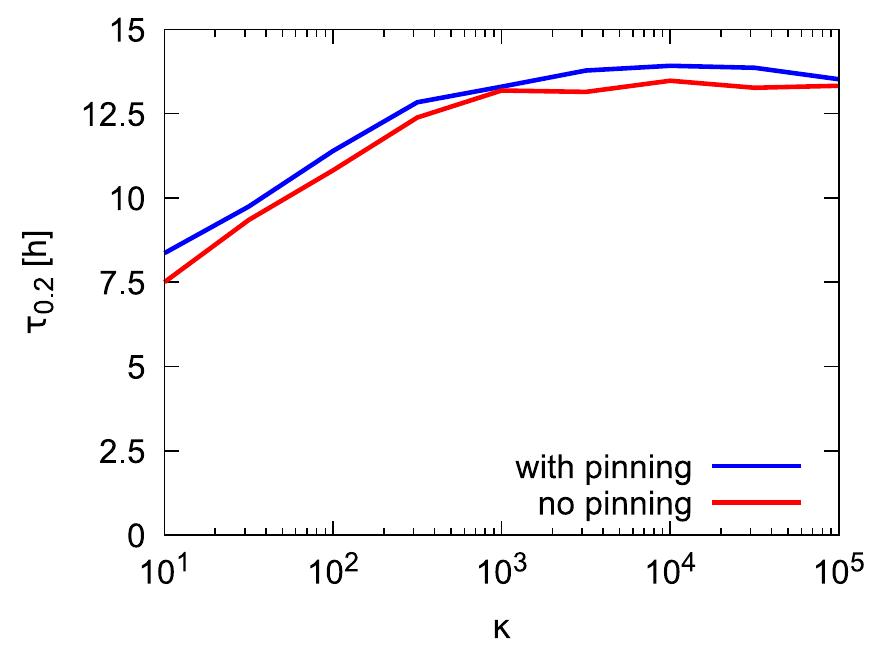}
    \label{Fig-GG-SI-DiffusionTimes}
  }
  \hspace*{0.465\textwidth} 
  \subfigure[][]{
    \includegraphics[width=0.45\textwidth]{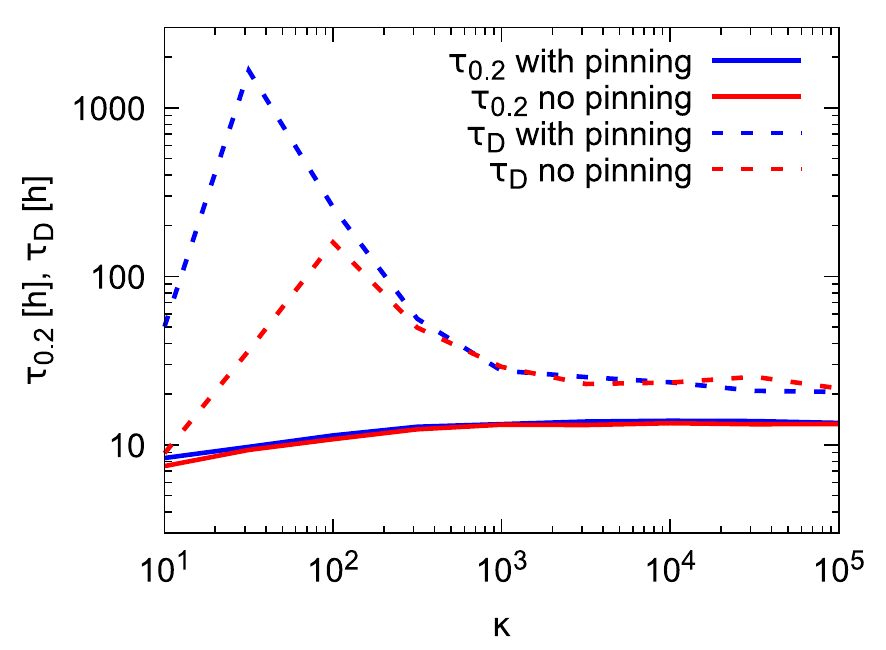}
    \label{Fig-GG-SI-DiffusionTimesPlusStab}
  }
  \caption{\textbf{Average phase space diffusion coefficients as a function of $\kappa$.} 
  \mysubref{Fig-GG-SI-DiffusionCoefficients} The diffusion coefficient of phase space trajectories in the $(\lambda_x, \lambda_y) = (\lambda_{AC}, \lambda_{BD})$ space are obtained from the overdamped Langevin analysis, see section \ref{Sec-SI-OverdampedLangevin}. The diffusion coefficients are averaged over the phase space region $R_P=[0.3,0.4]^2$, which is part of the diffusive plateau, for different repression strength ratios $\kappa$. The PSU stands for the phase space units.
  \mysubref{Fig-GG-SI-DiffusionTimes} The resulting approximate diffusion times from the phase space region of the five-stripe relaxed patterns, towards the edge of the diffusive plateau as a function of $\kappa$, assuming a distance of $0.2$ PSU for the initial phase space distance to the edge. The edge of diffusive plateau is defined as a region of $(\lambda_{AC}, \lambda_{BD})$ from which the states are quickly absorbed into the regions with one the domains lost.
  \label{Fig-GG-SI-Diffusion}
}
\end{figure}

The results of this analysis applied to our simulated systems are shown in Fig.~\ref{Fig-GG-SI-Diffusion}, where we plot the diffusion coefficients and the resulting expected diffusion times to the edge of the basin of initial patterns as a function of our main parameter, the repression strength ratio $\kappa$. Interestingly, we find that the diffusion constant remains almost constant for large $\kappa$, corresponding to weak nearest-neighbor repression (Fig.~\ref{Fig-GG-SI-Diffusion}A). In this regime, the estimated time of diffusion to the edge of the stable basin, $\tau_{\rm 0.2}$ is about 12~hours and comparable to the recorded average stability times ($\sim$~20~hours, panels B and C). When $\kappa$ is reduced, meaning that the nearest-neighbor repression is increased, we observe an increase of the diffusion constant and correspondingly a reduction of $\tau_{\rm 0.2}$ to values below 10 hours. Since in this regime the pattern stability increases dramatically, our analysis corroborates the finding that the stability increase is not due to a slow-down of the pattern boundary dynamics, but due to the emergence and deepening of a metastable basin that generates restoring forces to pattern perturbations.

\pagebreak

\section{Supplementary velocity field figures}
\begin{figure}[H]
  \centering
  \includegraphics[width=\textwidth]{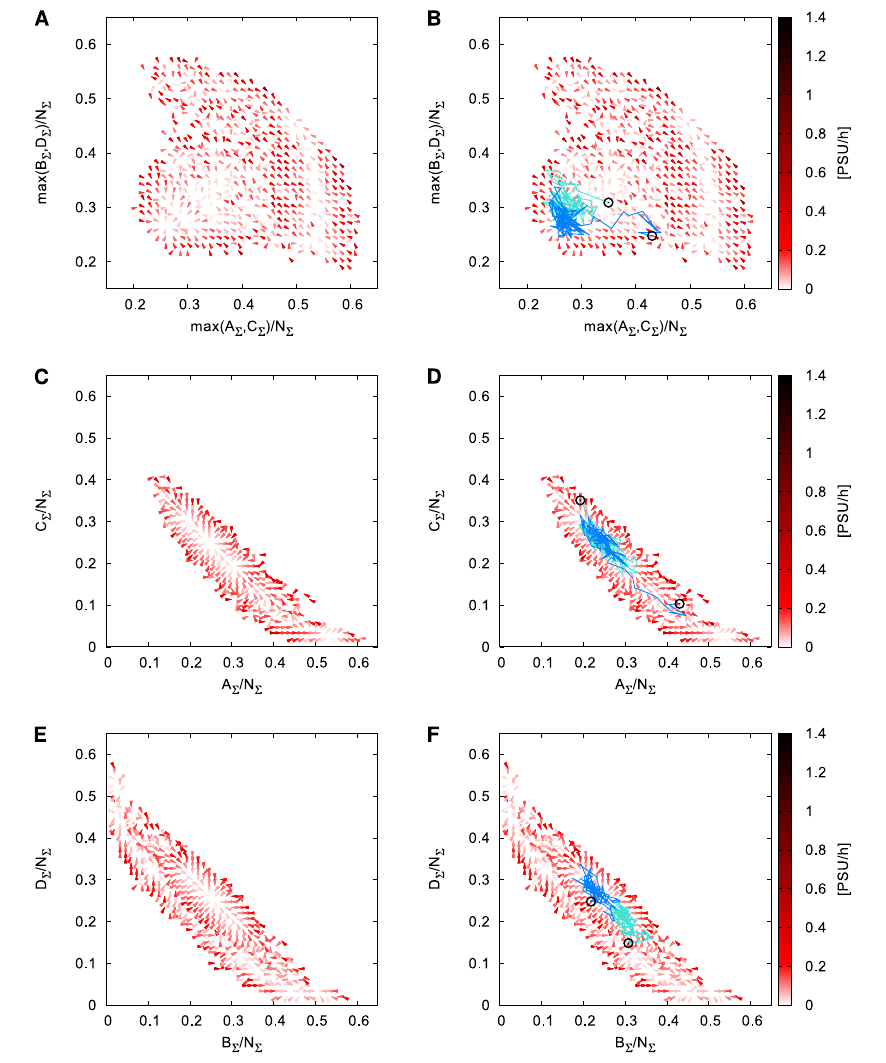}
  \caption{\textbf{Average phase space velocities for the maximally stable system ($\kappa\simeq31.6$) with pinning.}
      Left plots (A, C, E) show local average phase space velocities, right plots (B, D, F) additionally show 
      example trajectories for the two types of perturbations considered in the restoration experiments (blue = pert. from boundary, turquoise = pert. from center).
      Starting points are marked by black bullets.
  \label{Fig-GG-SI-VelocitiesPinnedOptimum}
 }
\end{figure}

\begin{figure}[H]
  \centering  
  \includegraphics[width=\textwidth]{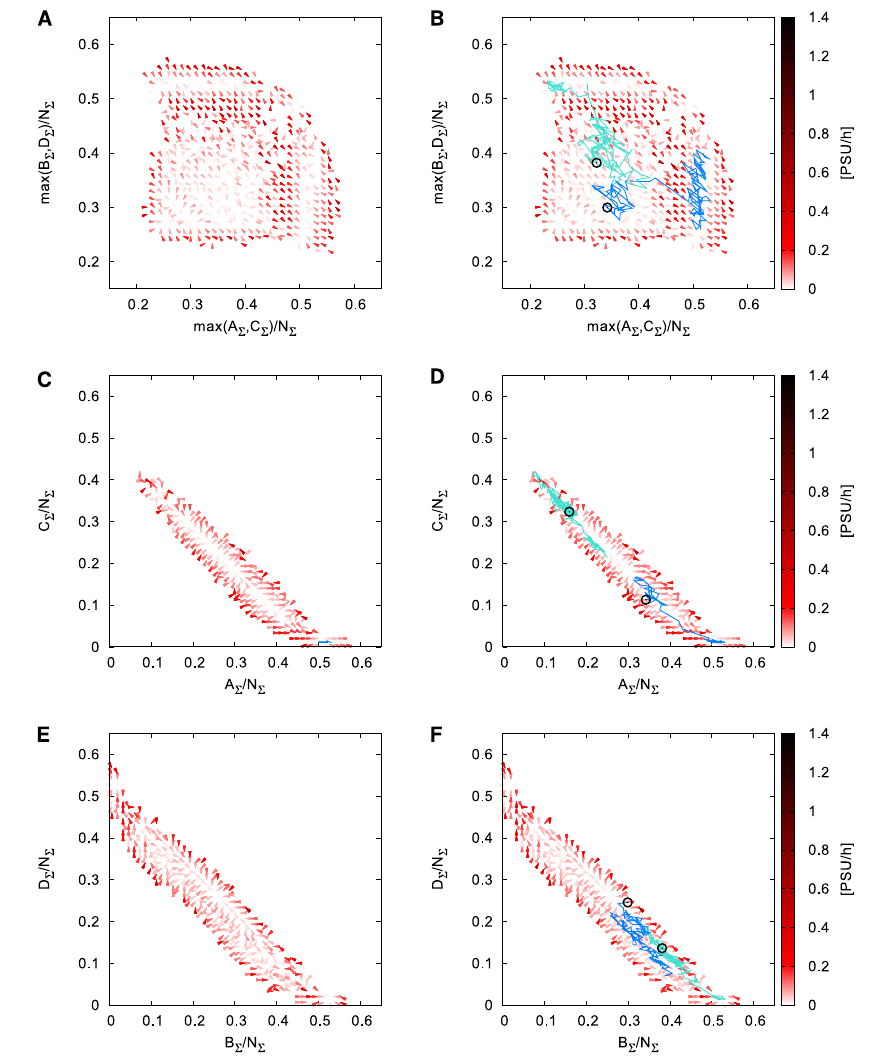}
  \caption{\textbf{Average phase space velocities for weaker NN interaction ($\kappa=1000$) in the system with pinning.}
      Left plots (A, C, E) show local average phase space velocities, right plots (B, D, F) additionally show 
      example trajectories for the two types of perturbations considered in the restoration experiments (blue = pert. from boundary, turquoise = pert. from center).
      Starting points are marked by black bullets.
    \label{Fig-GG-SI-VelocitiesPinnedKappa1000}
    }
\end{figure}

\begin{figure}[H]
  \centering
  \includegraphics[width=\textwidth]{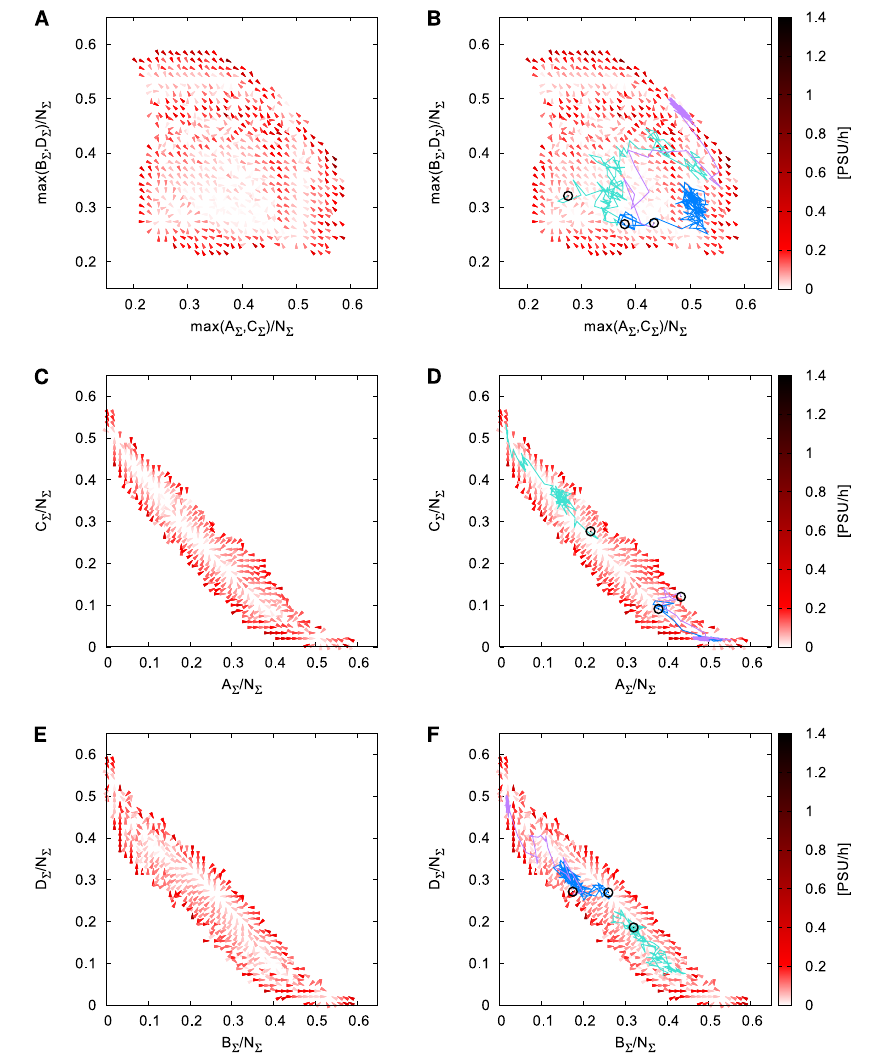}
  \caption{\textbf{Average phase space velocities for the maximally stable system ($\kappa=100$) without pinning.}
    Left plots (A, C, E) show local average phase space velocities, right plots (B, D, F) additionally show 
    example trajectories for the two types of perturbations considered in the restoration experiments (blue = pert. from boundary, turquoise = pert. from center; purple = pert. from center resulting in complete pattern destruction).
    Starting points are marked by black bullets.
    \label{Fig-GG-SI-VelocitiesUnpinnedOptimum}
    }
\end{figure}

\begin{figure}[H]
  \centering
  \includegraphics[width=\textwidth]{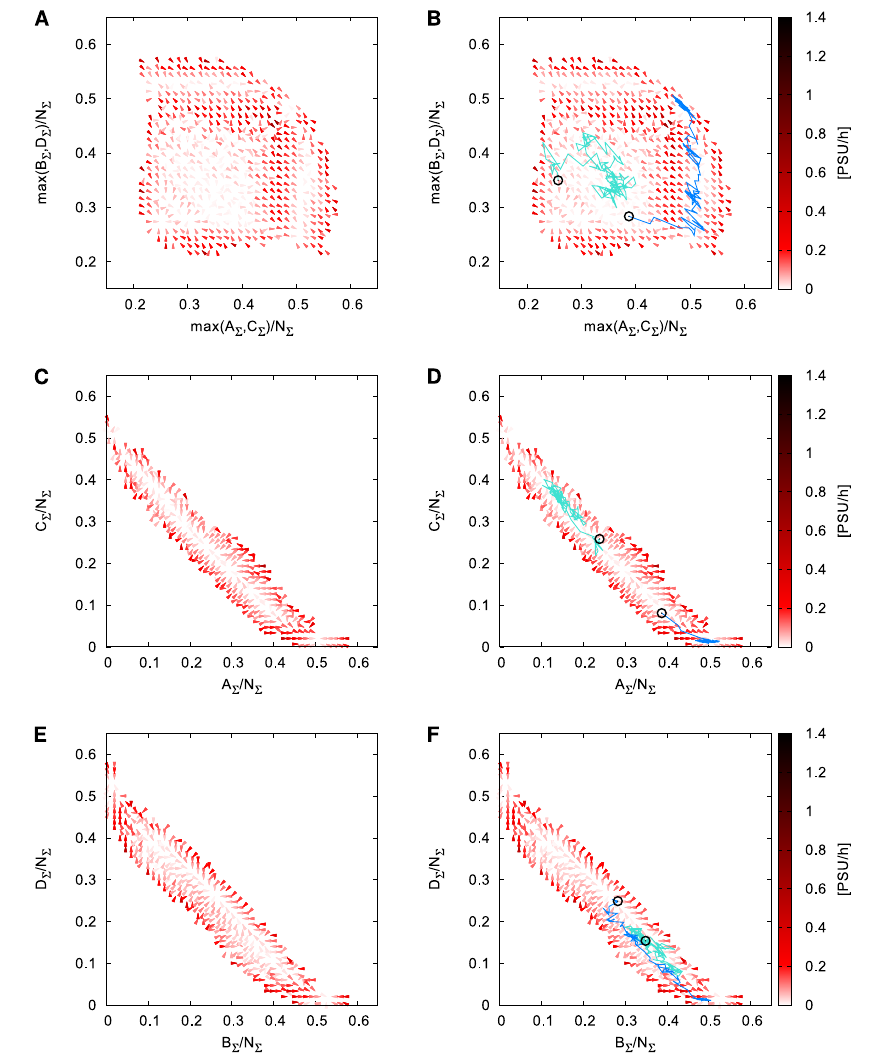}
  \caption{\textbf{Average phase space velocities for weaker NN interaction ($\kappa=1000$) in the system without pinning.}
    Left plots (A, C, E) show local average phase space velocities, right plots (B, D, F) additionally show 
    example trajectories for the two types of perturbations considered in the restoration experiments (blue = pert. from boundary, turquoise = pert. from center).
    Starting points are marked by black bullets.
    \label{Fig-GG-SI-VelocitiesUnpinnedKappa1000}
    }
\end{figure}

Figures \ref{Fig-GG-SI-VelocitiesPinnedOptimum}, \ref{Fig-GG-SI-VelocitiesPinnedKappa1000}, \ref{Fig-GG-SI-VelocitiesUnpinnedOptimum} and \ref{Fig-GG-SI-VelocitiesUnpinnedKappa1000}
show phase space velocity fields and trajectories starting from perturbed initial conditions for three different projections of the reaction coordinates,
for the following cases:
\begin{itemize}
\item for the repression strength ratio $\kappa\simeq31.6$ that optimizes pattern stability with pinning, see Fig. \ref{Fig-GG-SI-VelocitiesPinnedOptimum},
\item for weaker NN repression strength $\kappa=1000$ with pinning, see Fig. \ref{Fig-GG-SI-VelocitiesPinnedKappa1000},
\item for the optimal repression strength ratio $\kappa=100$ that optimizes pattern stability without pinning, see Fig. \ref{Fig-GG-SI-VelocitiesUnpinnedOptimum},
\item for weaker NN repression strength $\kappa=1000$ without pinning, see Fig. \ref{Fig-GG-SI-VelocitiesUnpinnedKappa1000}.
\end{itemize}

\begin{figure}[H]
  \centering
  \includegraphics[width=\textwidth]{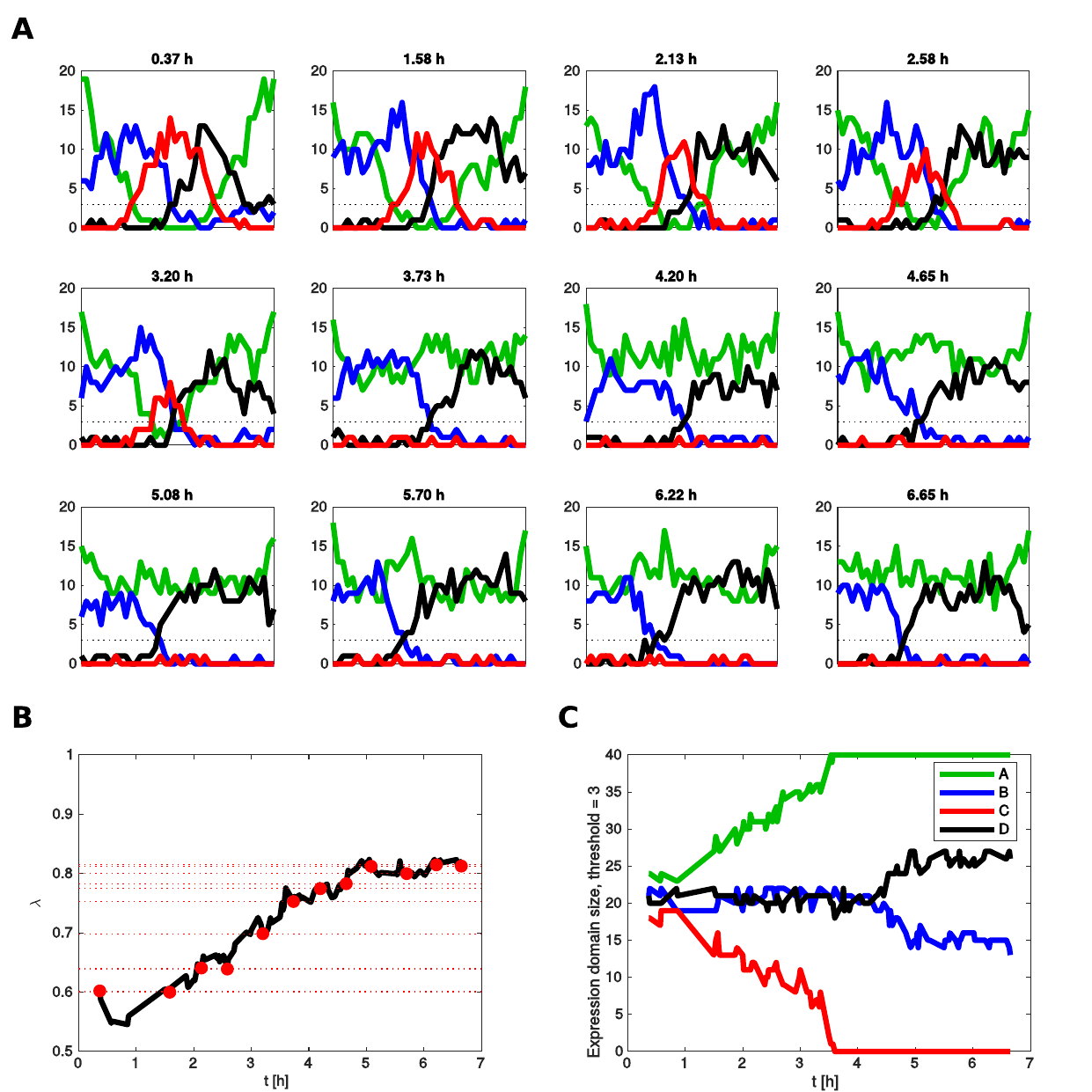}
  \caption{
  \changed{
  \textbf{Example destruction event: Destruction of the \GC domain at optimal NN repression strength ($\kappa = \kappa_{\rm opt}\simeq 30$) in the system with pinning.}
  Panel (A) shows snapshots of gene expression patterns (averaged along the circumference) at the indicated times of the biased simulation trajectory.
  Panel (B) shows the progress of the reaction coordinate $\lambda = \lambda_{AC} + \lambda_{BD}$ along the simulated time trajectory; red bullets correspond to the snapshots in (A) in the same temporal order, red dotted lines show the corresponding $\lambda$ levels.
  Panel (C) shows the corresponding evolution of the expression domain sizes of the four individual genes, here defined by the number of positions at which the (circumference-averaged) expression level is above 3 copies (dotted lines in (A) panels).
  The \GC domain (red) starts to disappear from 2.13~h onwards and is completely destroyed by the \GA domain (green) by 3.73~h. Notice the posterior movement of the boundary between the \GB and \GD domains that appear to precede \GA destruction in this case. Later the \GB-\GD boundary moves back towards the anterior again.
  }
  \label{Fig-Destruction-PinOpt-C}
 }
\end{figure}

\begin{figure}[H]
  \centering
  \includegraphics[width=\textwidth]{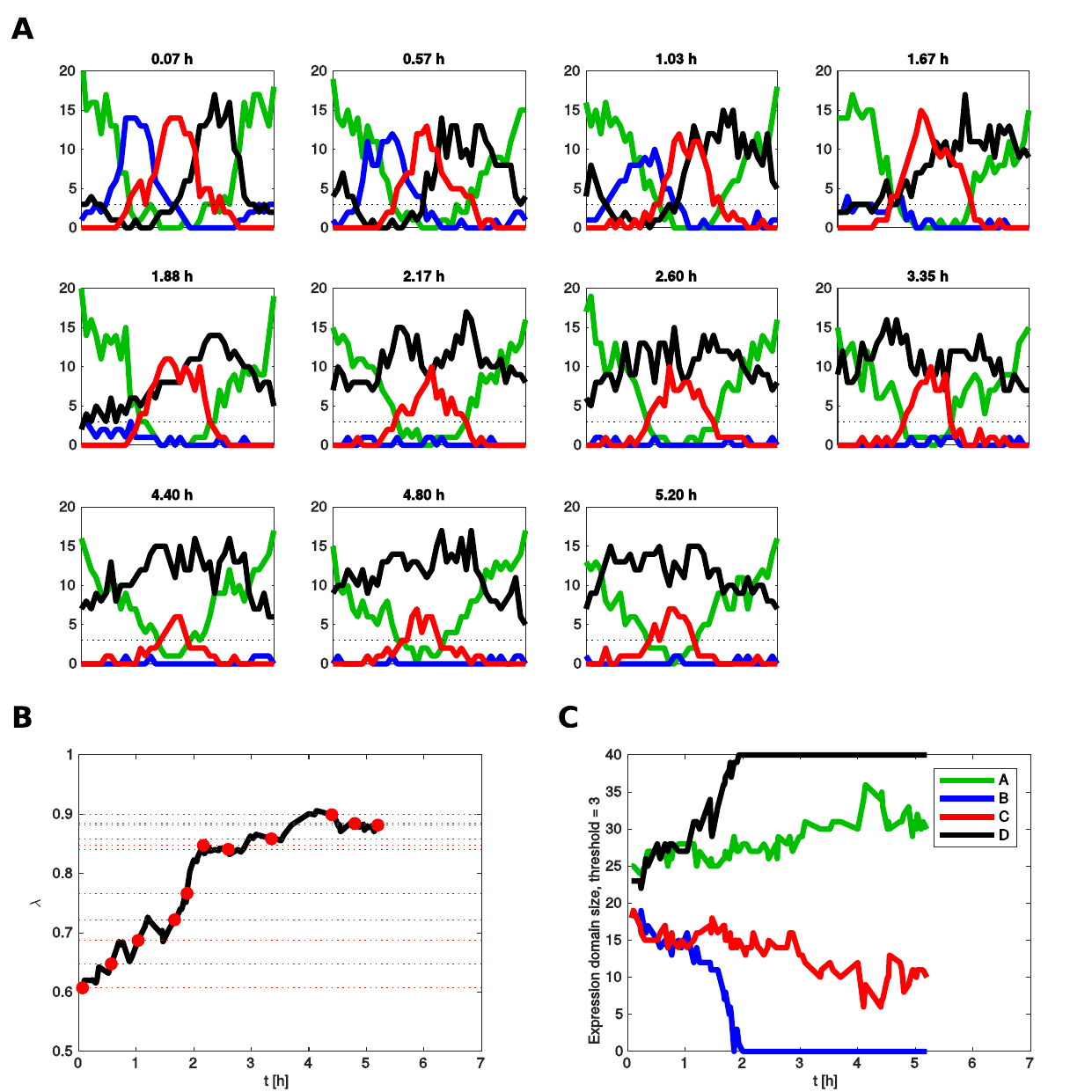}
  \caption{
  \changed{
  \textbf{Example destruction event: Destruction of the \GB domain at optimal NN repression strength ($\kappa = \kappa_{\rm opt}\simeq 30$) in the system with pinning.}
  Panel (A) shows snapshots of gene expression patterns (averaged along the circumference) at the indicated times of the biased simulation trajectory.
  Panel (B) shows the progress of the reaction coordinate $\lambda = \lambda_{AC} + \lambda_{BD}$ along the simulated time trajectory; red bullets correspond to the snapshots in (A) in the same temporal order, red dotted lines show the corresponding $\lambda$ levels.
  Panel (C) shows the corresponding evolution of the expression domain sizes of the four individual genes, here defined by the number of positions at which the (circumference-averaged) expression level is above 3 copies (dotted lines in (A) panels).
  The \GB domain (blue) starts to be suppressed by the \GD domain (black) already from 1.03~h onwards and is completely destroyed by 2.17~h. Notice that the destruction of the \GB domain appears to be accompanied by a compression of the \GC (red) domain by the posterior \GA (green) domain at 0.57~h.
  }
  \label{Fig-Destruction-PinOpt-B}
 }
\end{figure}

\begin{figure}[H]
  \centering
  \includegraphics[width=\textwidth]{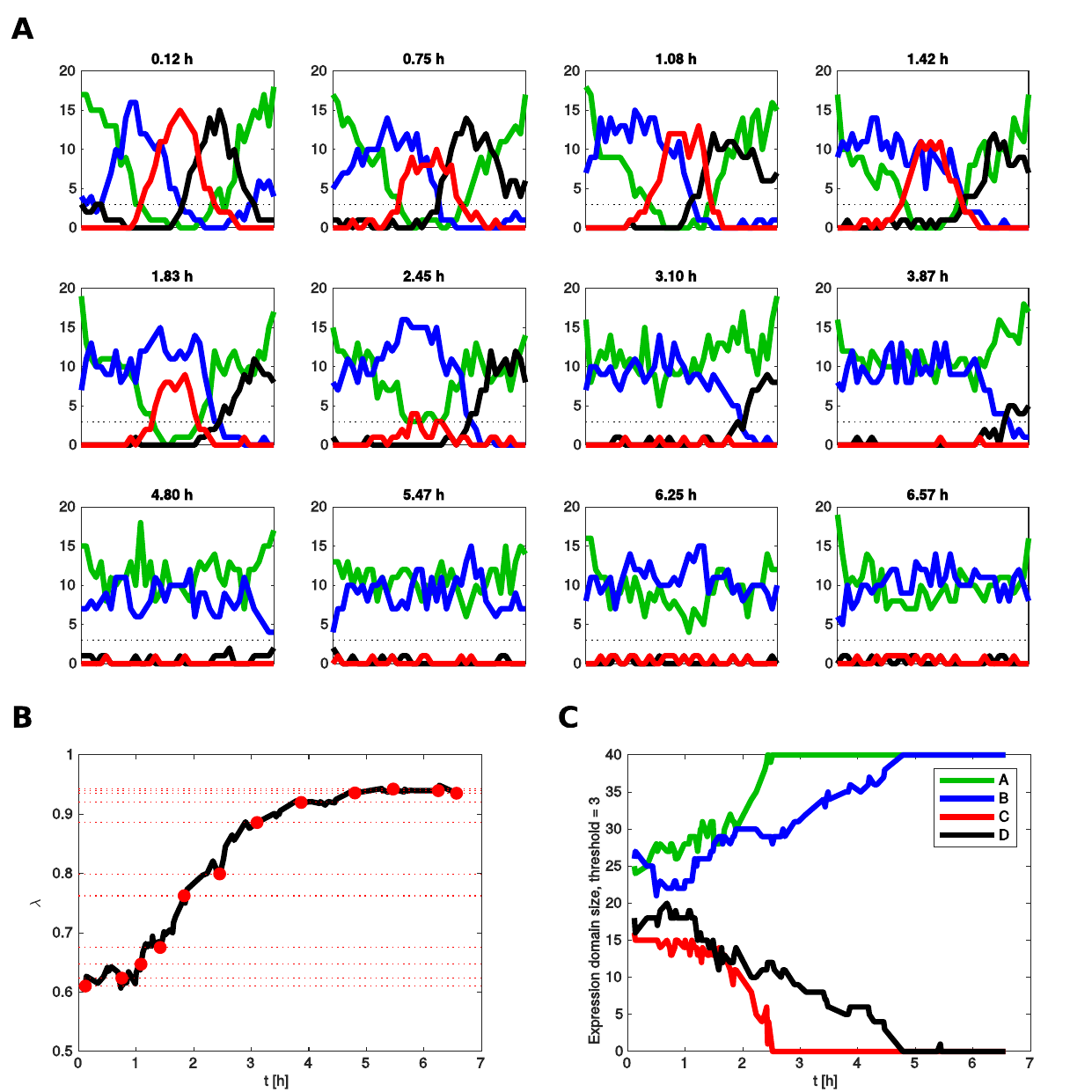}
  \caption{
  \changed{
  \textbf{Example destruction event: Destruction of both the \GC and \GD domains at optimal NN repression strength ($\kappa = \kappa_{\rm opt}\simeq 30$) in the system with pinning.}
  Panel (A) shows snapshots of gene expression patterns (averaged along the circumference) at the indicated times of the biased simulation trajectory. 
  Panel (B) shows the progress of the reaction coordinate $\lambda = \lambda_{AC} + \lambda_{BD}$ along the simulated time trajectory; red bullets correspond to the snapshots in (A) in the same temporal order, red dotted lines show the corresponding $\lambda$ levels.
  Panel (C) shows the corresponding evolution of the expression domain sizes of the four individual genes, here defined by the number of positions at which the (circumference-averaged) expression level is above 3 copies (dotted lines in (A) panels).
  Notice that in this example the \GC (red) domain is under pressure already at 0.75~h, but then rebounds again at 1.08~h, demonstrating the pattern's restoring capability. In parallel, the interface between \GB (blue) and \GD (black) moves posteriorly, which initiates the destruction of the \GC domain at 1.83~h, completed by 3.10~h. Subsequently, the \GB domain also expels the \GD domain across the posterior system boundary, enabled by the fact that \GD is not pinned there (in contrast to \GA).
  }
  \label{Fig-Destruction-PinOpt-C+D}
 }
\end{figure}

\changed{
\section{Example pattern destruction processes}
Figures \ref{Fig-Destruction-PinOpt-C}, \ref{Fig-Destruction-PinOpt-B} and \ref{Fig-Destruction-PinOpt-C+D} summarize typical pattern destruction processes traced in biased simulated time (within the NS-FFS scheme) as the reaction coordinate $\lambda = \lambda_{AC} + \lambda_{BD}$ increases, for the system with pinning and at optimal repression strength ratio $\kappa \simeq 30$ (maximally stable regime).

In all figures, panel (A) shows snapshots of the circumference-averaged expression patterns, initially consisting of four expression domains, some of which are destroyed as $\lambda$ increases. Panels (B) show the corresponding increase of $\lambda$. Panels (C) illustrate how the four gene expression domain sizes (defined as the number of positions at which the expression level of the respective gene is above a predefined threshold of 3 copies) change as biased simulation time and $\lambda$ increase--initially starting from approximately equal domain sizes, some domains are eliminated while their counterpart domains increase in size and ultimately can span the whole system.

More specifically, Fig.~\ref{Fig-Destruction-PinOpt-C} shows an example destruction process in which the expression domain of gene C (red) is destroyed first. Notice that once the C domain is eliminated, this appears to enable subsequent expansion of the D domain (black) at the expense of the opposing B domain (blue).

Fig.~\ref{Fig-Destruction-PinOpt-B} shows an example in which first the B domain (blue) is destroyed by the counterpart D domain (black). This is accompanied by a steady shrinkage of the C domain (red) by the A domain (green); however, the A domain does not manage to destroy the C domain completely by the end of the simulated time interval.

In contrast, Fig.~\ref{Fig-Destruction-PinOpt-C+D} shows a case in which two genes manage to destroy their opposing counterparts' expression domains by that time point: First the A domain grows at the expense of the C domain, while growth of B at the expense of D follows afterwards with reduced speed. Notice that the early asymmetry between B and D is transiently reduced before B starts to push out D ultimately completely, highlighting a (partial) pattern restoration event.
}

\newcommand{\GGSIVelocityFigureScale}{0.33\textwidth}

\begin{figure}[ht!]
  \centering
  \includegraphics[width=0.8\textwidth]{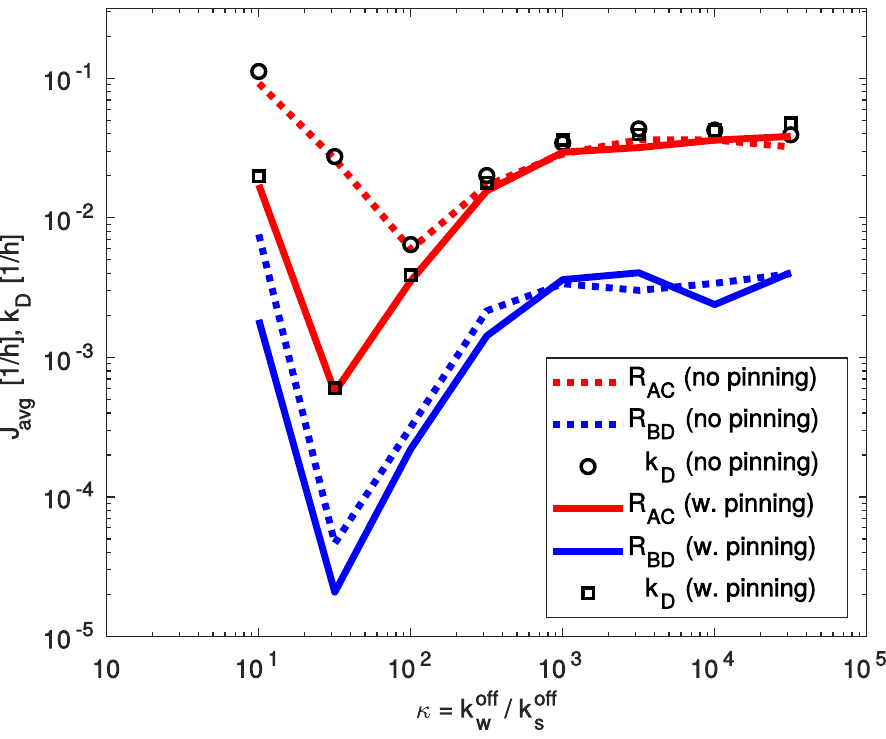}
\caption{\textbf{Pinning affects destruction pathways} 
    We plot here the average probability fluxes from the region of stable patterns $R_S$
    into the different remote basins identified in the ($\lambda_{AC},\lambda_{BD}$) phase space
    as a function of the repression strength ratio $\kappa$ for the systems with and without pinning.
    Here the flux is defined as the average increase per time of the total probability in the basin.
    Basin boundaries and flux quantities are described in detail in \MM.
    Shown are the flux into the basin $R^\dagger_{AC}$, corresponding to destruction of either the \GA or \GC domain (red lines),
    the flux into the basin $R^\dagger_{BD}$, in which either the \GB or \GD domain breaks down (blue lines),
    and the total outflux from $R_S$, which equals the pattern destruction rate $k_D$ (black bullets).    
    Solid lines and triangles show the data for the system with pinning, dashed lines and circles the values for
    the system without pinning.
    Clearly, in both with and without pinning and for all $\kappa$ considered here, $R^\dagger_{AC}$ is the dominant
    fraction of the flux, reflecting that the dominant pathway to destruction is the one that starts with the 
    disappearance of either the \GA or \GC domain.
    Pinning of \GA expression at the system boundaries leads to a pronounced reduction of the flux through this
    pathway for $\kappa=10 - 100$.
  \label{Fig-Fluxes}
}
\end{figure}

\clearpage

\begin{figure}[ht]
  \centering
  \subfigure{
      \label{Fig-FluxesACPinned}
      \includegraphics[width=0.66\textwidth]{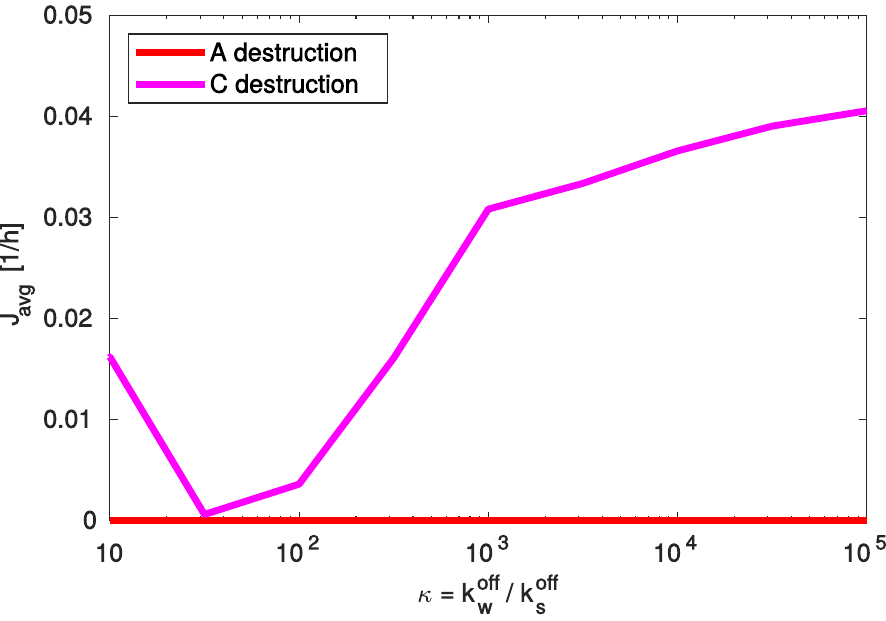}
  }
  \subfigure{
      \label{Fig-FluxesACUnpinned}
      \includegraphics[width=0.66\textwidth]{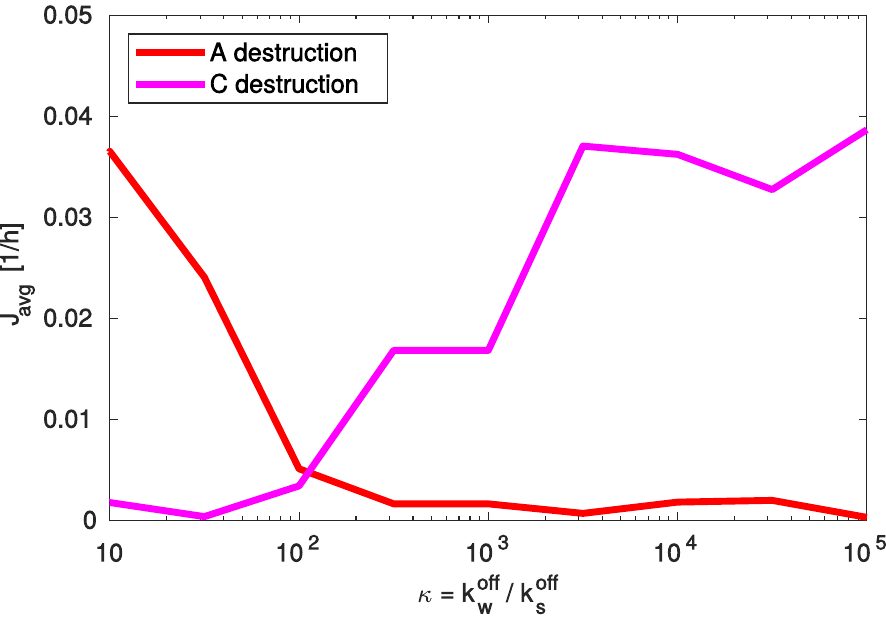}
  }
\caption{\textbf{Pinning shifts the destruction flux balance in the dominant (\GA-\GC) destruction pathway}
    The figure shows the contributions of the \GA-destruction and \GC-destruction pathways to the outflux from $R_S$
     as a function of the repression strength ratio $\kappa$ for the systems with \mysubref{Fig-FluxesACPinned}
    and without \mysubref{Fig-FluxesACUnpinned} pinning.
    See \MM for the definitions of basin boundaries and details of flux calculation.	  	  
    Without pinning and for strong NN repression, the preferred pathway to
    destruction is the one in which the \GA domains are destroyed first, while for weaker coupling (large $\kappa$)
    destruction begins via annihilation of the \GC domain.
    Interestingly, in both cases the flux through the \GC-destruction pathway is minimal at $\kappa\simeq31.6$.
    However, in the system without pinning this value falls into the regime in which the flux through the A-pathway markedly increases.
    Pinning forbids destruction via the \GA pathway and thus dramatically reduces the overall destruction flux
    for low $\kappa$ in the system with pinning, giving rise to the additional enhancement of optimal stability at $\kappa\simeq31.6$.
  \label{Fig-FluxesDominantPathway}
}
\end{figure}

\section{Pinning alters the proportion of destruction pathways}

Both with and without pinning of \GA expression at the system boundaries, pattern 
stability is maximal at an optimal strength of NN repression.
Stability times, however, are significantly higher in the system with pinning.
In order to understand whether this is simply due to the fact that pinning prohibits destruction of the \GA domains or 
due to other pinning-induced effects, we compared the different pathways to destruction by computing 
probability fluxes through distinct reaction pathways (see \MM for details).
The different reaction pathways are defined by the order in which gap gene domains are destroyed.
In our system there are two major pathways: the \GA-\GC destruction pathway (either the \GA or the \GC domain vanishes first) 
and the \GB-\GD destruction pathway (either the \GB or \GD domain vanishes first).
The phase space histograms in Figure~3 of the main text demonstrate that simultaneous destruction of two domains,
corresponding to trajectories that progress diagonally in ($\lambda_{AC},\lambda_{BD}$) space, is highly improbable.
We find that, while in general the \GA-\GC destruction pathway prevails, 
the fact that the \GA-destruction pathway is dominant for $\kappa\leq100$ in the system without pinning 
accounts for the strong enhancement of pattern stability due to pinning.

In Figure \ref{Fig-Fluxes} we plot for different repression strength ratios $\kappa$
the magnitude of average fluxes from the region of intact patterns $R_S$ in the ($\lambda_{AC},\lambda_{BD}$) space 
into the respective neighboring regions that correspond to states in which one expression domain vanished.
The figure reveals that for all $\kappa$ the flux through the \GA-\GC destruction pathway is approximately 
ten times higher than the flux through the \GB-\GD pathway, for systems both with and without pinning.
The figure also shows that pinning indeed reduces the flux through the dominant, 
i.e. \GA-\GC, pathway, most significantly for $\kappa\simeq 10-100$, i.e. around the optimal value $\kappa_{opt}$. 
This gives rise to the pronounced stability enhancement.
The simultaneous reduction of the flux through the \GB-\GD pathway is not relevant for overall stability.

We analysed further the detailed composition of fluxes through the dominant (\GA-\GC) pathway 
by computing the average flux into the regions of destroyed states in $([\GA]_{tot}/N_{tot},[\GC]_{tot}/N_{tot})$ space, see Fig.~\ref{Fig-FluxesDominantPathway}.
As expected, in the systems with pinning the entire flux through the dominant pathway goes into the \GC-destroyed state.
Interestingly, this is also the case for the weakly coupled systems without pinning.
Here the flux into the \GA-destroyed state is clearly dominant over the flux into the \GC-destroyed state for strong NN interaction.
This explains why pinning, which prohibits exit through the \GB-destruction pathway, increases stability in the $\kappa \lesssim 100$ regime.
While the flux through the \GC-destruction pathway is minimal at $\kappa=31.6$ with or without pinning,
in the system without pinning the accessibility of \GA destruction shifts the minimum of the combined flux through both pathways
towards $\kappa=100$, see Fig.~\ref{Fig-Fluxes}.

\FloatBarrier

\newpage
\captionsetup[table]{width=.6\textwidth}
\mbox{} \vfill
\begin{table}[H]
 \centering
 \begin{tabular}{|l|c|lr|}
  \hline
  Quantity & Symbol & Value & Unit\\
  \hline
  \textit{Geometry} &&&\\
  Nuclear radius		& $r_N$		& 2.5	& $\mum$ \\
  Nuclear volume		& $V_N$		& 65.4	& $\mum^3$ \\
  No. of nuclei in axial direction		& $N_z$		& 40	& \\
  - resulting system length			& $L$		& 340	& $\mum$ \\
  No. of nuclei in circumferential direction	& $N_\phi$	& 8	& \\  
  \hline
  \textit{Production / degradation} &&&\\
  Protein production rate	& $\beta$	& 0.20 	& $\unit{s^{-1}}$ \\
  Monomer degradation rate	& $\mu_M$	& 0.05 	& $\unit{s^{-1}}$ \\
  Dimer degradation rate	& $\mu_D$	& 0.005 & $\unit{s^{-1}}$ \\
  - resulting effective degr. rate & $\mu_{eff}$	& 0.0095 & $\unit{s^{-1}}$ \\
  \hline
  \textit{Binding / unbinding} &&&\\
  Intranuclear diffusion const. & $D_N$		& 3.2	& $\mum^2/s$\\
  Repressor target site radius	& $\sigma_R$	& 0.5	& $\mum$ \\    
  - resulting (diff. ltd.) repressor on-rate 	& $\kRon$	& 20.1	& $\mum^3/s$ \\
  Standard (strong) repressor off-rate		& $\kRoffS$	& 0.06 	& $\unit{s^{-1}}$ \\
  Weak repressor off-rate			& $\kRoffW$	& varied & $\geq k^{\rm R,s}_{off}$ \\
  Monomer protein radius			& $\sigma_M$	& 0.05  & $\mum$ \\
  - resulting (diff. ltd.) dimerization forward rate & $\kDon$	& 4.0	& $\mum^3/s$ \\
  Dimerization backward rate			& $\kDoff$	& 0.062	& $\mum^3/s$ \\
  \hline
  \textit{Internuclear diffusion} &&&\\
  Standard internuclear diffusion const.	& $D$	& 1.0	& $\mum^2/s$ \\
  Internuclear lattice distance			& $l$	& 8.5	& $\mum$ \\
  \hline
 \end{tabular}

 \caption{The standard parameters of the simulated model of four mutually repressing genes
  prearranged in the ``alternating cushions'' pattern on a cylindrical lattice of expressing
  nuclei.
  \label{Tab-GG-SI-Parameters}
}
\end{table}
\vfill \mbox{}
\captionsetup[table]{width=\textwidth}

\newpage
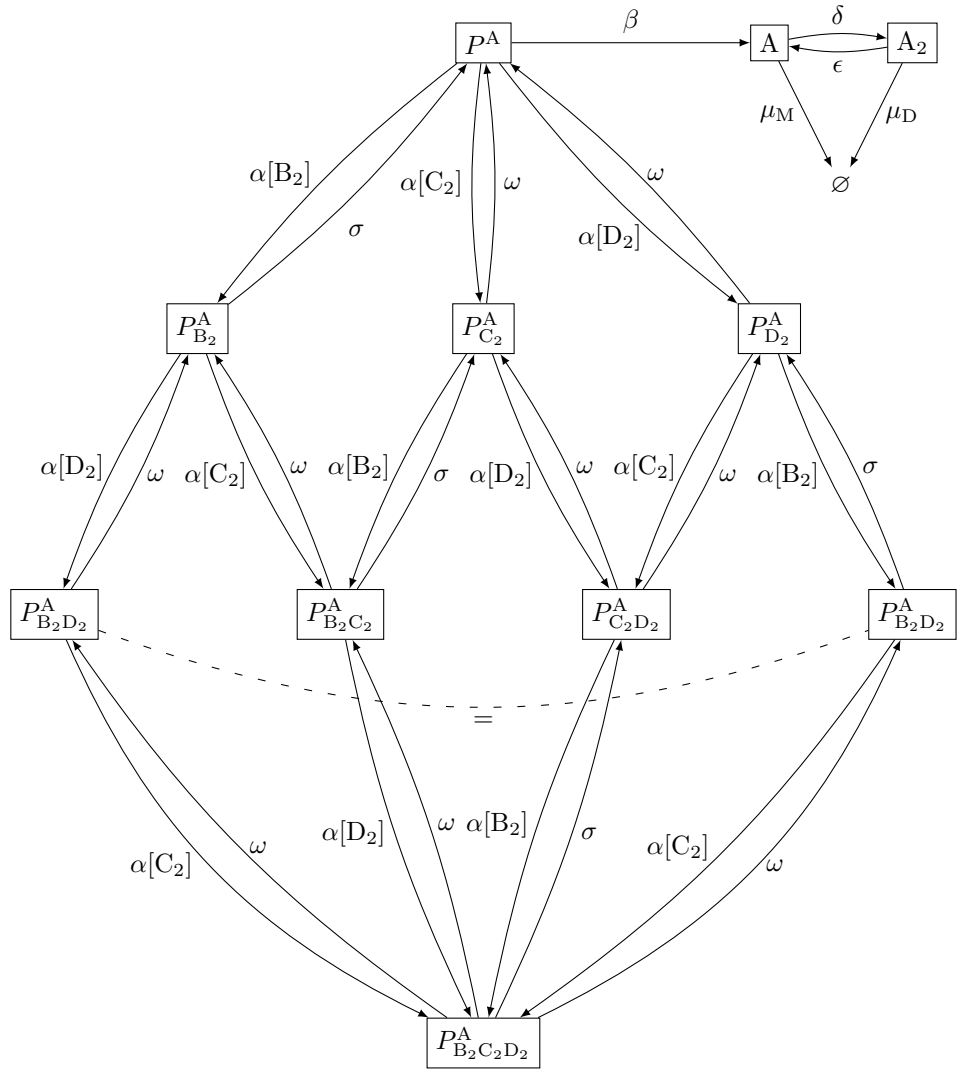
\begin{figure}[H]
 \centering
 \begin{tikzpicture}[align=center, xscale=1.9, yscale=1.9]
 \node[draw,rectangle] (P0) at (0,6)   {$\PromA$};

 \node[draw,rectangle] (A)  at (2,6)   {${\rm A}$};
 \node[draw,rectangle] (A2) at (3,6)   {$\Adim$};

 \node                 (NIL) at (2.5,5) {$\varnothing$};

 \node[draw,rectangle] (PB) at (-2,4)  {$\PromA_{\Bdim}$};
 \node[draw,rectangle] (PC) at (0,4)   {$\PromA_{\Cdim}$};
 \node[draw,rectangle] (PD) at (2,4)   {$\PromA_{\Ddim}$};

 \node[draw,rectangle] (PBDl) at (-3,2) {$\PromA_{\Bdim\Ddim}$};
 \node[draw,rectangle] (PBC)  at (-1,2) {$\PromA_{\Bdim\Cdim}$};
 \node[draw,rectangle] (PCD)  at (1,2)  {$\PromA_{\Cdim\Ddim}$};
 \node[draw,rectangle] (PBDr) at (3,2)  {$\PromA_{\Bdim\Ddim}$};

 \node[draw,rectangle] (PBCD) at (0,-1)  {$\PromA_{\Bdim\Cdim\Ddim}$};

 \draw[-latex] (P0) to[bend right=7] node[left=0.1cm]  {\kOn $[\Bdim]$}	(PB);
 \draw[-latex] (PB) to[bend right=7] node[below=0.3cm] {\kOffW}   	(P0);
 \draw[-latex] (P0) to[bend right=7] node[left]  {\kOn $[\Cdim]$}		(PC);
 \draw[-latex] (PC) to[bend right=7] node[right] {\kOffS}  		(P0);
 \draw[-latex] (P0) to[bend right=7] node[below=0.3cm]  {\kOn $[\Ddim]$}	(PD);
 \draw[-latex] (PD) to[bend right=7] node[right] {\kOffW}		(P0);

 \draw[-latex] (PB)   to[bend right=7] node[left]  {\kOn $[\Ddim]$} 	(PBDl);
 \draw[-latex] (PBDl) to[bend right=7] node[right] {\kOffW}		(PB);
 \draw[-latex] (PB)   to[bend right=7] node[left]  {\kOn $[\Cdim]$}	(PBC); 
 \draw[-latex] (PBC)  to[bend right=7] node[right] {\kOffS}		(PB);

 \draw[-latex] (PC)   to[bend right=7] node[left]  {\kOn $[\Ddim]$}	(PCD);
 \draw[-latex] (PCD)  to[bend right=7] node[right] {\kOffW}		(PC);
 \draw[-latex] (PC)   to[bend right=7] node[left]  {\kOn $[\Bdim]$} 	(PBC); 
 \draw[-latex] (PBC)  to[bend right=7] node[right] {\kOffW}		(PC);

 \draw[-latex] (PD)   to[bend right=7] node[left]  {\kOn $[\Bdim]$}	(PBDr);
 \draw[-latex] (PBDr) to[bend right=7] node[right] {\kOffW}		(PD);
 \draw[-latex] (PD)   to[bend right=7] node[left]  {\kOn $[\Cdim]$}	(PCD); 
 \draw[-latex] (PCD)  to[bend right=7] node[right] {\kOffS}		(PD);

 \draw[-latex] (PBDl) to[bend right=20] node[left=0.1cm] {\kOn $[\Cdim]$} (PBCD);
 \draw[-latex] (PBCD) to[bend right=-10] node[right] {\kOffS}		(PBDl);
 \draw[-latex] (PBC)  to[bend right=7] node[left]  {\kOn $[\Ddim]$}	(PBCD); 
 \draw[-latex] (PBCD) to[bend right=7] node[right] {\kOffW}		(PBC);

 \draw[-latex] (PCD)  to[bend right=7] node[left]  {\kOn $[\Bdim]$}	 (PBCD);
 \draw[-latex] (PBCD) to[bend right=7] node[right] {\kOffW}		 (PCD);
 \draw[-latex] (PBDr) to[bend right=-10] node[left=0.1cm] {\kOn $[\Cdim]$} (PBCD); 
 \draw[-latex] (PBCD) to[bend right=20] node[right] {\kOffS}		 (PBDr);

 \draw[loosely dashed] (PBDl) to[bend right=20] node[below] {$=$}	(PBDr);

 \draw[-latex] (P0) to node[above] {$\beta$} (A);
 \draw[-latex] (A)  to[bend left=10] node[above] {$\delta$} (A2);
 \draw[-latex] (A2) to[bend left=10] node[below] {$\epsilon$} (A);

 \draw[-latex] (A)  to node[left]  {$\mu_M$} (NIL);
 \draw[-latex] (A2) to node[right] {$\mu_D$} (NIL);

 \end{tikzpicture}

\caption{\textbf{Reaction network.} This schematic shows the set of reactions that affect production and degradation of a single
	gap gene species A. The strong repressor of A is denoted by B, the weak interaction partners by C and D. For each
        species, X denotes the monomer, $\Xdim$ the dimer. For easy readability here we abbreviate:
        $\alpha \equiv \kRon =$ diffusion limited repressor binding rate;
	$\sigma \equiv \kRoffS =$ next-nearest neighbor / strong repressor unbinding rate;
	$\omega \equiv \kRoffW =$ nearest-neighbor / weak repressor unbinding rate;
        $\delta \equiv \kDon =$ dimerization forward rate;
	$\epsilon \equiv \kDoff =$ dimerization backward rate.
    \changed{
	The schematic shows the reactions for the promoter of species A and
    its protein products, which we denote by ${\rm \left(A \vert\vert C,B,D\right)}$,
    but holds similarly for all other combinations of regulated and regulating species,
    ${\rm \left(B \vert\vert D,A,C\right)}$,  ${\rm \left(C \vert\vert A,B,D\right)}$
    and ${\rm \left(D \vert\vert B,A,C\right)}$; the order of the regulating species is not strictly alphabetical because the strongly repressing species is indicated first.}
	\label{Fig-GG-SI-ReactionNetwork}
}
\end{figure}


